\documentclass[twocolumn]{aastex62}

\usepackage{etoolbox}
\makeatletter\patchcmd{\@makechapterhead}{\vspace*{50\p@}}{}{}{}
\patchcmd{\@makeschapterhead}{\vspace*{50\p@}}{}{}{}
\makeatother

\newcommand{\kms}{km\,s$^{-1}$}

\def\ergs{erg\,s$^{-1}$}

\def\M{M$_{\odot}$}

 \def\Mej{$M_{\rm ej}$}

\def\ergs{erg\,s$^{-1}$}

\def\ca{[\ion{Ca}{2}]\,$\lambda$7300}
\def\o{[\ion{O}{1}]\,$\lambda$6300}
\def\mg{\ion{Mg}{1}]\,$\lambda$4571}
\def\oi{\ion{O}{1}\,$\lambda$7774}

\graphicspath{{./}{figures/}}

\shorttitle{Nebular spectra of SLSNe}
\shortauthors{Nicholl et al.}

\begin{document}

\title{Nebular-phase spectra of superluminous supernovae: physical insights from observational and statistical properties}

\correspondingauthor{Matt Nicholl}
\email{mrn@roe.ac.uk}

\author[0000-0002-2555-3192]{Matt Nicholl}
\affiliation{Harvard-Smithsonian Center for Astrophysics, 60 Garden Street, Cambridge, MA, 02138, USA}
\affiliation{Institute for Astronomy, University of Edinburgh, Royal Observatory, Blackford Hill, Edinburgh EH9 3HJ, UK}

\author[0000-0002-9392-9681]{Edo Berger}
\affiliation{Harvard-Smithsonian Center for Astrophysics, 60 Garden Street, Cambridge, MA, 02138, USA}

\author{Peter K.~Blanchard}
\affiliation{Harvard-Smithsonian Center for Astrophysics, 60 Garden Street, Cambridge, MA, 02138, USA}

\author{Sebastian Gomez}
\affiliation{Harvard-Smithsonian Center for Astrophysics, 60 Garden Street, Cambridge, MA, 02138, USA}

\author{Ryan Chornock}
\affiliation{Astrophysical Institute, Department of Physics and Astronomy, 251B Clippinger Lab, Ohio University, Athens, OH 45701, USA}

\begin{abstract}

We study the spectroscopic evolution of superluminous supernovae (SLSNe) later than 100 days after maximum light. We present new data for Gaia16apd and SN\,2017egm, and analyse these with a larger sample comprising 41 spectra of 12 events. The spectra become nebular within 2-4 $e$-folding times after light curve peak, with the rate of spectroscopic evolution correlated to the light curve timescale. Emission lines are identified with well-known transitions of oxygen, calcium, magnesium, sodium and iron. SLSNe are differentiated from other Type Ic SNe by a prominent \oi\ line and higher-ionisation states of oxygen. The iron-dominated region around 5000\,\AA\ is more similar to broad-lined SNe Ic than to normal SNe Ic. Principal Component Analysis shows that 5 `eigenspectra' capture $\gtrsim70$\% of the variance, while a clustering analysis shows no clear evidence for multiple SLSN sub-classes. Line velocities are 5000--8000\,\kms, and show stratification of the ejecta. \oi\ likely arises in a dense inner region that also produces calcium emission, while \o\ comes from further out until 300--400 days. The luminosities of \oi\ and \ion{Ca}{2} suggest significant clumping, in agreement with previous studies. Ratios of \ca/\o\ favour progenitors with relatively massive helium cores, likely $\gtrsim 6$\,\M, though more modelling is required here. SLSNe with broad light curves show the strongest \o, suggesting larger ejecta masses. We show how the inferred velocity, density and ionisation structure point to a central power source.

\end{abstract}


\section{Introduction} \label{s:intro}

In recent years, much progress has been made in characterising the new population of hydrogen-poor superluminous supernovae (SLSNe). These events first started to appear in wide-field, untargeted transient surveys \citep{qui2011,chom2011,gal2012}, and now comprise a few percent of the supernovae classified each year. SLSNe have a median luminosity $M \approx -21$\,mag \citep{nic2015b,lun2018,dec2018}, making them up to two orders of magnitude brighter than typical supernovae (SNe), and sparking intense interest in the unexpectedly diverse outcomes of massive star deaths.

SLSNe are now generally classified spectroscopically rather than photometrically. Their unique early-time spectra show a series of broad \ion{O}{2} absorption lines superposed on a blue continuum, indicating hot, ionized ejecta. As they expand and cool, the spectra evolve to resemble those of lower luminosity Type Ic SNe \citep{pas2010,ins2013}, though there may be subtle differences \citep{Liu&Modjaz2017,qui2018}. Thus SLSNe are best described as Type Ic SNe (explosions of stripped massive stars lacking hydrogen and helium) that manage to stay hot ($T \gtrsim 10,000$\,K) over several weeks or months, allowing them to attain higher luminosities.

The favoured source of additional heating is a central engine, such as the spin-down of a rapidly rotating magnetar \citep{kas2010,woo2010}. It has been shown that this model can reproduce both the light curves \citep{ins2013,cha2013,nic2017c} and early spectra \citep{des2012,how2013,maz2016} of SLSNe. Maximum-light observations, however, only probe the outer layers of the ejecta, because of a large optical depth to the centre. 

Over time, recombination reduces the optical depth, and by $t_{\rm neb} \sim 360\,{\rm d} (M_{\rm ej}/10$\,\M$)^{1/2} (v/10^4$\,\kms$)^{-1}$ after the explosion, where \Mej\ is the ejecta mass and $v$ is the expansion velocity, the ejecta become largely transparent \citep[see review by][]{jer2017b}. Once this so-called nebular phase is reached, it is possible to directly probe with spectroscopy the conditions at the centre of the explosion, to constrain the composition and distribution of material and to search for any hydrodynamic signatures of the explosion mechanism. However, such observations are challenging because the SN will have faded substantially over the time $t_{\rm neb}$, and hence nebular spectroscopy is only currently possible for SLSNe at $z \lesssim 0.2$.

\citet{nic2016c} and \citet{jer2017a}, following earlier work by \citet{mil2013}, showed that the few existing nebular spectra of SLSNe resemble those of broad-lined Type Ic SNe -- thought to be engine-driven explosions and sometimes accompanied by long gamma-ray bursts (LGRBs). This was interpreted as evidence for a similar internal structure for SLSNe and LGRB SNe, suggesting that they may arise from similar progenitors.
They also inferred large ejected masses of $\gtrsim 10$\,\M\ for some SLSNe. Since those initial nebular observations, the available sample of SLSNe with nebular phase observations has increased as a number of recent nearby SLSNe have evolved to sufficiently late times. For example, \citet{qui2018} recently published a large spectroscopic sample of SLSNe from the Palomar Transient Factory (PTF), including many late-phase spectra.

In this paper, we undertake a systematic observational study of SLSN nebular spectra. We present new data for Gaia16apd/SN\,2016eay ($z=0.1013$) and SN\,2017egm ($z=0.0307$), two nearby SLSNe that have been well-studied at earlier phases \citep{yan2016,nic2017,kan2016b,nic2017d,bos2018}, and combine this with all available published spectra of SLSNe obtained more than 100 days after maximum light.

We describe our observations and the processing of new and archival data in Section \ref{s:data}. In Section \ref{s:obs}, we present the spectral sequence and mean population properties, including line identifications and comparisons to Type Ic SNe. We then apply machine learning techniques to characterise the diversity of SLSNe in Section \ref{s:mach}. The line profiles are used to investigate the distribution of material in Section \ref{s:profiles}, and their luminosities and ratios to infer ejecta conditions in Section \ref{s:lums}. We discuss the implications of our findings in the context of SLSN models in Section \ref{s:diss} and summarise our conclusions in Section \ref{s:conc}.

\section{Data}
\label{s:data}

\subsection{Gaia16apd}

We observed Gaia16apd on 2017-05-17 (399 rest-frame days after peak luminosity) with GMOS on Gemini North \citep{hook2004}. We used the R150 grating with a central wavelength of 7000\,\AA, and the GG455 blocking filter to prevent second-order contamination. The data were reduced using the Gemini pipeline in \textsc{Pyraf}, to apply bias and flat-field corrections, determine a wavelength solution, and calibrate the relative flux with a standard star observed in the same setup. Narrow emission lines from the host galaxy were subtracted after fitting Gaussian profiles.


\begin{figure}
\centering
\includegraphics[width=8.25cm]{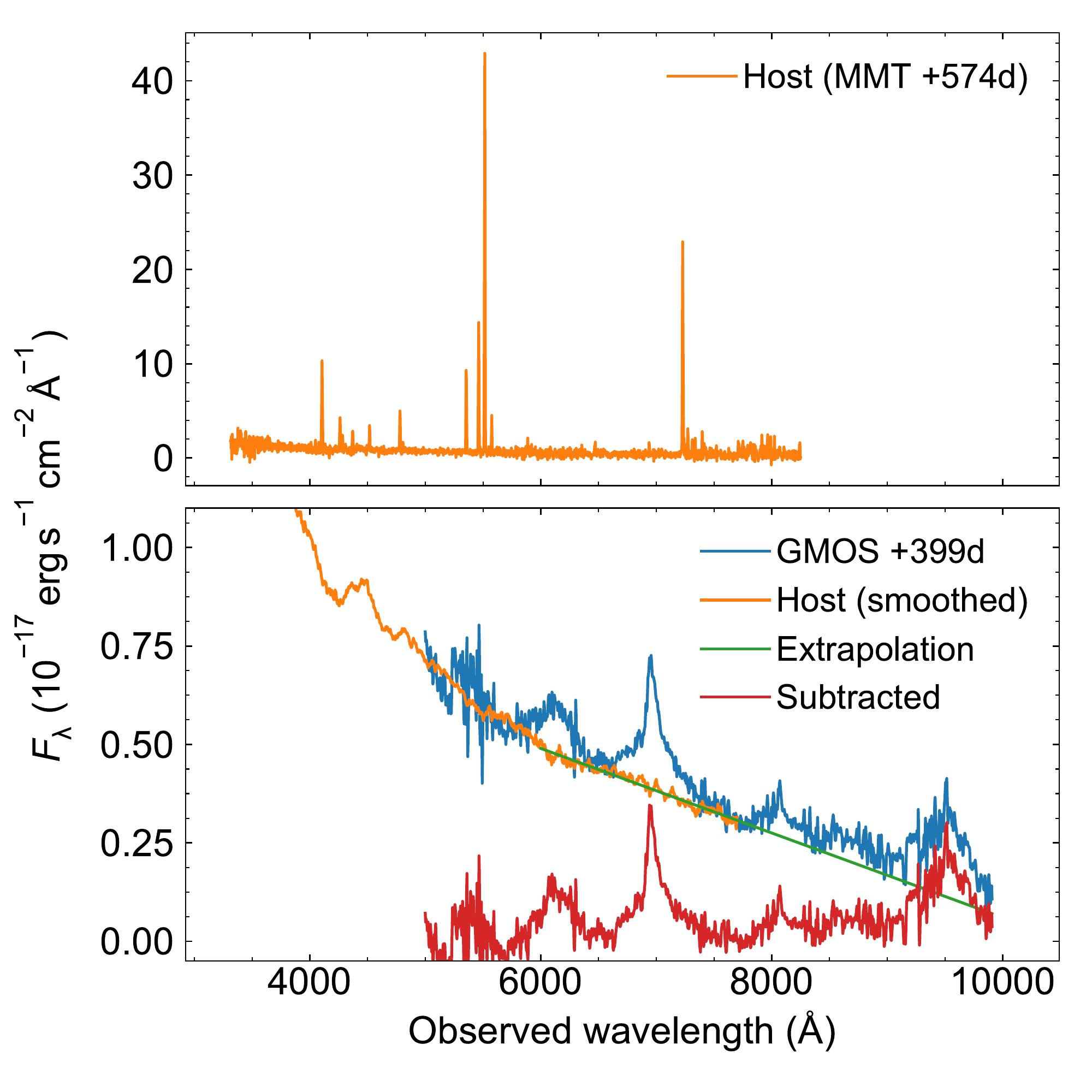}
\caption{Gemini/GMOS spectrum of Gaia16apd, a SLSN at $z=0.1013$, at 399 rest-frame days after maximum light. We subtract a host spectrum obtained with MMT/BlueChannel, with a linear extrapolation redward of 8000\,\AA.
}
\label{f:16apd-spec}
\end{figure}


The spectrum is shown in Figure \ref{f:16apd-spec}, and clearly contains contributions from both Gaia16apd (broad lines) and its host (continuum). 
After Gaia16apd had faded below detectability, as indicated by late-time imaging (P.~Blanchard et al., in preparation), we obtained a host galaxy spectrum on 2017-11-26 (574 rest-frame days) using the Blue Channel spectrograph at MMT \citep{schmidt1989}, with the 300 lines/mm grating. We reduced the spectrum using standard \textsc{Pyraf} packages. This spectrum is also plotted in Figure \ref{f:16apd-spec}, and shows no signs of residual SN contamination. 

Finally, we subtracted the host spectrum from the earlier SN spectrum.
The Blue Channel spectrum does not extend to the redder wavelengths covered by GMOS, so we used a simple linear extrapolation of the host galaxy light between 8000-10000\,\AA. The result of the subtraction is shown in Figure \ref{f:16apd-spec}.

\subsection{SN\,2017egm}

SN\,2017egm is the second-closest known SLSN, at $z=0.0307$, and is therefore ideal for a late-time study\footnote{The closest is SN\,2018bsz at $z=0.0267$ \citep{anderson2018}, but that event is too young for a nebular study at the time of writing}. We obtained multiple spectra spanning 126--353 days after maximum in the rest-frame. Observations were carried out using the Ohio State Multi-Object Spectrograph \citep[OSMOS;][]{martini2011} on the 2.4m Hiltner telescope at MDM observatory, Blue Channel and Binospec on MMT, and GMOS on Gemini North. Binospec\footnote{https://www.cfa.harvard.edu/mmti/binospec.html} is a newly-commissioned imaging spectrograph with dual fields of view. We used the 270 lines/mm grating, providing spectral coverage from 3900--9240\,\AA\ at $\sim 6$\,\AA\ resolution. 

OSMOS and Blue Channel data were de-biased, flat-fielded, wavelength- and flux-calibrated in \textsc{iraf}/\textsc{Pyraf}, while for the Gemini data we used the GMOS pipeline. Binospec data were reduced with a dedicated pipeline, based on the pipeline for the MMT Magellan Infrared Spectrograph \citep{chilingarian2013}.

Almost uniquely among SLSNe, SN\,2017egm occurred in a massive spiral galaxy, rather than the metal-poor dwarfs that typically host these explosions \citep{nic2017d,bos2018,izzo2018,chen2017}. Host galaxy light was removed when extracting the one-dimensional spectra using low-order polynomials along the direction of the slit, fitted to regions of the two-dimensional host spectrum on either side of the SN spectral trace. While the host light profile appears smooth on the spectrograph, we cannot exclude the possibility that a bright region underlying the SLSN location could be contributing some excess continuum. However, this can only be tested once SN\,2017egm has completely faded.

\begin{figure}
\centering
\includegraphics[width=8.25cm]{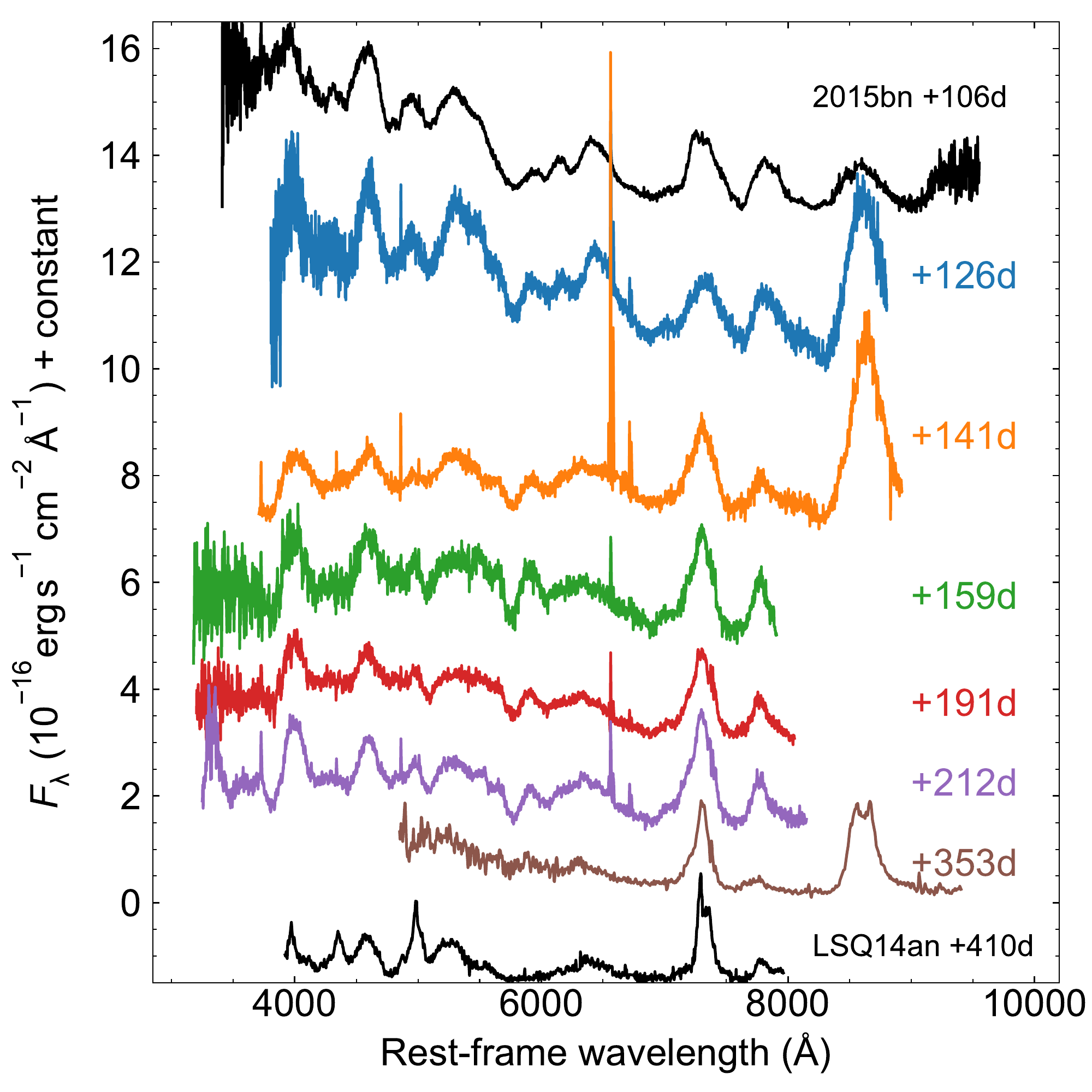}
\caption{Late-time spectra of SN\,2017egm, a SLSN at $z=0.0307$, obtained with MDM/OSMOS (126 days), MMT/Binospec (141 days), MMT/BlueChannel (159--212 days), and GeminiGMOS (353 days). Phases are given relative to maximum light and are in the SN rest-frame.}
\label{f:17egm-spec}
\end{figure}

The resultant spectra are shown in Figure 
\ref{f:17egm-spec}. The earliest spectrum at 126 rest-frame days still shows a relatively blue continuum. The spectrum evolves quickly between the first two epochs, but slowly thereafter. Initially, the spectrum is very similar to SN\,2015bn \citep{nic2016b}, but as it evolves it is better matched by LSQ14an \citep{jer2017a}.

\subsection{SLSNe from PTF}\label{s:ptf}

\citet{qui2018} recently published a spectroscopic sample of all SLSNe classified in PTF. While their rigorous analysis of the spectral properties and line evolution focuses only on the photospheric phase, their sample also includes many late-time spectra. The authors have made these spectra available via WISeREP \citep{yar2012}, with various levels of processing. We downloaded the spectra that have had host emission lines removed, but with no further smoothing. We carry out our own smoothing simply to ensure that all spectra in our analysis are subject to consistent processing (see section \ref{s:smooth}).

Many of the spectra of PTF events show clear evidence of host galaxy continuum, similar to that shown for Gaia16apd in Figure \ref{f:16apd-spec}. Fortunately, most of these galaxies were observed by \citet{per2016}, after the SLSNe had faded. These host spectra are also available from WISeREP. The galaxies are typically faint and the host spectra are often noisy. To avoid adding additional noise to the SLSN spectra, we median-filtered the host spectra over 100\,\AA\ windows (after manually removing any emission lines), leaving only smooth continua. While this heavy smoothing will result in a failure to remove any narrow stellar absorption features present in the SLSN spectra, no such features are visible above the noise, nor would they have a significant impact on our analysis. Moreover, we do not expect strong absorption features in the host spectra, since these galaxies are dominated by young stellar populations \citep{lun2014,lel2015,per2016,chen2016b,ang2016,schu2018}. Both the SLSN and host spectra were scaled to match photometry from \citet{dec2018} and \citet{per2016} before finally subtracting the host continua from the SLSN data.

\begin{table}[t!]
\caption{List of SLSNe and spectra used in this paper, with redshifts, light curve decline rates, spectroscopic phases and original data sources.}
\label{t:sample}
\centering
\begin{tabular}{ccccc}
SLSN  & $z$$^a$ & $t_d$$^b$ & $t$$^c$ & Reference \\
\hline
SN2017egm & 0.0307 & 60 & 126 & This work \\
 &  &  & 141 & This work \\
 &  &  & 159 & This work \\
 &  &  & 191 & This work \\
 &  &  & 212 & This work \\
 &  &  & 353 & This work \\
Gaia16apd & 0.1013 & 43 & 169 & \citet{nic2017} \\
 &  &  & 198 & \citet{kan2016b} \\
 &  &  & 399 & This work \\
SN2015bn & 0.1136 & 80 & 106 & \citet{nic2016b} \\
 &  &  & 243 & \citet{nic2016b} \\
 &  &  & 256 & \citet{nic2016c} \\
 &  &  & 295 & \citet{nic2016c} \\
 &  &  & 315 & \citet{jer2017a} \\
 &  &  & 343 & \citet{nic2016c} \\
 &  &  & 392 & \citet{nic2016c} \\
PTF12dam & 0.1075 & 73 & 171 & \citet{nic2013} \\
 &  &  & 221 & \citet{nic2013} \\
 &  &  & 269 & \citet{qui2018} \\
 &  &  & 324 & \citet{qui2018} \\
 &  &  & 509 & \citet{chen2015} \\
LSQ14an & 0.1637 & 85 & 111 & \citet{ins2017} \\
 &  &  & 149 & \citet{ins2017}  \\
 &  &  & 365 & \citet{jer2017a} \\
 &  &  & 410 & \citet{jer2017a} \\
SN2007bi & 0.1279 & 85 & 367 & \citet{gal2009}; \\
 &  &  & 471 & \citet{you2010} \\
PTF10hgi & 0.0982 & 36 & 241 & \citet{qui2018} \\
 &  &  & 315 & \citet{qui2018} \\
PTF10nmn & 0.1236 & 82 & 182 & \citet{qui2018} \\
 &  &  & 213 & \citet{qui2018} \\
 &  &  & 321 & \citet{qui2018} \\
 &  &  & 527 & \citet{qui2018} \\
PTF09cnd & 0.2585 & 75 & 121 & \citet{qui2018} \\
PTF10vwg & 0.190 & 40 & 147 & \citet{qui2018} \\
PTF11hrq & 0.0571 & 67 & 159 & \citet{qui2018} \\
 &  &  & 350 & \citet{qui2018} \\
PTF12hni & 0.1056 & 27 & 298 & \citet{qui2018} \\
\hline
\end{tabular}
\begin{flushleft}
$^a$Redshift\\
$^b$Light curve $e$-folding timescale, in rest-frame days\\
$^c$Time at which spectrum was obtained, in rest-frame days since light curve maximum
\end{flushleft}
\end{table}

\subsection{Other SLSNe from the literature}

In addition to our new data and the host-subtracted PTF SLSN spectra, we include in our sample all published late-time spectra of SLSNe available from WISeREP \citep{yar2012} and the Open Supernova Catalog \citep{gui2017}. These SLSNe (and the publications from which their spectra are taken) are: SN 2007bi \citep{gal2009,you2010}, PTF12dam \citep{nic2013,chen2015}, SN\,2015bn \citep{nic2016b,nic2016c,jer2017a}, LSQ14an \citep{ins2017,jer2017a}, and Gaia16apd \citep{nic2017,kan2016b}. In total, our sample comprises 41 spectra of 12 SLSNe. We did not include SLSNe for which no SN features were visible above host galaxy light: SN\,2011ke \citep{qui2018} and PS16aqy \citep{bla2018}. Two additional events with late-time spectra were not included because the wavelength range of their spectra did not cover the main lines of interest: PS1-14bj \citep{lun2016} and PTF09atu \citep{qui2018}. We also exclude the few SLSNe with a dominant H$\alpha$ line in their spectra. These have been extensively analysed by \citet{yan2015,yan2017}. We exclude these events because the hydrogen emission most likely probes circumstellar material, rather than the interior of the explosion, while obscuring important SN features. A list of all spectra is given in Table \ref{t:sample}.

\subsection{Processing}\label{s:smooth}

Prior to our analysis, we corrected all spectra for Galactic extinction and de-redshifted them to the rest-frame. We used the \citet{schlaf2011} calibration for dust extinction along the line of sight. Values for $E(B-V)$ and $z$ were obtained from the Open Supernova Catalog \citep{gui2017} or directly from the file headers of the \citet{qui2018} spectra. We assume that extinction in the SLSN host galaxies is negligible \citep{lun2014,lel2015,nic2017c}.

The signal-to-noise ratio varies significantly between the spectra in our sample. In some cases, noisy or over-sampled spectra make it difficult even to identify real spectral features. Additionally, it greatly facilitates a statistical analysis to have all spectra on a common wavelength grid. We therefore apply a two-step interpolation and smoothing procedure to our spectra. 

\begin{figure}
\centering
\includegraphics[width=8.25cm]{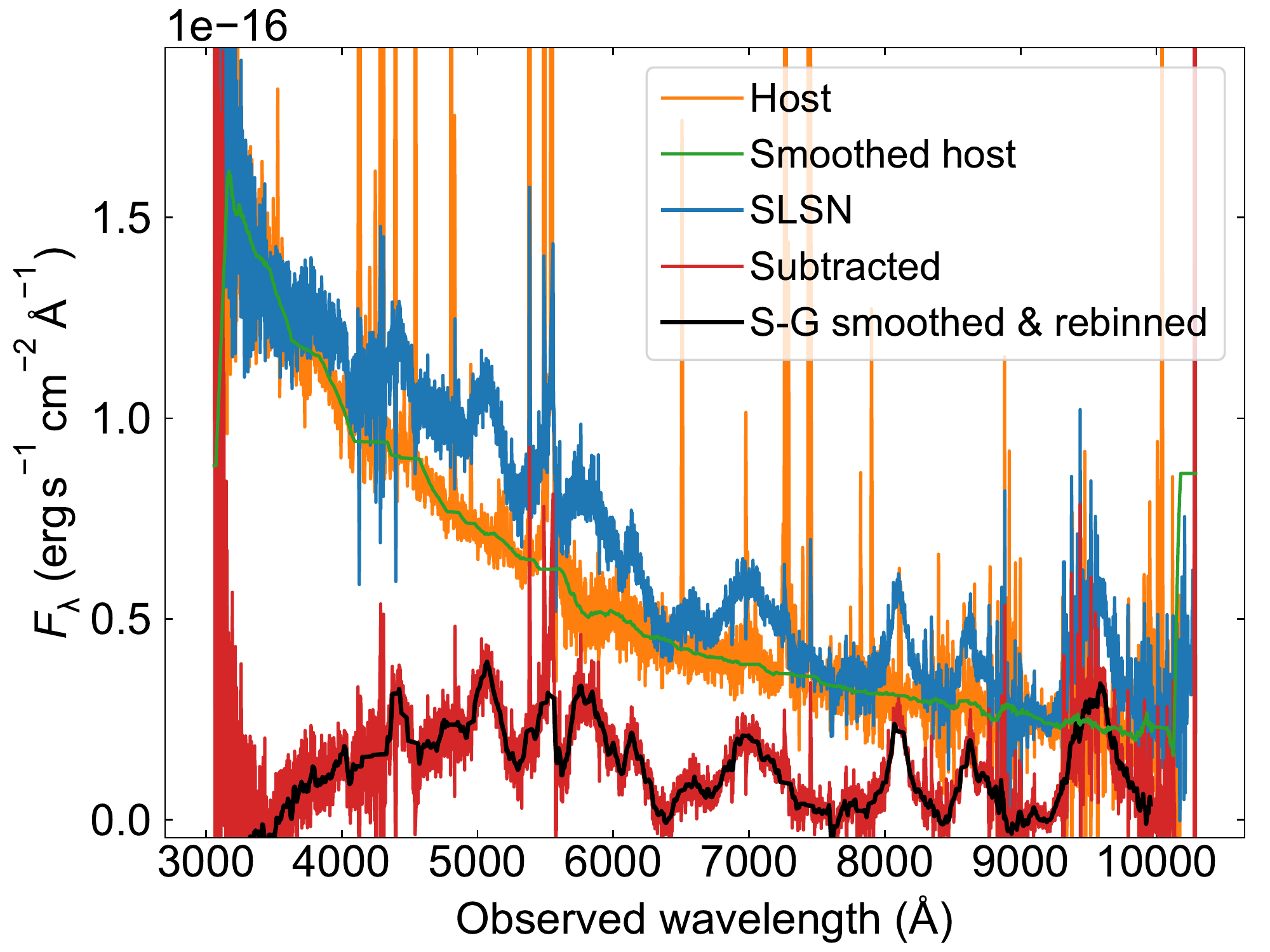}
\caption{Example of our data reduction procedure, for PTF12dam at 269 days. A median-filtered host spectrum is subtracted from each SLSN after scaling to photometry (section \ref{s:ptf}. We smooth all subtracted spectra with a Savitsky-Golay filter and rebin to a common 5\,\AA\ (rest-frame) pixel scale (section \ref{s:smooth}).}
\label{f:smooth}
\end{figure}

First, we smoothed each spectrum using a Savitsky-Golay filter implemented in \textsc{SciPy}. This algorithm replaces the flux in each pixel with an interpolated value based on a polynomial fit of order $n$ to the neighboring pixels within a window $\pm w$. After experimenting with a range of values, we found that using $w=19$ pixels (i.e. fitting to a region spanning 45\,\AA\ either side of the target wavelength) and $n=1-2$ gave satisfactory results. We examined each spectrum by eye to determine which value of $n$ achieved the best balance between accurately preserving the shape of real structure, and suppressing the noise. In practice we generally used $n=2$ for spectra with a clean signal, and $n=1$ for noisy data.

We then linearly interpolated the spectra to a common dispersion scale of $\Delta\lambda = 5$\,\AA, spanning a wavelength range 3000--9000\,\AA\ in the rest frame of each SLSN. This range was chosen to cover all spectral lines of interest, while avoiding any flux calibration uncertainties at the extrema of either the SLSN or host spectra. Finally, we produced normalised spectra by dividing each spectrum by its mean flux. An example of this smoothing (and prior host subtraction) is shown in Figure \ref{f:smooth}.

\section{Observed properties}
\label{s:obs}

\begin{figure*}
\centering
\includegraphics[width=18cm]{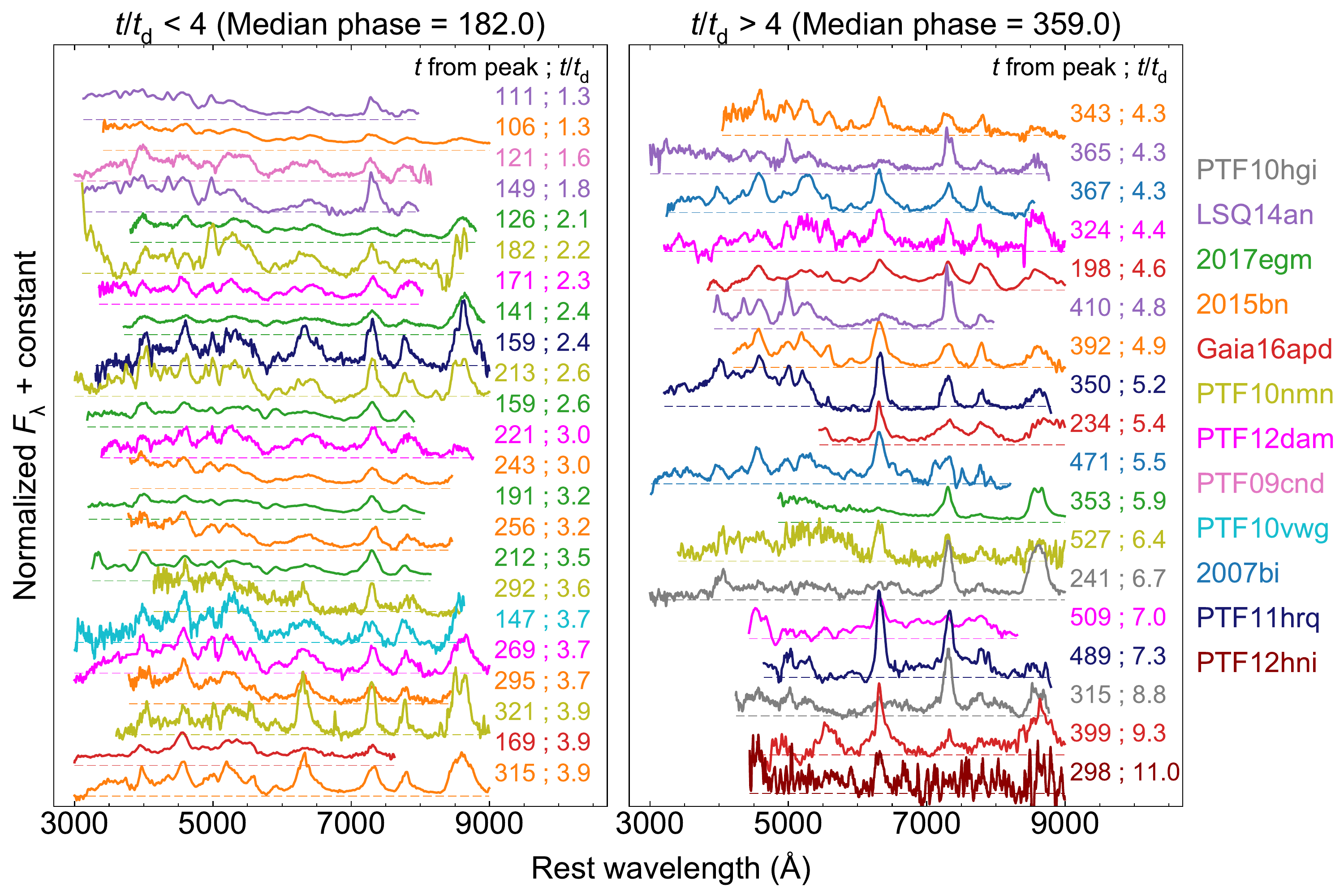}
\caption{Full series of SLSN nebular spectra, after host subtraction, smoothing, and correcting for redshift and extinction. Spectra are arranged in order of increasing phase, defined as the time after peak in rest-frame days, normalized by $t_d$, where $t_d$ is the characteristic light curve decline timescale \citep{nic2015b}. Dashed lines show the zero-flux level for each spectrum.
}
\label{f:sample}
\end{figure*}

\subsection{Spectral sequence}

All spectra in our sample are shown in Figure \ref{f:sample}, following the processing described in Section \ref{s:smooth} (a version with unsmoothed spectra is provided in the appendix). Each spectrum is labeled with the date on which it was obtained, measured in rest-frame days relative to when that SLSN reached its peak brightness. We use the dates of bolometric maximum as given in the literature.

SLSNe show a broad range of timescales in their photometric and spectroscopic evolution \citep{nic2015b}, so displaying spectra in order of time after peak is not necessarily a fair comparison. Instead, we rank our spectra using a normalised timescale $t/t_d$, where $t$ is the rest-frame phase from maximum light, and $t_d$ is the exponential decay time of the light curve for each event, which ranges over $\approx 25-90$ days. \citet{nic2015b} and \citet{dec2018}\footnote{\citet{dec2018} give the time to decline by 1 magnitude in $g$-band; we convert to $t_d$ using an empirical relation 
$t_d \approx 1.25 \,t_{g,{\rm 1 mag}}$, 
based on 3 events in common with \citet{nic2015b}.} provide tabulated values for most of this sample, while individual values for Gaia16apd and SN\,2017egm come from \citet{kan2016b} and \citet{bos2018}, respectively.

Virtually all of these spectra are at normalised phases $t/t_d>2$. We show on the left panel of Figure \ref{f:sample} those spectra with $t/t_d<4$ (23 spectra), and on the right panel those with $t/t_d>4$ (18 spectra). This immediately leads to an important insight. Spectra on the left often show residual continua, and have a weak or poorly developed feature at 6300\,\AA. The strongest line in these spectra is generally at 7300\,\AA, usually attributed to \ca\ (really a doublet with components at 7291\,\AA\ and 7324\,\AA). The later spectra in the right panel, on the other hand, show little continuum, and the strongest line is nearly always at 6300\,\AA, consistent with \o\ (also a doublet, at 6300\,\AA\ and 6364\,\AA). 

Therefore we conclude that SLSNe can reach a pseudo-nebular phase, with several prominent emission lines but some residual continuum, after $\approx 2$ light curve timescales from maximum light. By $\approx 4$ decline times, SLSNe tend to be fully nebular. The nature of the continuum is not certain -- it could arise from an internal photosphere, circumstellar interaction, or possibly a forest of overlapping broad lines. The fact that \o\ is not fully developed during the continuum phase suggests the ejecta are still relatively dense ($n_e\sim10^6$\,cm$^{-3}$), so a residual photosphere may be most likely (however, see section \ref{s:profiles}).

In seven of the ten SLSNe with spectra at $t/t_d>4$, the strongest line is the 6300\,\AA\ line, with ratios $6300/7300 \gtrsim 1-2$. However, there are three that show weak 6300\,\AA\ features and prominent 7300\,\AA\ emission, with $6300/7300<0.3$. These events are PTF10hgi, LSQ14an and SN\,2017egm, including their spectra at quite late phases of 241--410 days, or $4.3-8.8\,t_d$. \citet{jer2017a} and \citet{ins2017} showed that the strong 7300\,\AA\ feature in LSQ14an is likely due to contamination by [\ion{O}{2}]\,$\lambda$7320,7330. The case of SN\,2017egm is even more surprising, as its spectrum is initially almost identical to the prototypical SN\,2015bn, but as it evolves the line at 6300\,\AA\ never becomes dominant. We explore these outliers further in sections \ref{s:profiles} and \ref{s:lums}.

\subsection{Mean spectrum}
\label{s:mean}

\begin{figure}
\centering
\includegraphics[width=8.25cm]{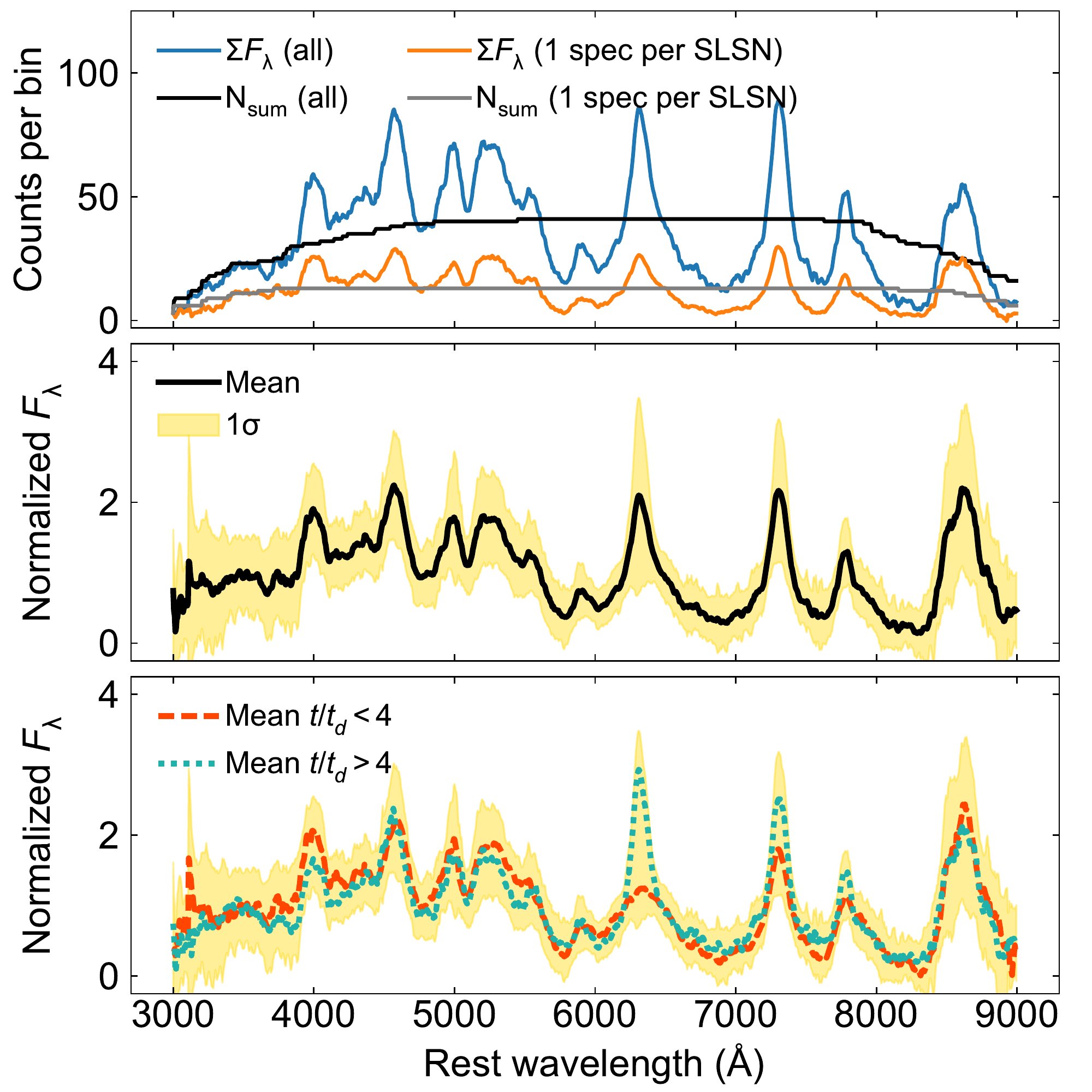}
\caption{Top: Sum of SLSN spectra and number of objects per wavelength bin. The blue line shows the full sample, while the orange line shows the sum of only one spectrum per event. Bottom: The mean nebular spectrum of SLSNe. The red dotted line shows the mean of spectra at $t/t_d<4$ (left panel of Figure \ref{f:sample}) while the teal line shows those with $t/t_d>4$ (right panel of \ref{f:sample}).
}
\label{f:ave}
\end{figure}

We begin by constructing the average nebular spectrum of SLSNe. We sum the normalised flux of all processed spectra in each 5\,\AA\ bin, then divide by the number of spectra contributing in that bin. The sums of fluxes and spectra per bin are shown in the top panel of Figure \ref{f:ave}. In order that the mean spectrum not be biased by the few events with the most data, we also constructed a sum taking only one spectrum per SLSN (choosing the spectrum with the best wavelength coverage). As shown in Figure \ref{f:ave}, this made no appreciable difference to the result, so we take the mean of the full sample as our average spectrum.

In the bottom panel of Figure \ref{f:ave}, we show the mean and standard deviation, as well as individual means for the spectra at $t/t_d<4$ and $t/t_d>4$. These two subsets are mostly similar, and lie within the 1$\sigma$ contours of the overall mean. The most significant difference is at 6300\,\AA, which is comparable in strength to the 7300\,\AA\ line when averaged over the whole sequence, but the ratio between these lines varies from $\sim 0.5$ in the earlier spectra to $\sim 2$ in the latter. The relative lack of evolution outside of the 6300\,\AA\ line, over significant periods of time, was noted for specific SLSNe by \citet{jer2017a} and \citet{ins2017}, here we show that this holds true for a much larger sample of the population.

\subsection{Line identifications}
\label{s:ids}

\begin{figure*}
\centering
\includegraphics[width=18cm]{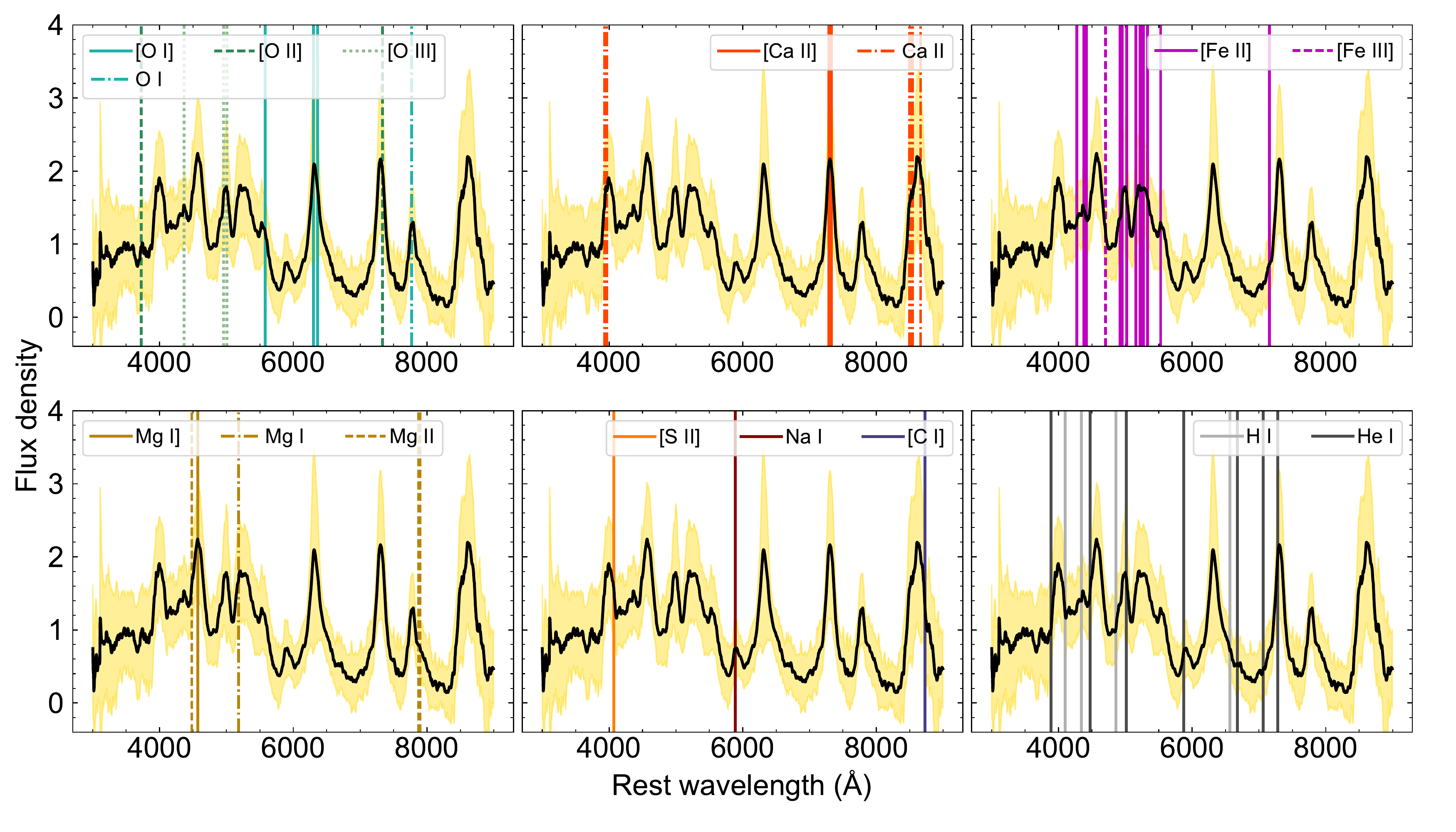}
\caption{Line identifications in SLSNe. All atomic data were retrieved from NIST. All of the strong lines can be associated with oxygen, calcium, magnesium, sodium and iron, with possible contributions from carbon and sulphur. No hydrogen or helium are observed.}
\label{f:id}
\end{figure*}

The main advantage of constructing the average spectrum is to identify significant features that are common in SLSNe, but may be difficult to detect in individual events at modest S/N. In Figure \ref{f:id}, we overplot spectral lines of common species observed in SNe. For each species, we searched the NIST Atomic Spectra Database \citep{nist} for transitions between low-lying energy levels, with a particular focus on forbidden lines close to the ground state. 

Most lines are matched using only 5 elements: oxygen, calcium, iron, magnesium and sodium. We also detect possible contributions from carbon and sulphur. While the strongest oxygen lines are from neutral \ion{O}{1}, there are weaker features consistent with \ion{O}{2} and \ion{O}{3}. The higher ionisation features overlap with iron lines, so their identification is not always clear -- we investigate this further in the next section. 

Calcium and iron appear singly-ionised, while oxygen and magnesium appear neutral. This is mostly consistent with their relative ionisation potentials: oxygen has a much greater ionisation potential (13.6\,eV) than these other elements ($6-8$\,eV). We do not see \ion{Fe}{3} (the second ionisation potential of iron is 16.2\,eV), which is responsible for the strongest line in the nebular spectra of Type Ia SNe \citep{axelrod1980}. Although magnesium has a lower ionisation potential close to that of neutral iron, \ion{Mg}{2} has few distinct lines in the optical. Asymmetry in the \oi\ line profile may be an indication of contamination from \ion{Mg}{2}\,$\lambda$7877,7896, as we will discuss in section \ref{s:profiles}. \mg\ is both a recombination line and an efficient coolant, so it is often strong even when most magnesium is ionised \citep{jer2017b}.


None of the species listed above predict additional strong lines not seen in our spectra. We apply a similar consistency check to test for weak lines of hydrogen or helium. While some features match the locations of these lines, they predict many unseen features, so it is not possible to interpret any of the lines as hydrogen or helium in a self-consistent manner. Thus we rule out any substantial contribution from these two species. We note that there does exist a subset ($\sim 15$\%) of known SLSNe, which did not display significant hydrogen lines during their photospheric phases but eventually show nebular hydrogen emission. These have been analysed extensively by \citet{yan2015,yan2017}. 

Since SLSNe seem to show either strong hydrogen or no hydrogen at all, we posit that any nebular hydrogen lines likely arise through late-time interaction with pre-expelled, hydrogen-rich material, rather than from some residual hydrogen in the SLSN atmosphere, which might be expected to lead to more of a continuum in properties. \citet{yan2017} also favoured interaction as the explanation, and calculated a distance to the hydrogen-rich material of $\sim 10^{16}$\,cm. The SLSNe in our sample clearly have not reached any such material by the time they enter the nebular phase, so if it exists it must be significantly further away or of significantly lower density in our events. This suggests a diversity in the timing of the stellar envelope loss, and/or the velocity with which it is expelled.

\begin{figure}
\centering
\includegraphics[width=8.25cm]{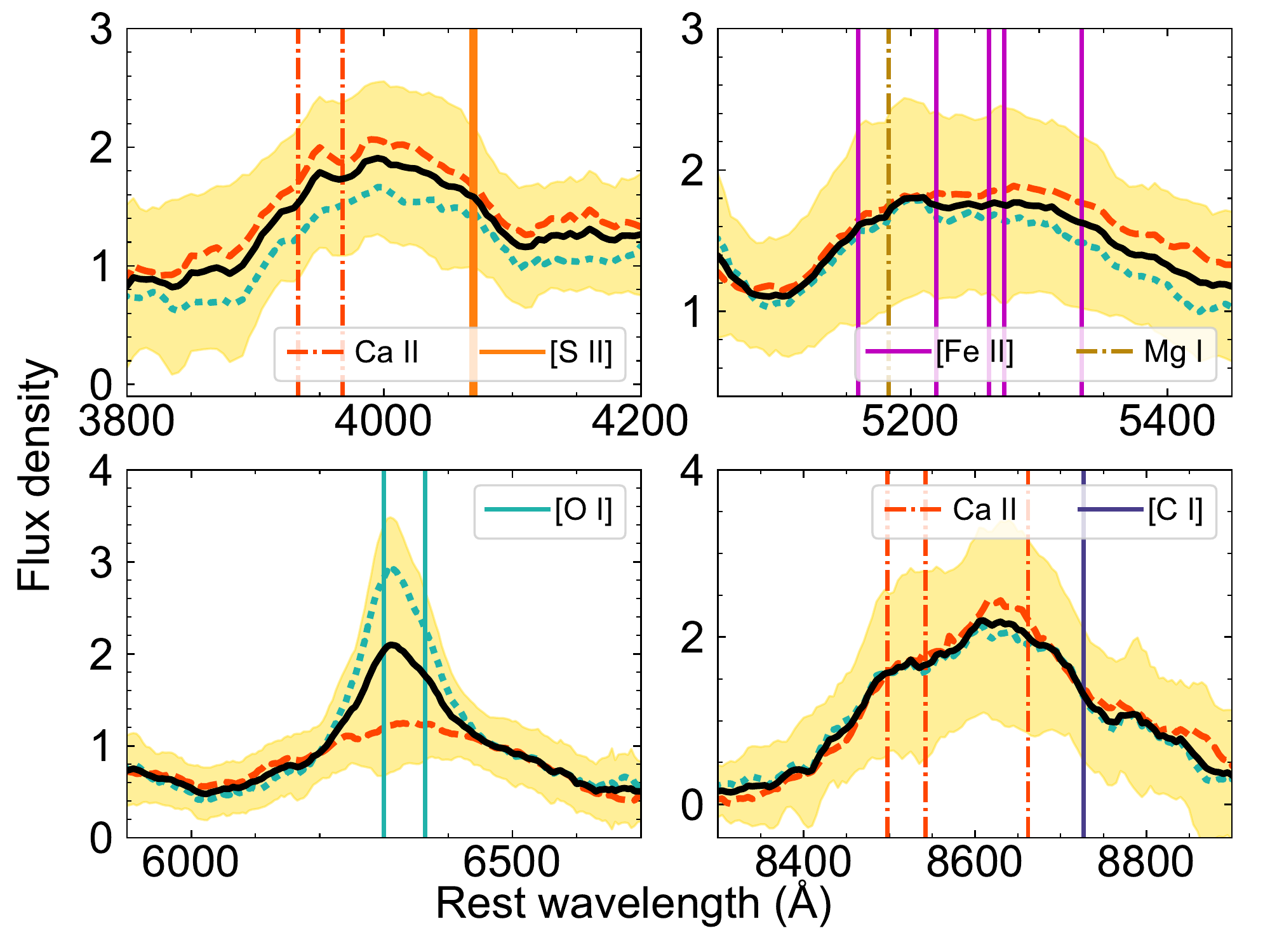}
\caption{Zoom-in around blended lines from Figure \ref{f:id}. Red dashed line corresponds to $t/t_d<4$ while the blue-green dotted line corresponds to $t/t_d>4$. The peak wavelengths shift over time as the ratios between components evolve.}
\label{f:evolve}
\end{figure}

We also investigate any subtle differences in line identifications as a function of time, by comparing the mean spectra at $t/t_d<4$ and $t/t_d>4$. Figure \ref{f:evolve} shows close-ups around several features identified above. The line at $\approx 4000$\,\AA\ can be attributed to a blend of \ion{Ca}{2} H \& K and the [\ion{S}{2}]\,$\lambda$4069 doublet. The line centre shifts slightly towards [\ion{S}{2}] at later phases. The [\ion{Fe}{2}] blend at $\sim 5250$\,\AA\ likely contains a contribution from \ion{Mg}{1}\,$\lambda$5183 \citep{nic2013,jer2017a} -- this line also changes as the spectra evolve, indicating that the relative contribution from magnesium likely increases over time.

The line centre of \o\ also shifts over time. This line is a doublet, with wavelengths of 6300\,\AA\ and 6364\,\AA. When the line is optically thick, these components have approximately equal strengths, but in the optically-thin regime their ratio is 3:1. Thus the shift of the central wavelength to the blue shows that as the line becomes stronger after $t/t_d \sim 4$, it also becomes optically thin. The final panel of Figure \ref{f:evolve} shows the \ion{Ca}{2} NIR triplet, which may be blended with [\ion{C}{1}]. However, we see no significant evidence for a change in the relative contribution from [\ion{C}{1}] between the early and late nebular phase.


\subsection{Comparison to SNe Ic}
\label{s:compare}

We conclude this section with a comparison to normal and broad-lined Type Ic SNe, to which SLSNe are known to share similar spectroscopic properties in both the photospheric \citep{pas2010,ins2013,Liu&Modjaz2017,qui2018} and nebular \citep{gal2009,nic2016b,jer2017a} phases. We retrieve from the Open Supernova Catalog \citep{gui2017} all spectra of Type Ic SNe obtained more than 100 days after maximum light. We exclude any spectra with clear evidence for host galaxy contamination in the continuum. Because SNe Ic exhibit a faster evolution that SLSNe, as we will show, we further include any additional spectra of these same SNe Ic obtained $\gtrsim 50$ days after maximum. This results in 80 spectra of 15 SNe Ic, which are shown in the appendix.
We construct a mean spectrum for SNe Ic following the method in Section \ref{s:mean}. To facilitate a fair comparison, we also tried omitting SN Ic spectra with $t/t_d>11$, the latest observed normalised phase in our SLSN sample (we assume $t_d \approx 20$ days for typical SNe Ic and 57 days for the slowly-evolving SN\,2011bm; \citealt{val2012,nic2015b}). However, excluding this data made no significant difference to the mean spectrum. 

\begin{figure}
\centering
\includegraphics[width=8.25cm]{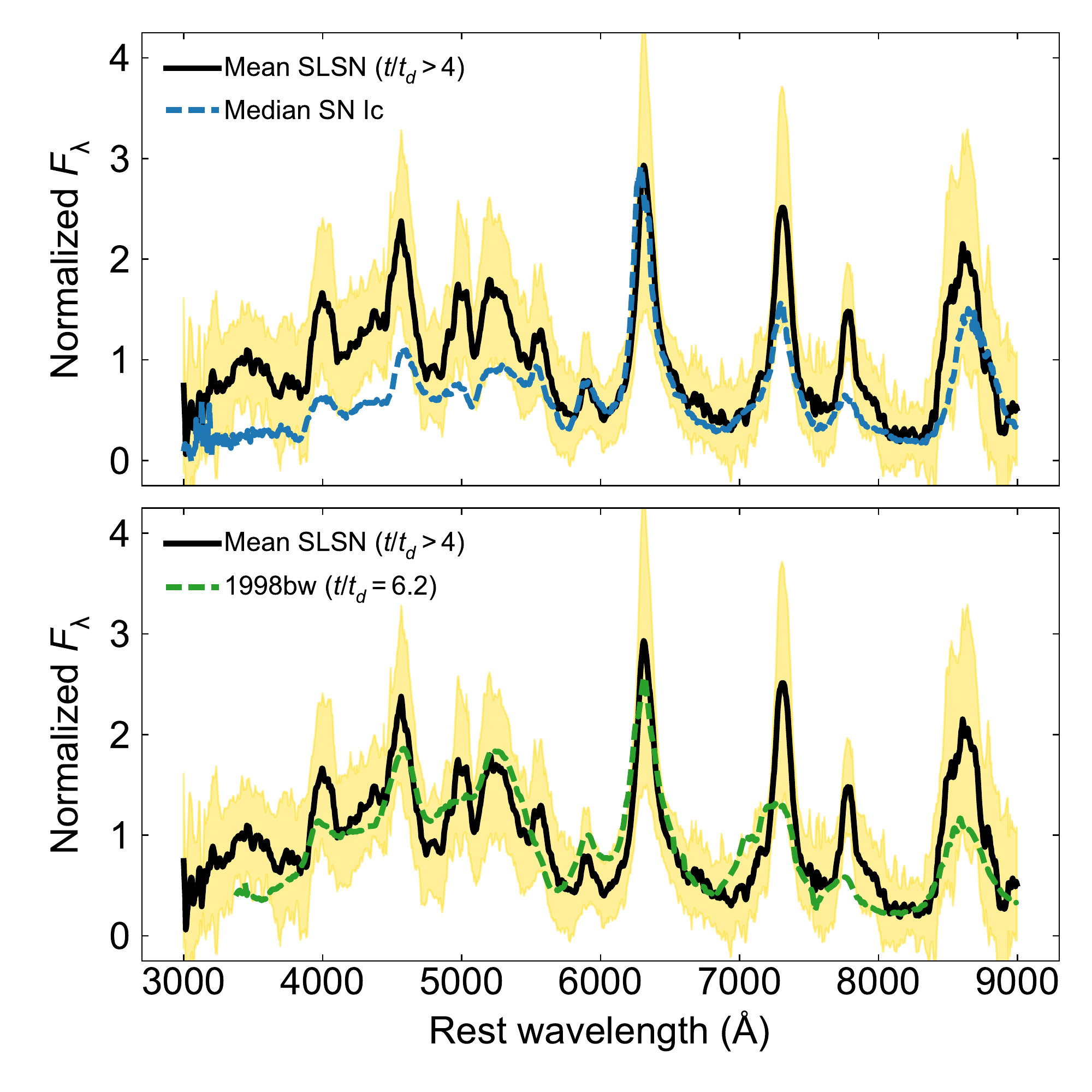}
\caption{Top: Mean nebular spectrum of SLSNe compared to the mean spectrum of SNe Ic at a comparable phase from explosion. SLSNe show a more pronounced \oi\ line, [\ion{O}{3}] lines not seen in SNe Ic, and an elevated flux over the region around 5000\,\AA, where [\ion{Fe}{2}] is thought to dominate. Bottom: Comparison to the broad-lined SN Ic, SN\,1998bw, shows that the iron line strength in such events can be similar to SLSNe.
}
\label{f:comp}
\end{figure}

We show a comparison of the mean SN Ic spectrum to that of SLSNe in Figure \ref{f:comp}. Type Ic SNe in the nebular phase generally exhibit a ratio of \ca/\o\,$\sim 0.5$, so we use the mean of SLSN spectra at $t/t_d>4$ when the ratio is similar. That the median age of the SN Ic spectra is only 150 days, compared to 359 days for SLSNe, immediately highlights the slower spectroscopic evolution of SLSNe compared to other stripped-envelope explosions.

While there is virtually a one-to-one correspondence in identified spectral lines, we note three key differences between the two classes. First, SLSNe exhibit a much stronger \oi\ line, often with a shoulder on the red side that is not seen in lower luminosity SNe Ic \citep[see also][]{mil2013,nic2016b}. Second, SLSNe have a strong line at 5000\,\AA, the only feature that appears in one class but not the other. In section \ref{s:ids}, we interpreted this line as a possible blend of [\ion{Fe}{2}] and [\ion{O}{3}]. The fact that the line does not appear in SNe Ic, which do display the other strong iron lines, indicates a substantial contribution from [\ion{O}{3}] in the SLSNe. This could be explained if SLSNe have higher temperature and ionisation in the oxygen-rich ejecta than do SNe Ic, despite the fact that the SLSNe here have been expanding for twice as long as the SNe Ic at the epochs of observation. This line was also identified as [\ion{O}{3}] by \citet{lun2016}, \citet{nic2016b}, \citet{jer2017a}, and \citet{ins2017}.

Finally, SLSNe are clearly elevated in flux, on average, over the iron-dominated region between $\approx 4000-5500$\,\AA. This was first pointed out by \citet{gal2009} and \citet{mil2013} in the context of SN\,2007bi, the first SLSN with a nebular spectrum \citep[though this was partly attributable to host galaxy contamination;][]{jer2017a}. Recently, \citet{moriya2018} showed that if some SLSNe were powered by fallback accretion they would need to accrete $>1$\,\M\ onto the central compact remnant to sustain their high luminosities. An observable consequence should be a deficiency of iron-group elements in the late-time spectra. The fact that SLSNe exhibit an excess, rather than a deficiency, in the iron lines therefore disfavours a fallback model.

\citet{nic2016b} and \citet{jer2017a} showed that the strength of the `iron plateaus' in SN\,2007bi and SN\,2015bn were comparable to some broad-lined Type Ic SNe. Figure \ref{f:comp} also shows the spectrum of SN\,1998bw, normalised to match the \o\ and \ca\ line of the SLSNe and SNe Ic. On average, the SLSNe have a relative flux between $\approx 4000-5500$\,\AA\ that is similar to SN\,1998bw. Thus, while SLSNe do exhibit an excess in this region compared to average SNe Ic, there is no evidence for an extreme fraction of iron-group elements above those seen in broad-lined SNe Ic \citep[see also recent work by][on the late-time light curves of SLSNe]{bla2018}.

We conclude this section by noting that these differences between SLSNe and other SNe Ic are in some ways quite subtle, despite SLSNe being more luminous than SNe Ic by up to two orders of magnitude even at nebular times. If a large contribution to the luminosity of SLSNe was from circumstellar interaction, it is difficult to imagine how one would hide all obvious interaction features in the nebular spectrum (this is in contrast to the subset that show late-time hydrogen emission; \citealt{yan2017}).

\section{Machine learning analysis}
\label{s:mach}

Machine learning algorithms are becoming increasingly valuable in astrophysical research. Recently, \citet{ins2018} applied some of these techniques to investigate a statistical basis for classifying SLSNe during the photospheric phase. In this section, we will analyse the nebular spectra of SLSNe using tools from the \textsc{Python} package \textsc{scikit-learn} \citep{ped2011}.

\subsection{Principal component analysis}

We begin by applying principal component analysis (PCA) to our spectra. Fundamentally, PCA transforms a set of observations (in our case, spectra) consisting of a number of variables (the flux in each wavelength bin), which are typically correlated (the fluxes in neighbouring bins are of course highly correlated), into a set of \emph{uncorrelated} vectors, i.e.~the `principal components' \citep{pea1901,hot1933}. These form an orthogonal basis for the space: any observation $\mathbf{O}_i$ can be represented by 
\begin{equation}
\mathbf{O}_i=\sum_j a_{ij} \mathbf{C}_j,
\label{e:pca}
\end{equation}
where $\mathbf{C}_j$ is the $j$th principal component and $\mathbf{a}_{ij}$ the coordinates of the observation in the PCA basis. 

\begin{figure}
\centering
\includegraphics[width=8.25cm]{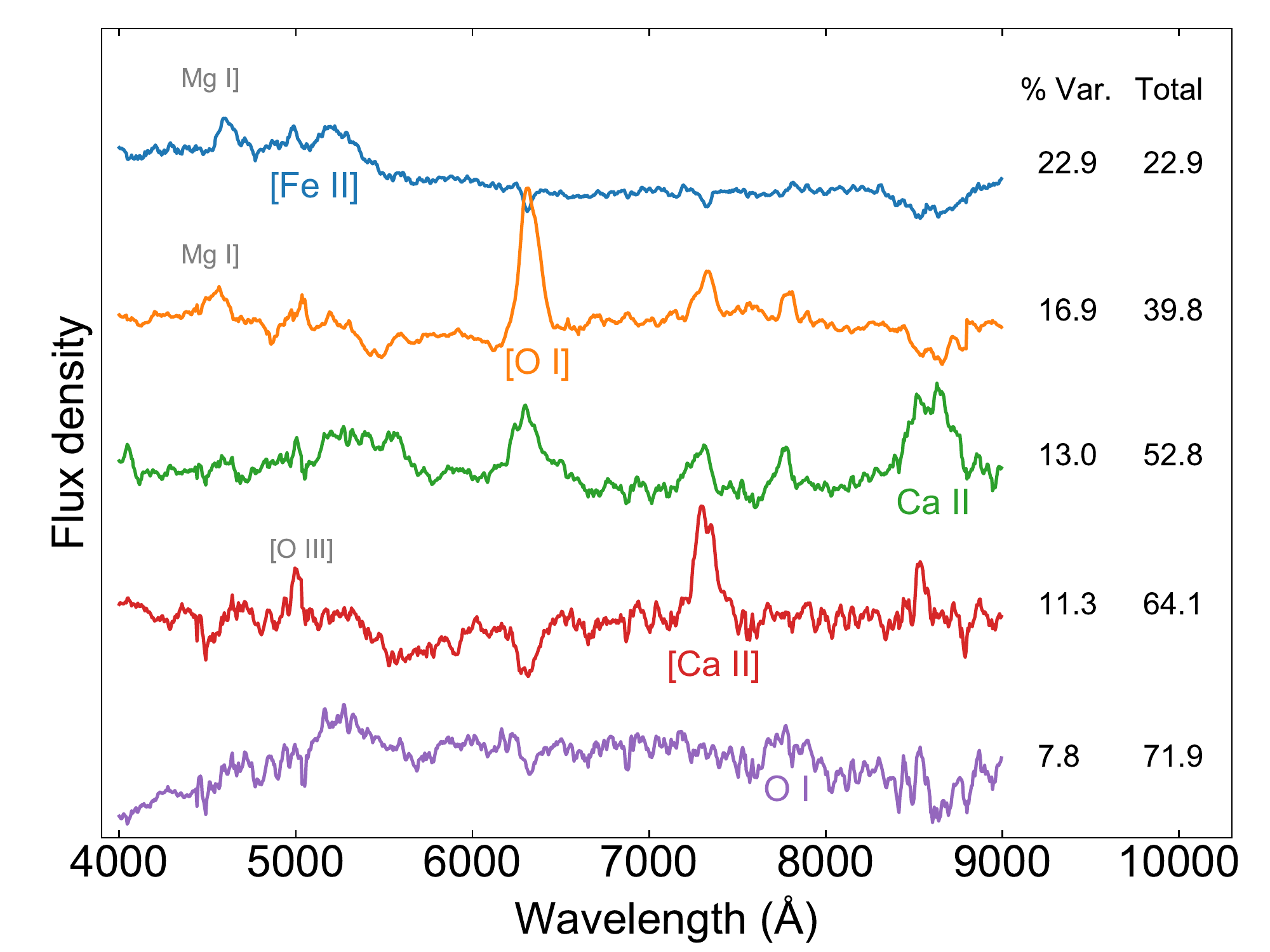}
\caption{PCA decomposition of SLSN spectra. 70\%\ of the variation in the sample can be explained with five components, which correspond to identified spectral lines -- showing that the evolution between different lines is not strongly correlated. The greatest variability is in the [\ion{Fe}{2}] lines.}
\label{f:eigen}
\end{figure}

For $n$ observations (41 spectra) with $p$ variables (1200 wavelength bins), the number of components is the lesser of $n-1$ and $p$ (i.e., 40). However, all components are not equally informative. The first component is chosen to explain the maximum amount of variance in the observations, and each successive component explains as much of the remaining variance as possible while remaining uncorrelated with the previous components. Thus, later components capture progressively weaker features, and eventually only noise, in the data.

We use the \textsc{PCA} method as implemented in the package \textsc{scikit-learn.decomposition} to transform our set of observed spectra into a basis of principal components. We find that over 70\% of the sample variance can be explained using only the first five `eigenspectra'. These are shown in Figure \ref{f:eigen}. The components largely correspond to distinct spectral features. Almost 25\% of the sample variance is explained by a component that primarily consists of [\ion{Fe}{2}] lines around 5000\,\AA. Perhaps this is to be expected, since this is the most complicated region of the spectrum consisting of many blended lines, but it may also suggest a diversity in iron-group element production between different SLSNe.

The next three components are dominated successively by \o, the \ion{Ca}{2} NIR triplet, and \ca, together accounting for over 40\% of the variance. The fifth component does not have a single line that is quite so dominant, but does exhibit the strong \oi\ line. All strong lines identified in our spectra are included among these five eigenspectra, and we exclude any further components from our analysis, leaving a five-parameter model following equation \ref{e:pca}, where the free parameters are the coefficients, $a_j$, for $j = 1-5$.

\begin{figure}
\centering
\includegraphics[width=8.25cm]{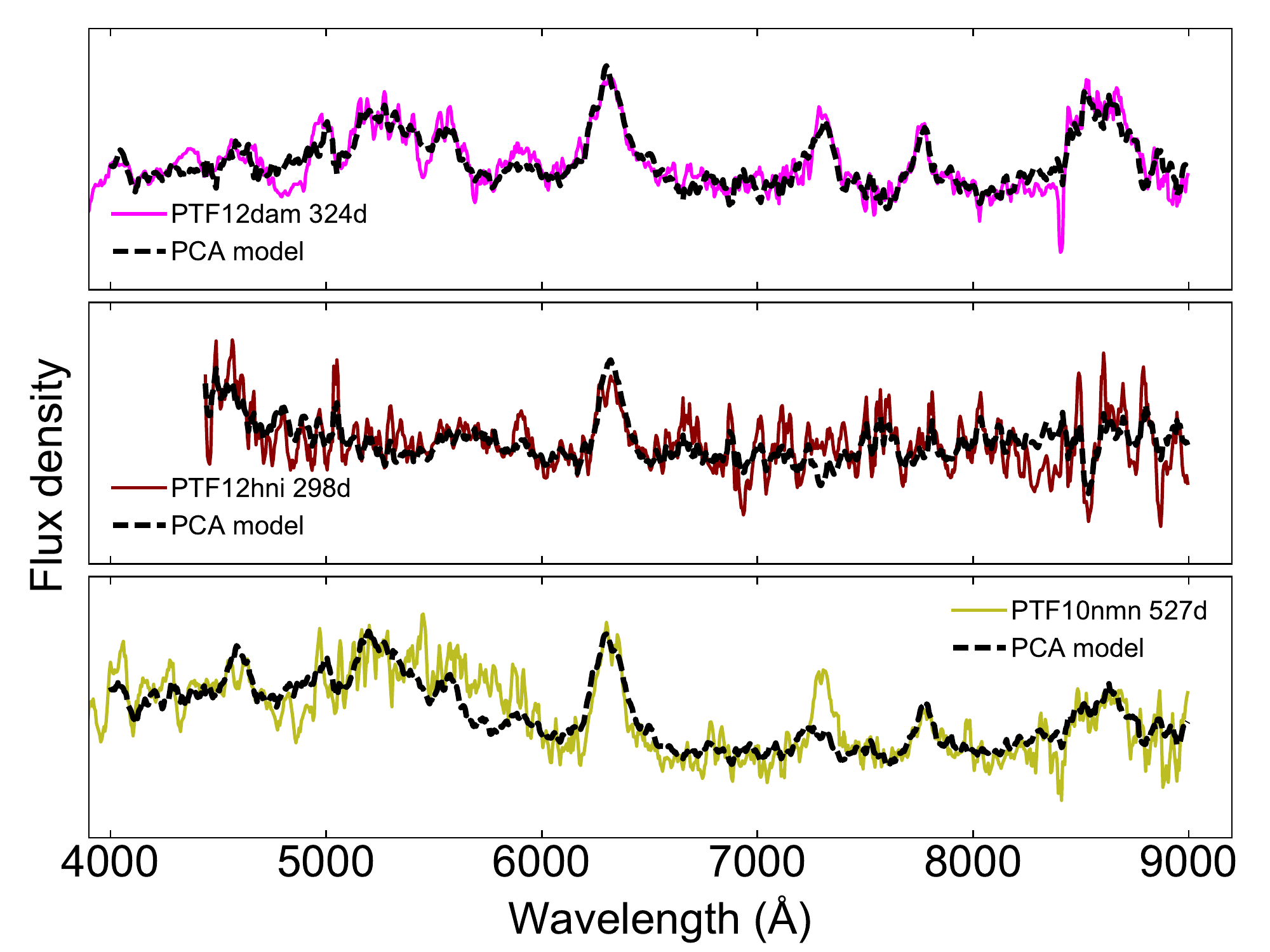}
\caption{Fits to three randomly selected SLSN spectra using the five-component PCA model, confirming this decomposition can well capture the key features.}
\label{f:model}
\end{figure}

We verify this model by randomly excluding three spectra as a test sample, reconstructing the PCA basis from the remaining training sample (the results are indistinguishable compared to using the full sample), and fitting the test sample with the eigenspectra. The results are shown in Figure \ref{f:model}. We find that the PCA basis effectively captures the key features of the spectra.

\subsection{Clustering}

A key question in the study of SLSNe is whether all of these events form a continuous class, or if there are separate sub-populations with different progenitors and/or power sources \citep{gal2012,nic2013,nic2015b,nic2017c,lun2018,dec2018,ins2018,qui2018}. An interesting recent result from \citet{ins2018} is the possibility that SLSNe with slower light curve evolution may show a flatter velocity gradient, though the sample of events with sufficient measurements remains small. 

The velocity gradient probes the density structure in the outer layers of the ejecta. In the nebular phase, we have access to the whole ejecta, including the innermost regions that are most sensitive to differences in nucleosynthesis or hydrodynamic effects of different explosion mechanisms. Thus it provides a valuable test of diversity in these events. Here we employ the same machine learning algorithm used by \citet{ins2018} to look for separate populations in our data.

K-means clustering partitions data into $k$ clusters of equal variance, where each element belongs to the group whose mean is nearest to that element \citep{mcq1967,ste1956,llo1982,for1965}. The algorithm requires that $k$ be specified in advance -- it is therefore important to include some additional metric to determine the optimal number of clusters.

\begin{figure}
\centering
\includegraphics[width=8.25cm]{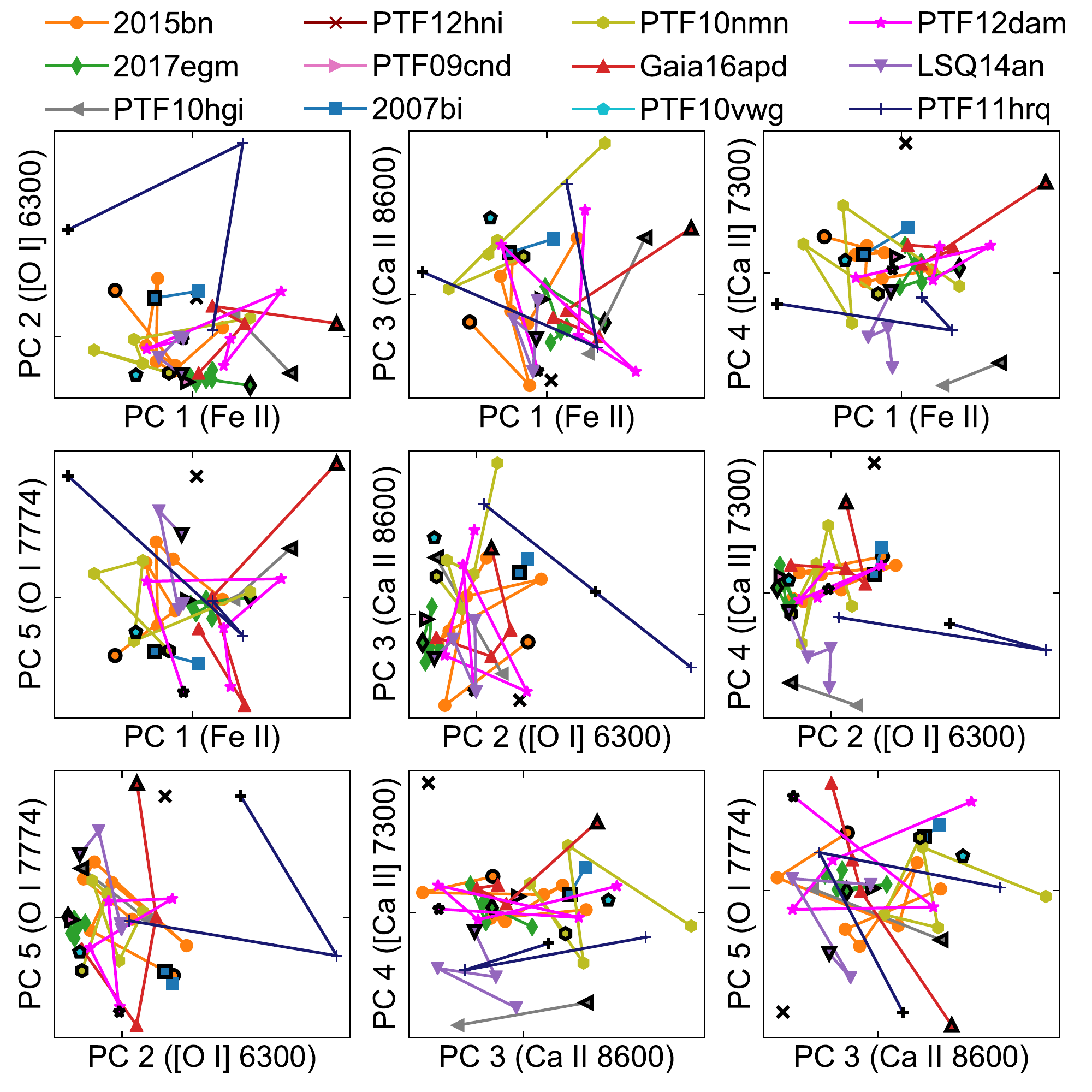}
\caption{Representation of SLSN nebular spectra in the PCA parameter space. A thick black border indicates the earliest spectrum of each event, and lines show the progression through the parameter space over time. Few clear trends are present, though the third component (\o) generally becomes more important with time.}
\label{f:pca}
\end{figure}

We reduce the dimensionality of our spectra from 1200 to 5 using the PCA decomposition described above, and then apply the \textsc{KMeans} method in \textsc{scikit-learn.cluster}. The decomposition is plotted in the PCA parameter space in Figure \ref{f:pca}. We also tested the algorithm on the un-reduced dataset, finding that the assignment of spectra to clusters was essentially unchanged.

\begin{figure}
\centering
\includegraphics[width=8.25cm,trim= 0 0 0 1.9cm,clip]{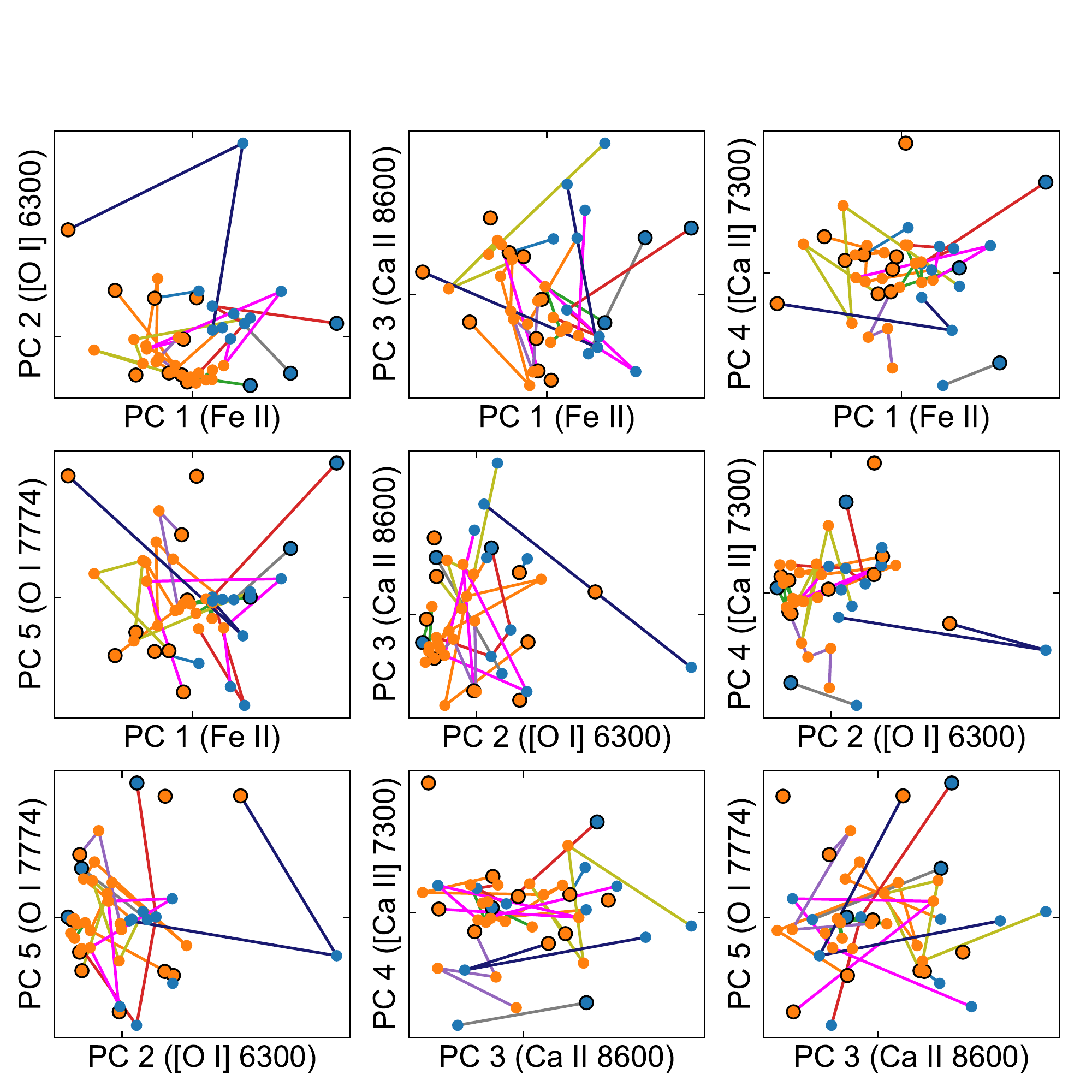}
\caption{Same as Figure \ref{f:pca}, but with points coloured by K means cluster membership assuming $k=2$. Individual SLSNe move between the clusters as their spectra evolve, i.e.~there is no evidence for distinct sub-populations of SLSNe differentiated by their nebular-phase spectra.}
\label{f:cluster}
\end{figure}

We search for the optimal number of clusters using three methods: the Bayesian information criterion \citep[BIC;][]{sch1978}, the elbow criterion, and silhouette analysis. The details can be found in the Appendix, but overall there is no strong evidence for clustering ($k>1$) in the data.

In the interest of completeness, we plot the clustering solution for $k = 2$ in Figure \ref{f:cluster}. The colour coding indicates cluster membership, but the plot is otherwise identical to Figure \ref{f:pca}. Clearly, the `clusters' are not well separated. Most significantly, individual SLSNe evolve between the two clusters; thus, cluster membership is primarily just a function of temporal evolution rather than physical differences between SLSNe. We therefore find no evidence for multiple sub-populations.

\subsection{Prospects for classification from late-time spectra}

Spectroscopic classification of transients tends to be geared towards young events whenever possible, since this generally provides a path to greater scientific returns. However, due to factors such as unfavourable sky location, uneven survey cadence, and weather-related gaps, it is not uncommon for SNe to be typed quite late after explosion. If the light curve peak is missed by photometric surveys, it is not possible to classify a target as a SLSN using the original $M<-21$\,mag definition, the early \ion{O}{2} absorption lines, or even the statistical definition proposed by \citet{ins2018}. To avoid missing events entirely, classification using late-time spectra will be important, particularly in the era of LSST, for example to improve our knowledge of the SLSN volumetric rate.

\citet{qui2018} showed that despite the strong similarity between SLSNe and normal Type Ic SNe during some phases of their evolution, there were sufficient subtle differences that they could be distinguished statistically. They did this through a rigorous process of template comparison, quantifying the best matches and determining threshold matching scores to separate SLSNe from the general SN population. Here we suggest an alternative (or complementary) approach, using PCA. This method is particularly suitable for late times, when emission lines are present in the spectra, but it may be appropriate for earlier phases too given an appropriate training set.

The method is simple, as we demonstrate using a test case. PS17aea was discovered by the Pan-STARRS Survey for Transients \citep{hub2015} on 2017 January 3, and we flagged it as a bright source in a faint galaxy. However, there were no observations of the field for over 200 days preceding this initial detection. We obtained a spectrum on 2017 January 28 using the Inamori-Magellan Areal Camera and Spectrograph \citep[IMACS;][]{dressler2011} on the 6.5-m Magellan Baade telescope. The spectrum showed narrow emission in H$\alpha$, H$\beta$ and [\ion{O}{3}] from the host galaxy, from which we measured a redshift $z=0.104$. The observed absolute magnitude at this distance is $-18.8$\,mag, which is not exceedingly bright for many well known SN classes.

\begin{figure}
\centering
\includegraphics[width=8.25cm]{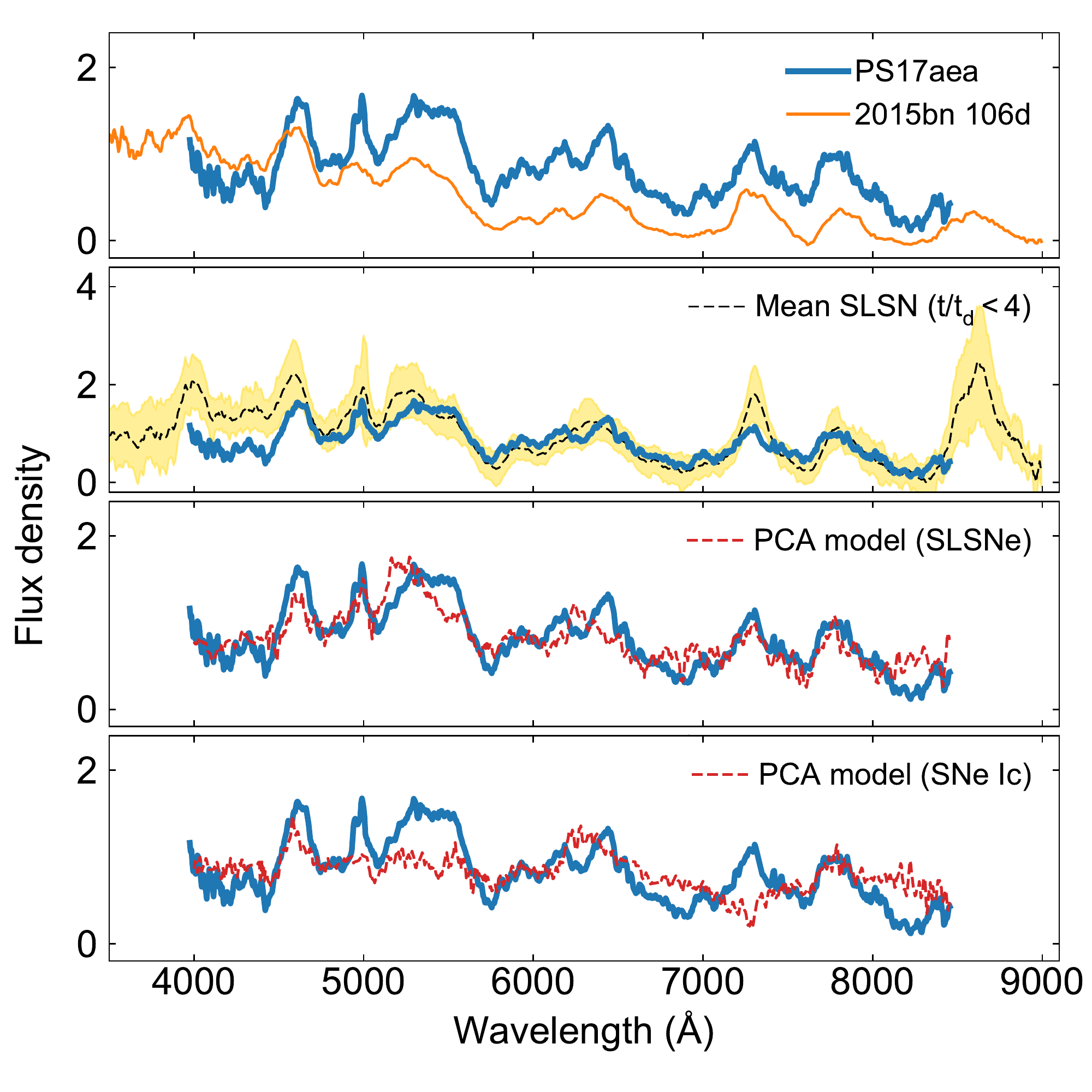}
\caption{Classification of PS17aea as a SLSN discovered $\gtrsim 100$ days after peak. Top: comparison to SN\,2015bn. Upper middle: Comparison to mean SLSN spectrum. Lower middle: Fit with PCA model for SLSNe. Bottom: Fitting with PCA model for SNe Ic gives an inferior match.}
\label{f:class}
\end{figure}

The spectrum is shown in Figure \ref{f:class}, after applying the same processing steps described in section \ref{s:smooth}. The top panel shows a comparison to SN\,2015bn at 106 days after maximum \citep{nic2016b}, to which PS17aea displays an intriguing degree of similarity. However, when trying to classify this spectrum using \textsc{Gelato}\footnote{https://gelato.tng.iac.es}, the top ten matches include not only SN\,2015bn, but a number of normal and broad-lined SNe Ic, SNe IIn, and even a peculiar SN Ia. 

The presence of several emission features suggests a nebular character to PS17aea. Comparing to the mean spectrum of SLSNe at $\approx 100-200$ days, we find that PS17aea lies within the $1\sigma$ contours across essentially the full wavelength range, and is therefore a good candidate for a SLSN evolving into the nebular phase. This would imply that the explosion occurred well before the first PSST detection. 

Comparison to the mean spectrum is a good start, but to proceed further we can appeal to PCA. We fit the spectrum of PS17aea with the eigenspectra in Figure \ref{f:eigen}. We find that the model fits PS17aea as well as it fits the SLSN spectra in Figure \ref{f:model}. With only five components, the model does not capture the complexity seen around the \o\ line (see also SN\,2015bn), but otherwise does an excellent job. For comparison, we construct an identical model for SNe Ic, using the comparison sample described in Section \ref{s:compare} and limiting the model to the first five eigenspectra as for SLSNe (this captures 91\% of the variance in the sample). Unlike the SLSN model, the SN Ic model does not satisfactorily capture some key features of the data, particularly the \ca\ line and the strength of the [\ion{Fe}{2}] bump. The reduced $\chi^2$ scores are $\sim 2$ for the SLSN model, and $\sim 5$ for the SN Ic model. To ensure that this is not simply a consequence of different normalised phases between the SLSN and SN Ic samples, we also tried fitting the data with a SN Ic PCA model restricted to spectra at $t/t_d<11$ (matching SLSN spectra), and found that using younger SNe Ic did not improve the fit quality. We therefore conclude that PS17aea was indeed a likely SLSN at relatively low redshift.

This test case shows the power of using PCA to resolve ambiguous SN classifications, even with a relatively modest training sample. In the future, further progress can be made by compiling large samples of all transient types, for example from the Open Supernova Catalog \citep{gui2017}, and using these to implement more advanced machine learning algorithms.

\section{Line profiles and ejecta distribution}
\label{s:profiles}

Having investigated the statistical properties of nebular-phase SLSN spectra, we now move on to analyse the properties of individual events and spectral lines.
The shapes of emission lines are an important diagnostic of the spatial distribution of material. \citet{jer2017b} summarised the predicted line profiles for simple geometries such as uniform spheres and disks. Most notable is a flat-topped line if the emission comes from a hollow shell.

A double-peaked line profile would suggest a bipolar explosion viewed at a large angle from the dominant axis; this has been seen in several stripped-envelope SNe \citep{maz2005,mae2008}. \citet{tau2009} conducted a detailed study of \o\ emission profiles in SNe Ic, and found considerable diversity between events, including symmetric Gaussian profiles with and without an additional narrow core; symmetric double peaks from a bipolar geometry; and more complex asymmetric multi-peaked profiles.

\subsection{Continuum subtraction}
\label{s:cont}

Before making velocity and flux measuremants for spectral lines, any continuum must be removed. For each line in every spectrum, we manually identified 50--100\,\AA-wide regions to the blue and red of the line wings, and approximated the continuum by fitting a linear function to these regions. A typical example is shown in the appendix. For isolated lines (essentially anything redwards of 6000\,\AA), this process is fairly straightforward, but in the case of blended lines such as \mg, the `pseudo-continuum' consists of many blended lines, making it difficult to estimate its true level. In this case, flux measurements are likely a lower limit. To test the sensitivity of our velocity and flux measurements to the choice of continuum estimate, we additionally employed an automated procedure where we defined two 100\,\AA\ regions centered at the local minimum flux on each side of the line to serve as the continuum level. We found that differences in derived properties were $\lesssim 10$\% compared to when we examined each SLSN manually.

\subsection{Individual SLSNe}

\begin{figure*}
\centering
\includegraphics[width=16cm]{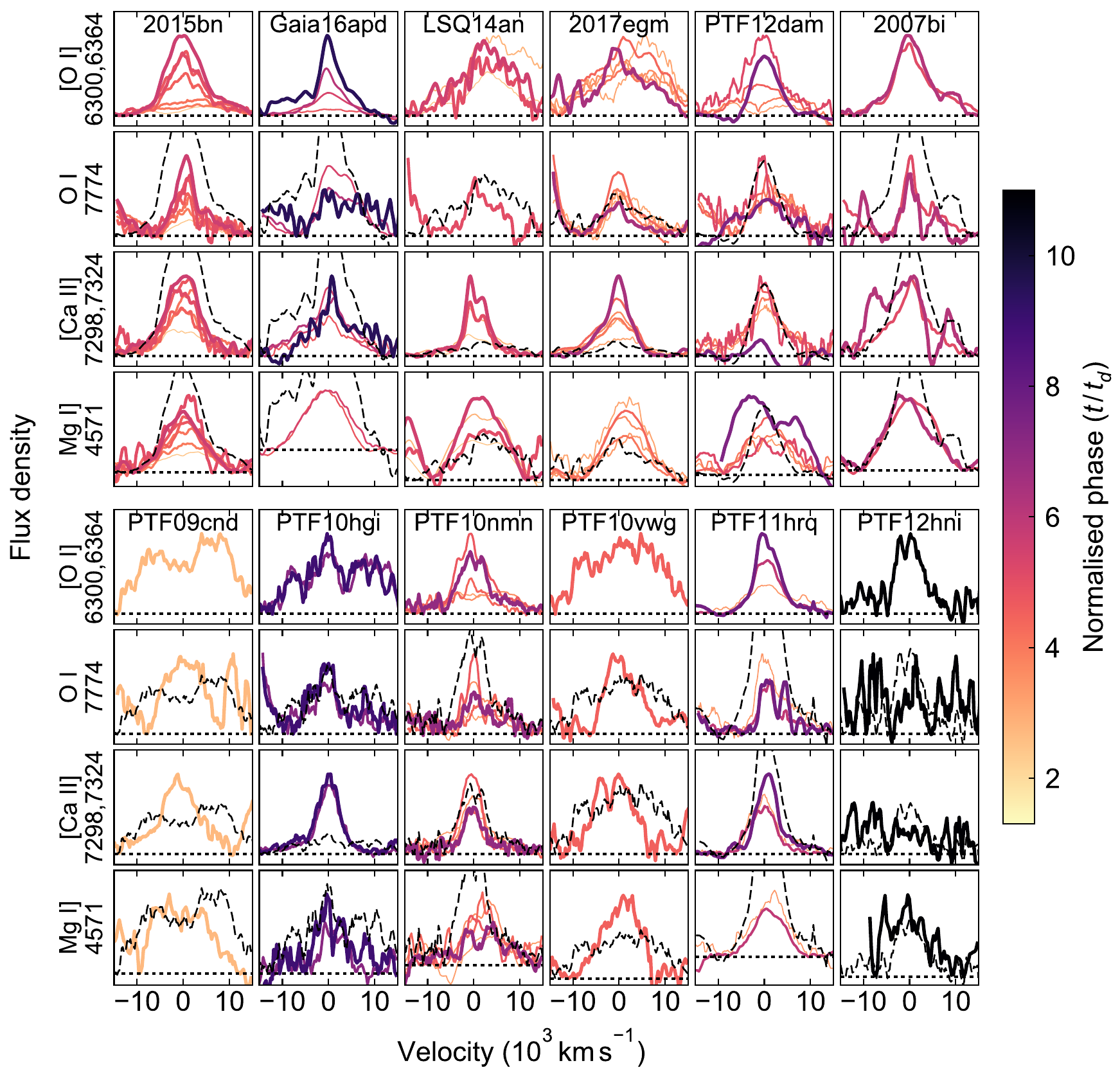}
\caption{Evolution of the line profiles for 
\o, \oi, \ca, and \mg.
Spectra have been continuum-subtracted and converted to velocity space. The color and thickness of each curve indicates the spectral phase, with the latest spectrum for each SLSN plotted with the darkest and heaviest line. Dashed lines indicate the zero-flux level of each subplot. The \o\ 
profile from the final epoch is overplotted as a dashed line on each panel as a visual aid to show the relative flux scales. At the latest phases, this line has a similar velocity to the other lines, but appears broader on this plot as it is a widely-separated doublet (6300,6364\,\AA).
}
\label{f:profiles}
\end{figure*}

Figure \ref{f:profiles} shows cutouts (in velocity space) around four of the strongest lines, \o, \oi, \ca, and \mg, for all 12 SLSNe in the sample. We discuss these on a case-by-case basis below.

s

\subsubsection{SN\,2015bn}

SN\,2015bn serves as an excellent baseline for comparison, with high signal-to-noise spectra well sampled from $\approx 100-400$ days. The line profiles are approximately Gaussian at all times. The most interesting characteristic is the narrowness of \oi\ relative to \o. \oi\ is primarily a recombination line, and is stronger at higher densities. Thus \citet{nic2016c} suggested the narrow \oi\ emission originated in a high-density region at low velocity coordinate, i.e.~close to the centre of the ejecta.

\subsubsection{Gaia16apd}

The line profiles in Gaia16apd are more complicated. Especially at later times, \o\ and \ca\ show some indication of a broad base and slightly narrower core. \citet{tau2009} found that this morphology is common in SNe Ic, and can be interpreted as either an axisymmetric explosion viewed edge-on (with the emitting material in a torus around the equator) or an ejecta distribution with a dense core.

\oi\ shows an even more interesting profile, with a shoulder or possible double-peak on the red side, offset by $\sim 4000$\,\kms\ from the centre. Emission lines with secondary peaks been seen previously in SNe Ic, and \citet{tau2009} propose that it could be attributable to large-scale clumping, the ejection of massive blobs of material, or even a unipolar jet -- though the latter interpretation seems unlikely when the other lines do not share a similar profile. In particular contrast to the other lines, \mg\ shows a simple Gaussian profile.

\subsubsection{LSQ14an}

LSQ14an shows a large contrast between the flux at 7300\,\AA\ and that at 6300\,\AA. While \mg\ exhibits a simple Gaussian shape, the other lines show a high degree of asymmetry. The peaks of the oxygen lines are consistent with zero velocity, but the profiles cut off sharply on the blue side and exhibit a broad, red shoulder out to several thousand \kms.

The very strong line at 7300\,\AA\ has sufficient S/N that it is clearly resolved into a double-peak, with the components separated by $\approx 2300$\,\kms. This feature is almost always attributed to \ca, but given the strength of the line, and other lines of ionised oxygen in LSQ14an, \cite{ins2017} and \cite{jer2017a} suggested that this is a blend with [\ion{O}{2}]\,$\lambda$7320,7330. The similarity in shape between this line and the \o\ and \oi\ lines confirms that the blended feature is most likely dominated by [\ion{O}{2}]. 

The oxygen-emitting region of LSQ14an thus appears to be highly asymmetric, and with higher ionisation than most SLSNe.

\subsubsection{SN\,2017egm}

SN\,2017egm initially exhibits very similar nebular-phase properties to SN\,2015bn (this despite the much faster evolution in the light curve, and unusual metal-rich host galaxy; \citealt{nic2017d,bos2018}). However, the relative strength of the \o\ line in SN\,2017egm does not increase with time, and as with LSQ14an we suggest that the strongest feature at 7300\,\AA\ is heavily contaminated by [\ion{O}{2}]. 

There is also a hint of a narrow core to \o, and in the final spectrum it appears somewhat asymmetric, being broader on the red side, though less so than LSQ14an. The other lines are well developed and appear largely Gaussian, with the \oi\ line narrower than \o.

\subsubsection{PTF12dam}

The \o\ line in PTF12dam strengthens slowly, and starts out very broad, but reaches a Gaussian profile by 324 days. The \mg\ line is broader than the others, and develops a hint of a flat top as the spectrum evolves.

\subsubsection{SN\,2007bi}

SN\,2007bi shows fairly straightforward line evolution (noting that the \ca\ in the later epoch is at low S/N and the implied line shape is likely contaminated by a spurious noise feature). The \oi\ line is again conspicuously narrow.


\begin{figure*}
\centering
\includegraphics[width=16cm]{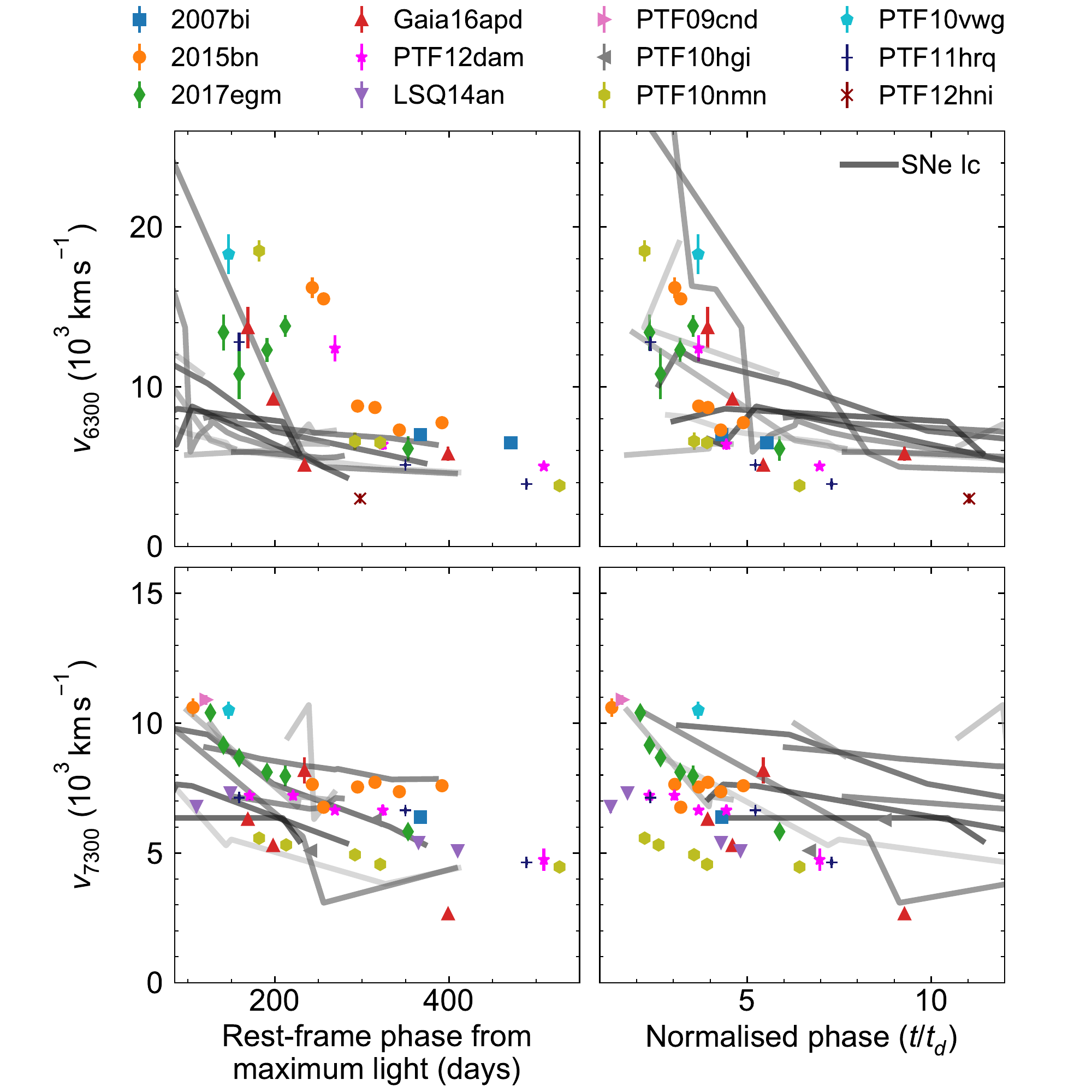}
\caption{Velocity evolution of the \o\ (top) and \ca\ (bottom) lines. The left panels show phases with respect to maximum light, while on the right phases are normalised by the light curve decline timescale $t_d$. Gray lines show the comparison sample of SNe Ic, which have been measured in the same way. While SLSNe initially show larger \o\ velocities than SNe Ic, this discrepancy disappears when accounting for the difference in light curve timescales.}
\label{f:v-ic}
\end{figure*}

\subsubsection{PTF09cnd}

The \o\ line in PTF09cnd has a possible flat-topped profile, which would indicate the emission comes from a shell. However, given that there is only one epoch of spectroscopy at 121 days, it is also possible that we are seeing a transitional phase before the line fully develops. This latter interpretation is consistent with the high \ca/\o\ ratio and overall very broad \o\ profile. Support for a genuine flat-top comes from \mg, which also shows a hint of such a profile, while even \oi\ is broader and less Gaussian than in the rest of the sample. The peak of \ca\ appears to be blueshifted, which could arise if there is a large optical depth blocking emission from the receding part of the ejecta.

\subsubsection{PTF10hgi}

Like LSQ14an and SN\,2017egm, the PTF10hgi spectra are dominated by the feature at 7300\,\AA, which at the observed strength is most likely a blend of \ca\ and [\ion{O}{2}]. If most oxygen is ionised, this would help explain why \o\ is so weak (barely above the noise at the earlier epoch, 241 days). At later times, \o\ and \oi\ appear to have similar strengths and profiles, with a possible shoulder on the red side, however in this case the 7300\,\AA\ feature is a smooth Gaussian.

\subsubsection{PTF10nmn}

PTF10nmn shows features largely typical of the class, with Gaussian profiles and a narrow \oi, once again with an asymmetric red shoulder. \mg\ in this case is weak and appears contaminated by spurious noise features.

\subsubsection{PTF10vwg}

The lines in PTF10vwg are on the broad side and appear to depart from simple Gaussian profiles, though we caution that as with PTF09cnd, the only spectrum is at a fairly early phase (147 days) and has been heavily smoothed. Nevertheless, \o\ shows a possible flat-top, and it is the only event that shows this in the \ca\ profile too. There is also a hint of the same in \mg. The \oi\ line is more typical, with a weak but detectable excess on the red side.

\subsubsection{PTF11hrq}

PTF11hrq has a prototypical line evolution with smooth Gaussian profiles, including a narrow \oi\ line.

\subsubsection{PTF12hni}

The only clearly detected emission line in the spectrum of PTF12hni is \o. There is a plausible narrow \oi\ centred at the correct wavelength, but this is equally consistent with noise. There is no obvious \ca\ emission at 298 days, while \mg\ falls outside the observed wavelength range.

\subsection{Summary of line shapes}

Overall, most SLSNe show smooth Gaussian line profiles without strong evidence for double peaks or flat tops, particularly in the \o\ line that is most commonly used to diagnose geometry \citep{maz2005,mae2008,tau2009}. This suggests that ejecta asymmetry is most likely modest. Polarisation has been detected in the emission from SN\,2015bn \citep{ins2016,leloudas2017} and SN\,2017egm \citep{bos2018}, implying that despite the smooth line profiles, there is certainly some asymmetry in these events, comparable to broad-lined SNe Ic.
LSQ14an stands out in our analysis as a candidate for a particularly asymmetric explosion.

The line profiles may also offer a clue to the nature of the spectral continuum at $t/t_d<4$. From an observer's point of view, an internal photosphere covers more of the receding side of the ejecta, predicting early blue-shifted line profiles that move towards the rest wavelength with time as more of the ejecta becomes optically thin. The fact that we do not see significant blue-to-red movement in the emission profiles for any SLSNe suggests either that the source of continuum is external (i.e.~interaction), or that the photosphere is already below a velocity coordinate $\lesssim 500$\,\kms\ (the typical size of our uncertainties) by 100 days after maximum light.

The most complicated line is \oi, which is often significantly narrower than the other strong oxygen line, \o. It also shows a shoulder or secondary maximum in at least half of the SLSNe, but always on the red side. This cannot be explained in terms of an oxygen-rich ejecta blob or unipolar jet, since in either case we expect the excess would be equally likely to be blue-shifted. 

Thus the most likely explanation is contamination by another line. As shown in Figure \ref{f:id}, there is a doublet of \ion{Mg}{2} at 7877, 7896\,\AA, which could contribute to the red shoulder. Since this asymmetry is not typically seen in SNe Ic, stronger \ion{Mg}{2} emission may constitute another difference between SLSNe and SNe Ic. As this line arises from an excited state of \ion{Mg}{2}, a possible explanation is higher line excitation in SLSNe, either due to a higher temperature or some non-thermal effect \citep[e.g.][]{maz2016}. A future test will be to compare \ion{Mg}{2} lines in the UV and NIR between SLSNe and SNe Ic. UV lines at 2795, 2802\,\AA\ may contribute a large fraction of the cooling \citep{jer2017a}, while NIR lines at 9218\,\AA\ and 10914\,\AA\ result from the lower level of the 7877,7986\,\AA\ transition.

\subsection{Velocities}

We now use these same lines to infer the velocity structure of the SLSN ejecta. Velocity measurements during the nebular phase have the advantage of probing deeper into the ejecta, compared to the photospheric phase, and hence measure the bulk velocities of different emitting layers. To proceed, we assume that all lines can be represented by Gaussian profiles, which works well in the majority of cases. In a few cases (typically at $t\lesssim 150$ days), the \o\ line does not exhibit a profile that can be reasonably represented with a Gaussian (see appendix). We exclude such cases from our analysis here.

We fit to \o, \ca, \oi, and \mg, accounting for the doublet natures of lines when appropriate. We assume a ratio 3:1 between the 6300\,\AA\ and 6364\,\AA\ components of [\ion{O}{1}], i.e.~that the line is optically thin (assuming a 1:1 ratio made no significant changes to the inferred velocities). We fit to the unsmoothed spectra, in order to better quantify the error in the fit. We report the full-width at half-maximum (FWHM) of each Gaussian, taking as our uncertainty the variance returned by \textsc{SciPy.curve\_fit}.

An alternative measure of line velocity is the full-width at zero intensity (FWZI). This has a number of advantages over the FWHM: it is less sensitive to the shape of the line profile, and better captures the full velocity extent of each ion. However, the FWHM is preferred here since it is less sensitive to the exact placement of the continuum. Taking the zero flux level as the point where the line intensity falls below 1\% of maximum, FWZI\,$\simeq2$\,FWHM for a Gaussian profile. The FWHM contains 90\% of the emitted flux, but we caution that 10\% of each ionic species may be located at velocities extending to twice the values reported here.

The velocity evolution of the \o\ and \ca\ lines is shown in Figure \ref{f:v-ic}. We plot against the rest-frame phase, with and without normalising to $t_d$, and in both cases compare to SNe Ic. Using both observed and normalised timescales allows for a fair comparison, and we see that while SLSNe initially exhibit much greater \o\ velocities than SNe Ic, this discrepancy largely disappears once we account for the slower light curve evolution of the SLSNe.

While the overall trend of initially decreasing and then relatively flat velocity is exhibited by each line, the absolute scales are quite different. The \o\ velocities start out much higher than \ca\ (and the other lines; see discussion below), up to $\sim 16,000$\,\kms. The width drops quickly (and smoothly) as the line becomes more prominent and peaked; a typical plateau velocity for \o\ is $\sim7000$\,\kms. \ca\ is the easiest line to measure in general, since it is strong and typically Gaussian-shaped starting from our earliest spectra. This line shows a very smooth evolution from around $10,000$\,\kms\ at 100 days, to $5000-7000$\,\kms\ when its velocity plateaus. 

\begin{figure}
\centering
\includegraphics[width=8.25cm]{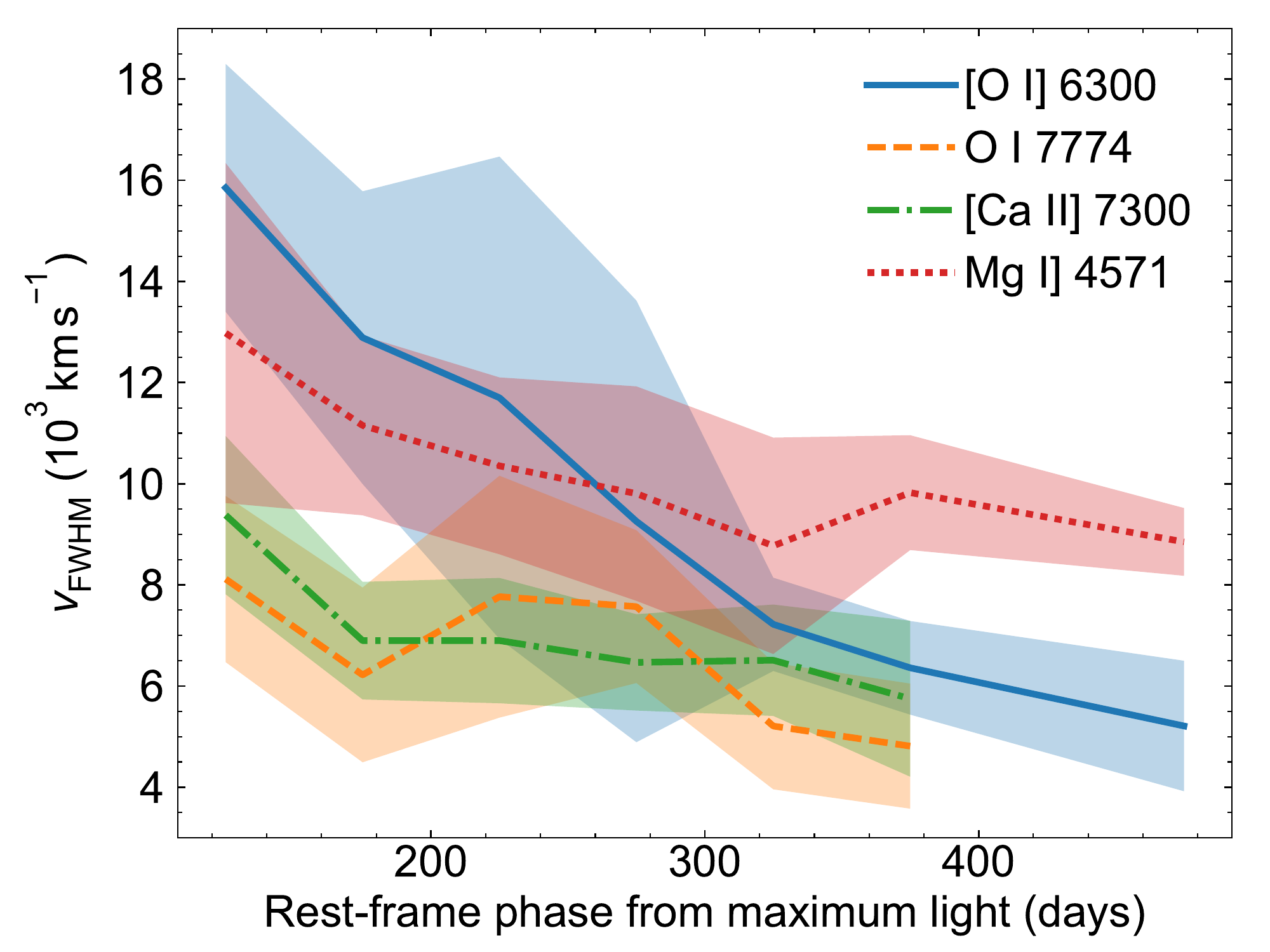}
\includegraphics[width=8.25cm]{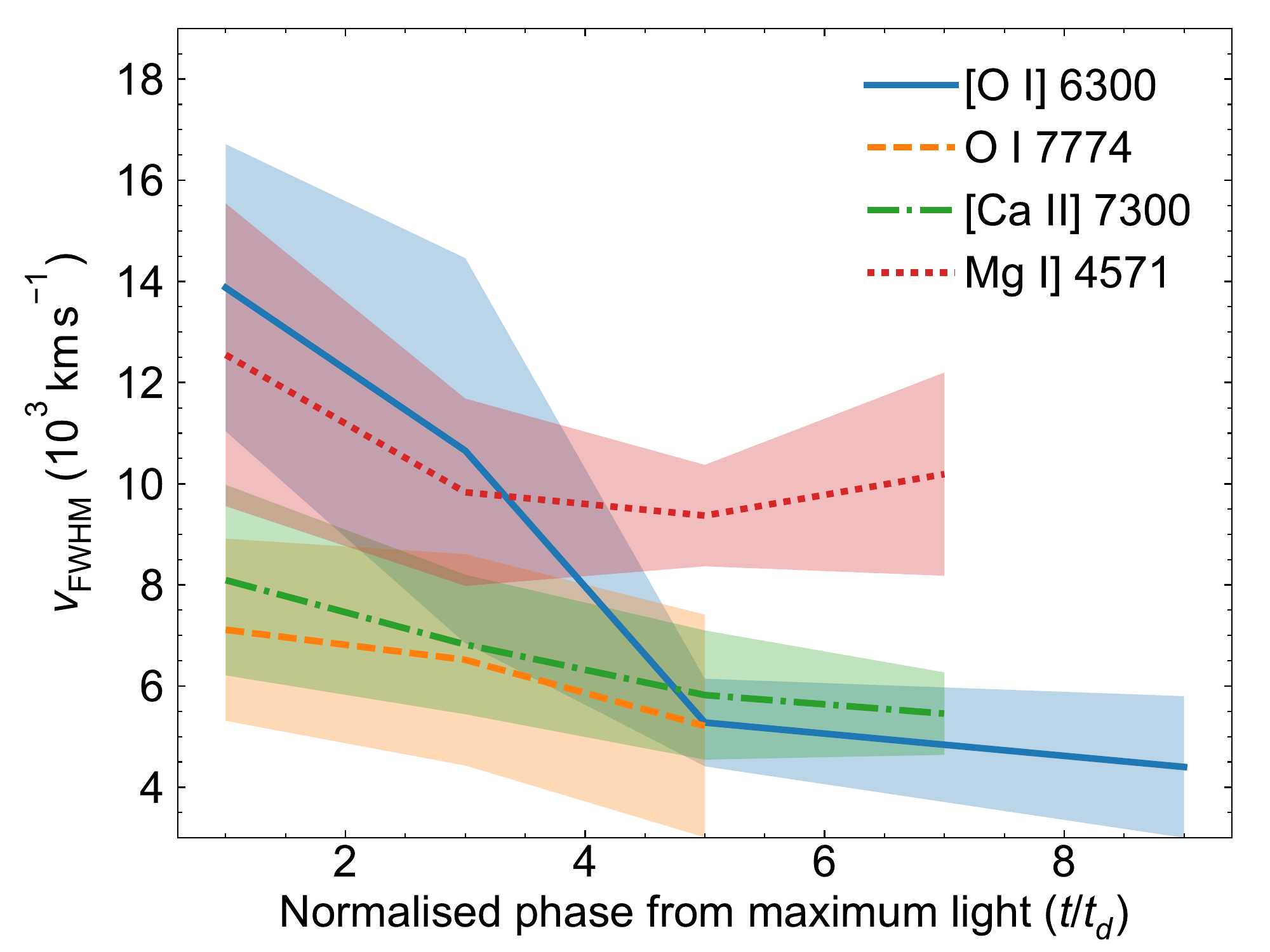}
\caption{Top: Velocities of each emission line averaged over the SLSN sample, binned in 50 day intervals. Shaded regions indicate one standard deviation. Bottom: Same as top, but in this case phases are normalised by the light curve timescale, and velocities are binned in intervals of $\Delta(t/t_d)=2$.}
\label{f:vave}
\end{figure}

We calculate the mean velocities of \o, \ca, \oi\ and \mg\ in 50 day bins (rest-frame phases) and bins of $\Delta(t/t_d)=2$ (normalised phases); these are plotted in Figure \ref{f:vave} (all individual measurements for these 4 lines are shown in the appendix). This clearly shows that \o\ and \oi\ emission arises in regions with significantly different velocity, in many cases until 400 days (or $\sim 5 t_d$) after maximum.
\oi\ may approach $10,000$\,\kms\ at early times, but quickly settles at $\sim 5000$\,\kms.
\ca\ emission appears to come from the same velocity zone.
\mg\ shows the flattest velocity evolution among the strong lines, remaining between $\approx 7000-12,000$\,\kms\ at all times.

The observed hierarchy in velocities is consistent with explosion models of massive stars. Examining the models of stripped SNe used by \citet{jer2014}, oxygen is distributed across a wide range of zones of different elemental composition: it is abundant in the O/Si/S zone (Ne-burning ashes), O/Ne/Mg zone (C ashes), and above this we expect unburned oxygen from the progenitor and/or an O/C zone (explosive He-burning). Calcium is present in the Si/S (O-burning) and O/Si/S zones, and \ca\ is expected to be the primary coolant of the Si/S zone \citep{jer2017a}. Magnesium is mostly in the O/Ne/Mg zone. 

Figure \ref{f:vave} therefore represents a consistent picture in which \o\ initially arises in the outer layers, but as time progresses we see a larger contribution from the deeper O/Ne/Mg zone, apparent when the \o\ velocity crosses the \mg\ velocity at $\sim 300$ days. Later, we see emission from the O/Si/S zone after $\sim 400$ days, when the average velocity of \o\ becomes comparable to \oi\ and \ca. The \oi\ emission, in contrast, arises primarily from the O/Si/S zone at all times, and is therefore a useful diagnostic of the density in this layer. These results will be important for our analysis in section \ref{s:diss}.

\section{Line luminosities and physical constraints}
\label{s:lums}

\begin{figure*}
\centering
\includegraphics[width=16cm]{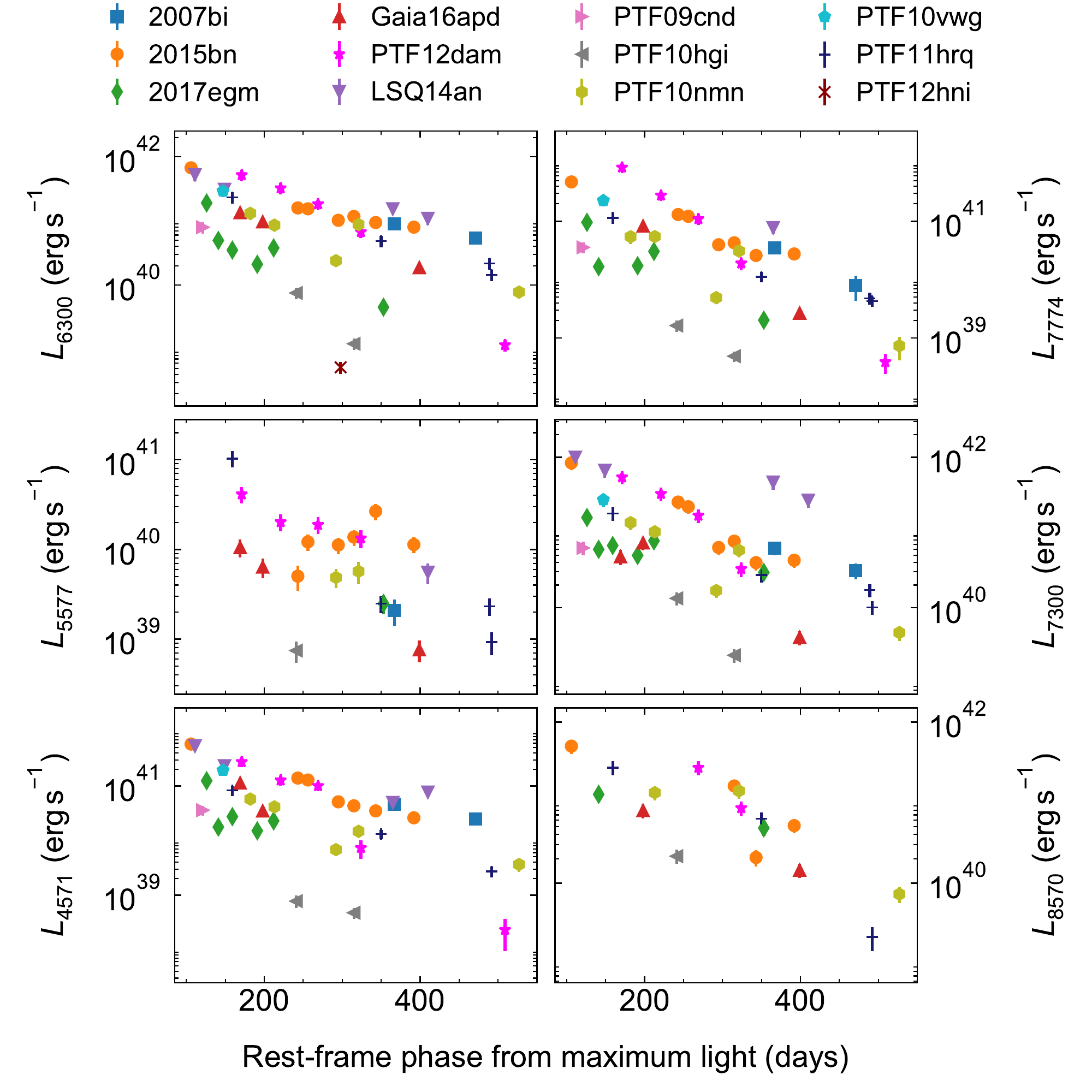}
\caption{Integrated luminosities in emission lines as a function of time. All lines show a gradually decreasing flux as the SLSNe fade.}
\label{f:lums}
\end{figure*}

\begin{figure*}
\centering
\includegraphics[width=16cm]{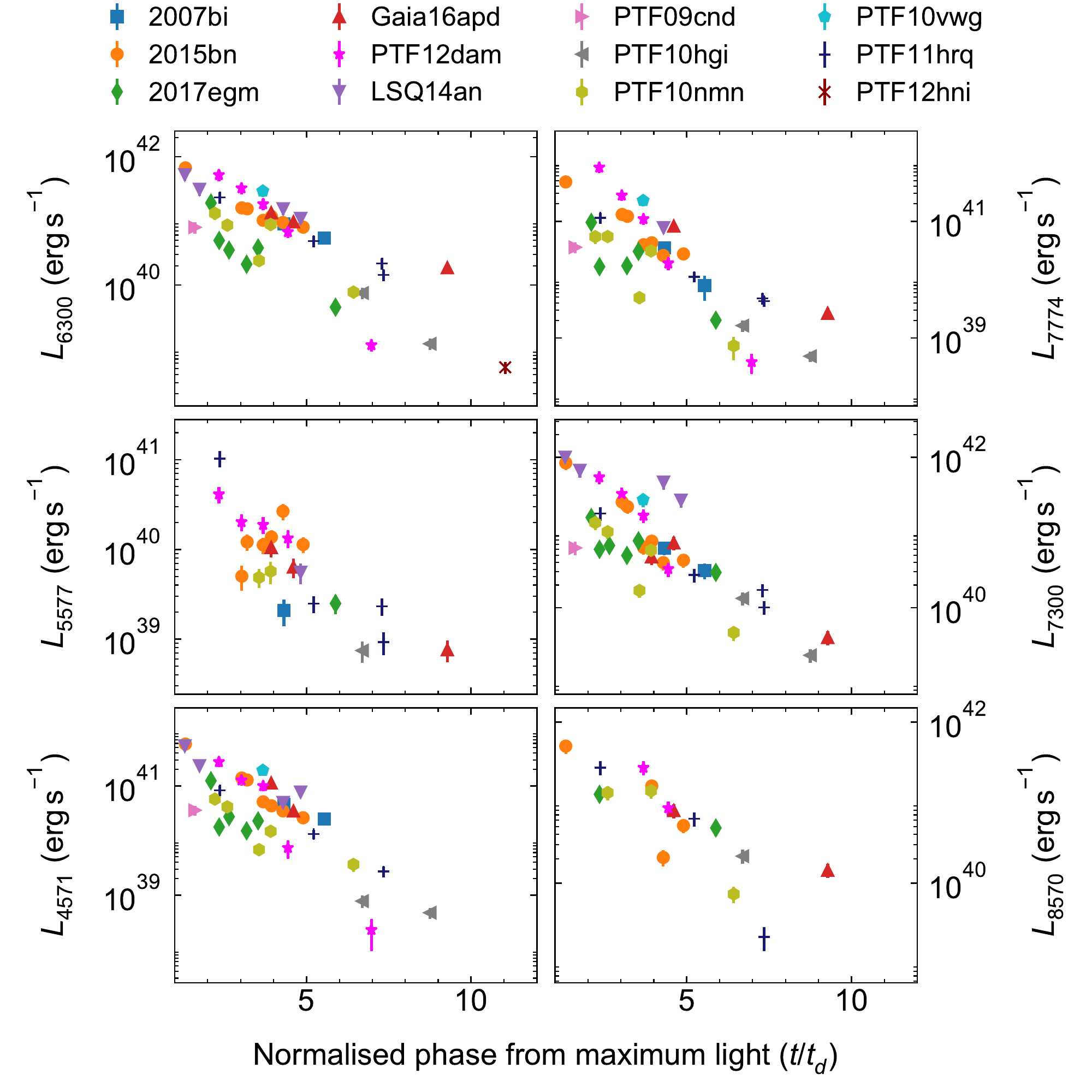}
\caption{Same as Figure \ref{f:lums}, but with phases normalised by the light curve decline timescale $t_d$. The line luminosities for all SLSNe now cluster tightly around relatively well-defined decay slopes.}
\label{f:ltd}
\end{figure*}

The luminosities of -- and ratios between -- spectral lines can be used to derive important constraints on the physical conditions in the ejecta, including the temperature, density and mass of the emitting material. Clearly, such insights offer clues about the progenitors and power sources of SLSNe. Progress has been made through constructing detailed models for the line emission in SLSNe \citep{gal2009,jer2017a}, and by employing analytic relations to convert integrated luminosities into physical parameters \citep{nic2016c,jer2017a}. In both cases, the analysis points towards rather massive explosions, with inferred oxygen masses of $\gtrsim 10$\,\M\ for some slowly-evolving SLSNe. These events are thought to most likely originate from the high-mass end of the SLSN population \citep{nic2015b,nic2017c}, whereas here we include SLSNe with a range of light curve properties. Full radiative transfer modeling is beyond the scope of this work, but we can make progress by analysing line ratios in a moderate sized sample of SLSNe for the first time. We will make heavy use of relations provided by \citet{jer2014,jer2015,jer2017a,jer2017b}.

\begin{figure*}
\centering
\includegraphics[width=16cm]{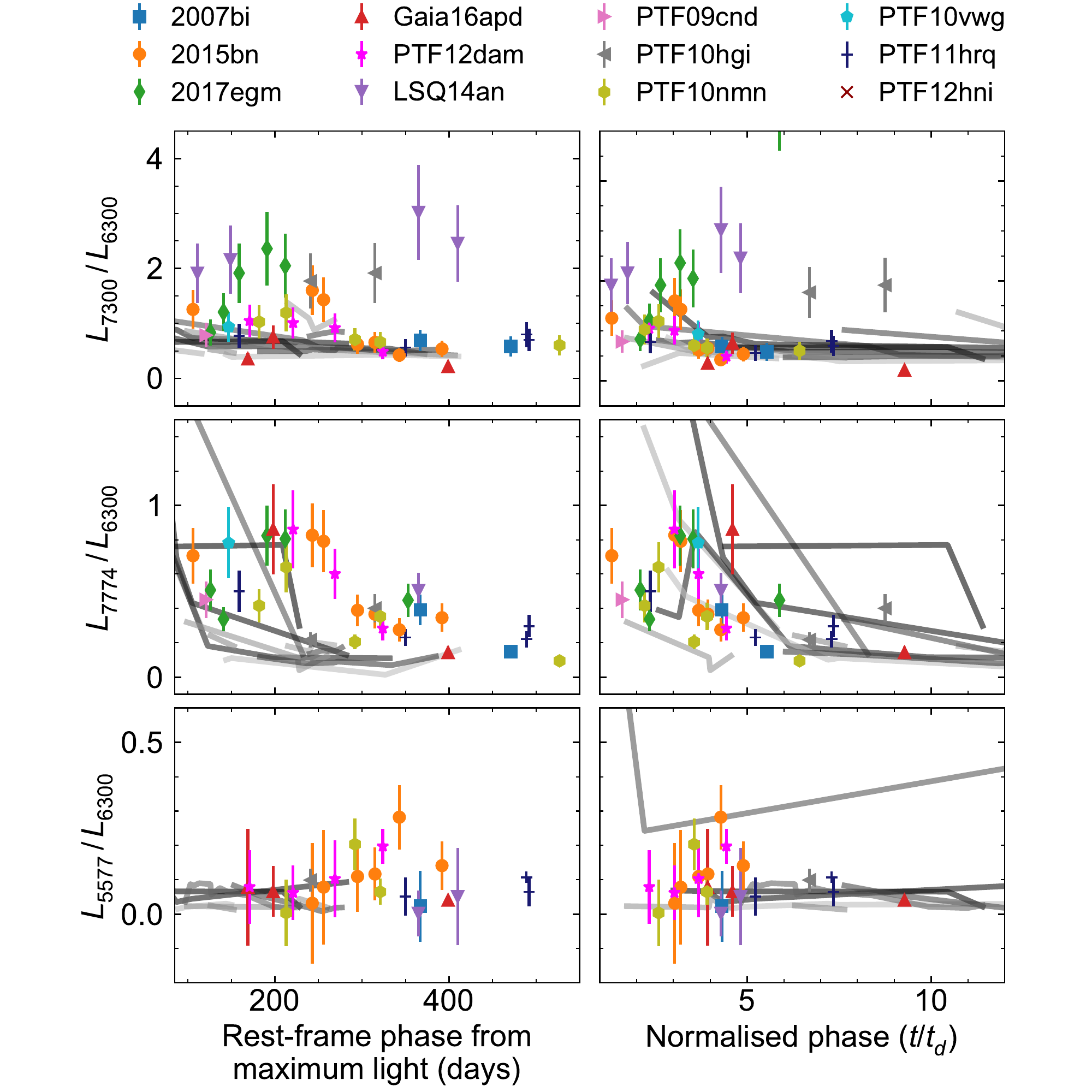}
\caption{Temporal evolution in the ratios of selected spectral lines to \o, used for diagnosing physical properties. The left column shows phase in the SLSN rest-frames, while the right column shows the same phases normalised by the light curve decline timescale $t_d$.}
\label{f:ratio}
\end{figure*}

\begin{figure}
\centering
\includegraphics[width=8.25cm]{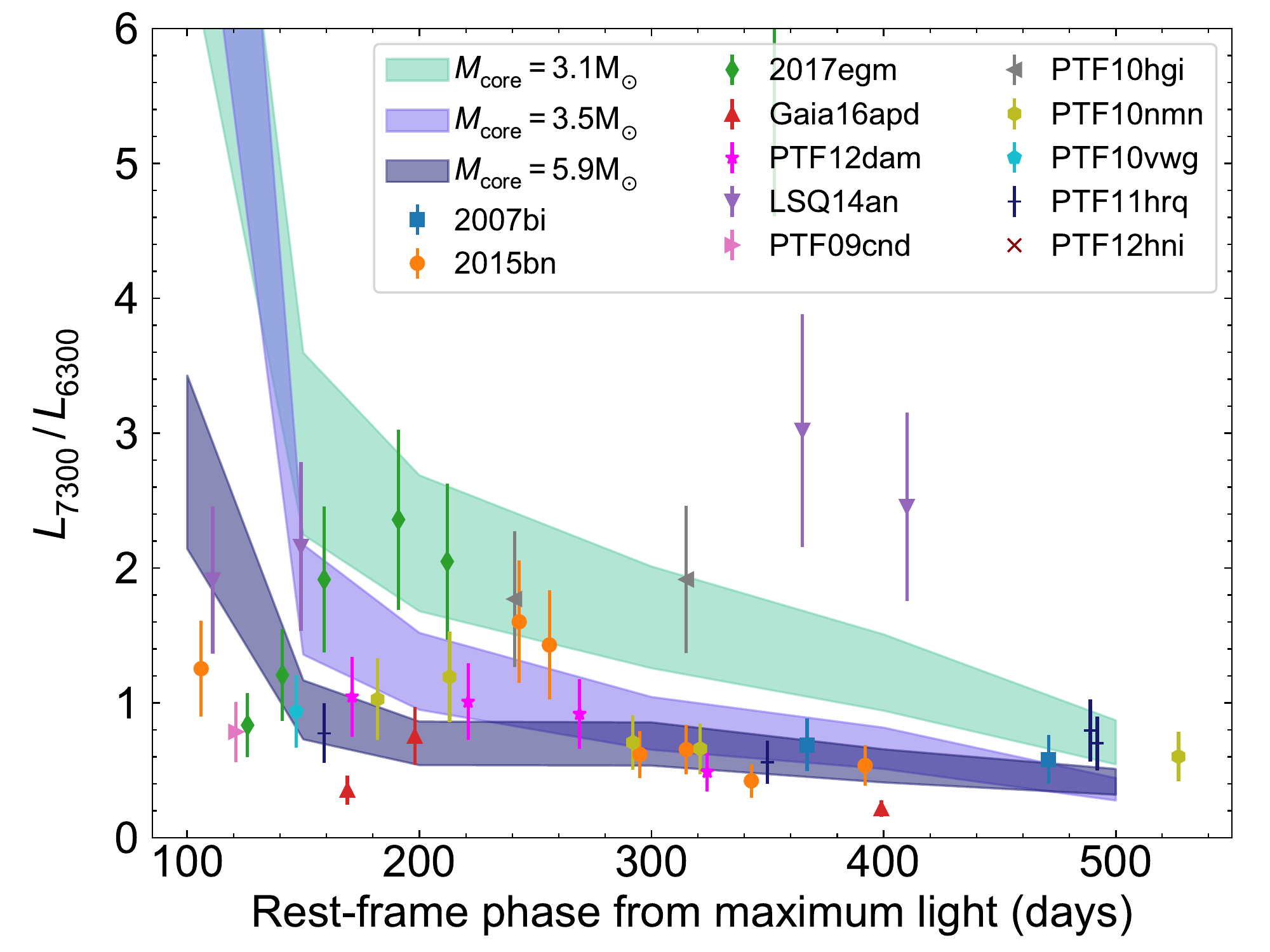}
\caption{\ca/\o\ ratio compared with stripped-envelope SN models from \citet{jer2015}, rescaled to a velocity range 5000-8000\,\kms. At late times (when the lines originate from similar velocity coordinates), the ratio is best matched for larger helium core masses.}
\label{f:core}
\end{figure}

\begin{figure}
\centering
\includegraphics[width=8.25cm]{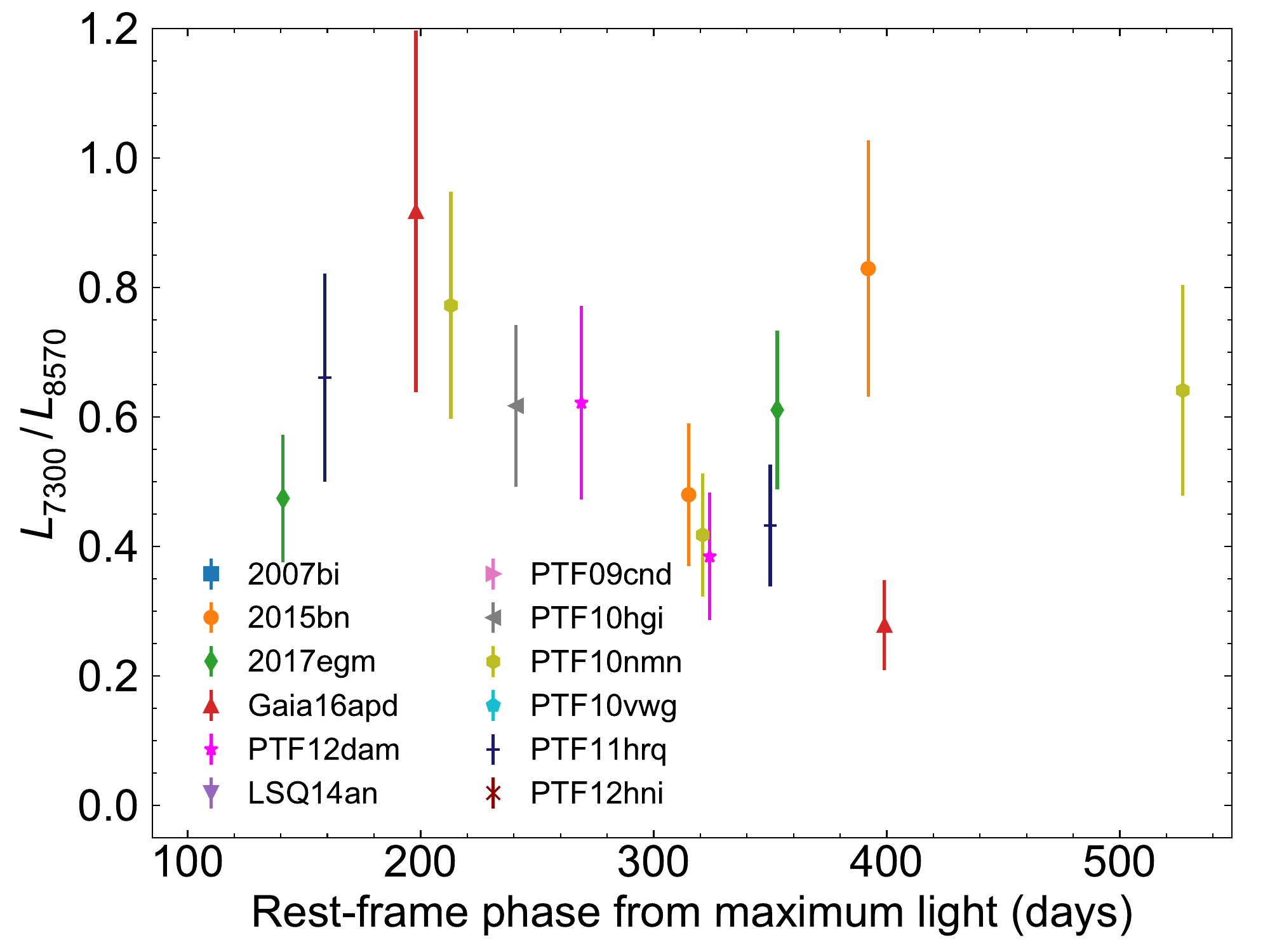}
\caption{Ratio of \ca\ to the \ion{Ca}{2} NIR triplet.}
\label{f:ca}
\end{figure}

\subsection{Integrated fluxes}

We measure the integrated luminosities of various diagnostic lines: \o, \oi, [\ion{O}{1}]\,$\lambda$5577, \mg, \ca, and the \ion{Ca}{2} NIR triplet (8498, 8542, and 8662\,\AA). Most of these are reasonably well isolated and can be measured by direct numerical integration, after subtracting a linear fit to the continuum as in Section \ref{s:profiles}. The exception is [\ion{O}{1}]\,$\lambda$5577, which is weak and blended with [\ion{Fe}{2}]\,$\lambda$5528. In this case we estimate the luminosity using a double Gaussian fit. The centres of the Gaussians are fixed at the rest wavelengths of these two lines, and their widths are assumed to be equal. 

Again we analyse the original unsmoothed spectra, and use the observed S/N to calculate the uncertainties. While we have endeavoured to scale all spectra to contemporaneous photometry before making measurements, in some cases large interpolations/extrapolations in light curves were required. We therefore impose an error floor of 20\%, added in quadrature to the statistical error, to account for systematic errors in the absolute flux calibration or placement of the continuum. Note that uncertainties in continuum level affect our flux measurements at the $\lesssim 10$\% level; see section \ref{s:cont}.

The results are plotted in Figures \ref{f:lums} and \ref{f:ltd}. In the first case we show luminosities plotted against rest-frame phase from maximum, whereas in the latter we normalise the times by $t_d$. We only show measurements for which the luminosity is measured at S/N\,$>3$ (i.e., the line is detected in the spectrum at $>3\sigma$ significance, including the additional systematic error in flux calibration/continuum placement).

The luminosities of the strongest lines can be as high as $\sim 10^{42}$\,\ergs\ at $\sim 100$ days, dropping to $\sim 10^{40}$\,\ergs\ by 400-500 days. In general this decrease is quite smooth in time, though this is less clear in the case of the noisy [\ion{O}{1}]\,$\lambda$5577 line. The smooth evolution is particularly striking when accounting for the differences in light curve decline rates: when plotted against $t/t_d$, the measurements for all SLSNe cluster quite tightly around the same exponential slope (Figure \ref{f:ltd}). This indicates that the rate of fading in the spectrum is very closely tied to that of the light curve. Comparing to late-time SLSN bolometric light curves \citep{chen2015,nic2016b}, a strong line such as \o\ typically accounts for around 5\% of the total SLSN luminosity.

More instructive is to examine the ratios between line luminosities. This also has the advantage of being independent of the absolute flux normalisation of the spectra, facilitating comparison with SNe Ic. In Figure \ref{f:ratio}, we show the luminosities of key diagnostic lines \ca, \oi\ and [\ion{O}{1}]\,$\lambda$5577, normalised to the luminosity of \o. With the exception of [\ion{O}{1}]\,$\lambda$5577, the luminosity in each line drops from $\sim 1-2$ to $\lesssim 0.5$ times the \o\ luminosity (the same is true for \mg; see appendix).

Compared to SNe Ic, the observed ratios of \o/ \ca\ and \oi/\o\ appear significantly enhanced in SLSNe. Indeed, the early prominence of \ca\ relative to \o, and unusually strong \oi, are some of the defining features of SLSN nebular spectra (Section \ref{s:obs}). Figure \ref{f:ratio} shows that this is partially attributable to the overall slow spectroscopic evolution of SLSNe compared to SNe Ic. When the phase is expressed relative to the light curve timescale, we see that most of the SN Ic observations are at $t/t_d>5$, by which time the line ratios in SLSNe have decreased to comparable values to the SNe Ic. Accounting for the relative light curve timescales mitigates for the fact that SLSNe stay hotter for longer than SNe Ic, and the eventual similarity in the line ratios may indicate that the structure and composition of SLSNe and SNe Ic are not dissimilar.

\subsection{[\ion{O}{1}]\,$\lambda$5577/\o\ ratio and temperature}

The [\ion{O}{1}]\,$\lambda$5577/\o\ ratio appears fairly constant in time, at a typical value of 0.1--0.2. Assuming these lines arise in the same zones, the ratio is sensitive only to the temperature and the optical depths in the lines. After $\approx 200$ days, the centroid of the \o\ line is always close to 6300\,\AA, implying that the line is optically thin (Figure \ref{f:evolve}). The optical depth in [\ion{O}{1}]\,$\lambda$5577 is lower than \o; therefore both lines should be optically thin at late times \citep{jer2014}, and we can use their ratio as a thermometer. There are significant caveats to this approach \citep{jer2017b} -- for example, [\ion{O}{1}]\,$\lambda$5577 may not form in local thermodynamic equilibrium (LTE) at late times, and the ratio can also depend on clumping.

Assuming that the lines are optically thin and form in LTE, the observed ratios $\lesssim 0.2$ imply temperatures $T \lesssim 5000$\,K, using equation 2 from \citet{jer2014}, with an overall plausible range of 4000--6000\,K for ratios of 0.05--0.3.

\subsection{\o\ and ejecta mass}

Since oxygen is expected to account for 60--70\% of the ejecta in stripped-envelope SNe \citep[e.g.][]{mau2010}, inferring the oxygen mass gives a good handle on the total mass. \citet{jer2017a} found that the luminosity in the \o\ line in some SLSNe could only be reproduced by models with $\gtrsim 10$\,\M\ of oxygen. In the case of SN\,2015bn, \citet{nic2016c} used a scaling relation to estimate an oxygen mass $\sim 9$\,\M\ for a temperature of 4900\,K.

Such analytic relations are fraught with danger however, as the mass is an exponential function of the temperature, which is itself uncertain due to the weakness of [\ion{O}{1}]\,$\lambda$5577 and the difficulty of deblending this line with [\ion{Fe}{2}]. SN\,2015bn was an unusual case in that the spectra were of very high S/N, and the [\ion{O}{1}]\,$\lambda$5577 line was well resolved. Applying equation 3 from \citet{jer2014} results in ejecta masses spanning many tens of \M, and with comparably large error bars. This ignores the effect of clumping, which can further skew mass estimates by factors of several.

\citet{jer2017a} show how the line luminosity varies as a function of mass, energy deposition and clumping, and argue that luminosity comparable to SN\,2015bn, SN\,2007bi, and LSQ14an requires at least $\sim 10$\,\M\ of ejecta (assuming a carbon-burning composition). One solid conclusion we can draw is that none of the SLSNe in our sample have \o\ luminosities that significantly exceed these three events (Figure \ref{f:lums}). Thus the ejecta masses estimated for these slowly evolving SLSNe (see also PTF12dam) likely represent the upper end of the SLSN mass distribution. This agrees with results from light curve modelling \citep{nic2017c}.

\subsection{\ca/\o\ ratio and core mass}

The ratio \ca/\o\ is sometimes used to estimate the helium core mass of the progenitor, with a larger luminosity in \o\ indicating a more massive core. However, this method has many caveats: the ratio depends sensitively on the distribution of calcium in the ejecta, the phase of observation, and the velocity of expansion, with higher velocities increasing \ca\ relative to \o\ \citep{fra1989,hou1996,jer2017b}.

While keeping the above in mind, it is still illustrative to compare our observed ratios to spectral models. In Figure \ref{f:core} we plot the ratio from the stripped-envelope SN models of \citet{jer2015}. Those models were calculated for a velocity of 3500\,\kms, lower than that observed in SLSNe. In the models of \citet{fra1989}, \ca/\o\,$\propto v$, so we use that scaling here, adopting a velocity range of 5000--8000\,\kms\ appropriate for SLSNe. 

Because mixing between zones can strongly affect the line ratios, the comparison is more reliable at times $\gtrsim 300$ days, when the velocity profiles indicate that oxygen and calcium emission are coming from a similar region in the ejecta. This is reflected in the spread of observed ratios at early times, before the eventual convergence to $\sim 0.5$. While absolute values should not be taken literally, we find that the highest mass model, with a $5.9$\,\M\ helium core, gives the best match to most events, but the range in the observations and the simplicity of the model mean that there is at least a factor $\gtrsim 2$ uncertainty in this estimate.

Plotting the ratios as a function of $t/t_d$, SN\,2017egm, LSQ14an and PTF10hgi appear even more discrepant in their \ca/\o\ ratios, compared to the other SLSNe and to the model predictions. This strengthens the conclusion that the 7300\,\AA\ feature is dominated by [\ion{O}{2}] in these events. We note that if the \ca\ line is contaminated by [\ion{O}{2}] (but at a lower level) in other SLSNe, the true ratio of \ca\ to \o\ must be even lower, which would favour larger core masses.

\subsection{\ca/\ion{Ca}{2} NIR and electron density}

In Figure \ref{f:ca}, we show the ratio between the strongest calcium lines, \ca\ and the \ion{Ca}{2} NIR triplet. \citet{jer2017a} pointed out that the triplet was unusually strong in SN\,2015bn relative to \ca, yet the lines do appear to originate in the same velocity zone \citep{nic2016c}. If the 7300\,\AA\ line has a significant component from [\ion{O}{2}], the true ratio of the NIR triplet to \ca\ is even lower and thus more discrepant with normal SNe Ic.

For our sample, we measure typical values \ca/ \ion{Ca}{2}\,NIR\,$\sim 0.5$. Comparing to the detailed models from \citet{jer2017a}, this indicates an electron density $n_{\rm e} \gtrsim 10^8$\,cm$^{-3}$ and a temperature $> 4000$\,K. \citet{jer2017a} also showed that to attain this electron density with a reasonable total mass for the emitting zone, the ejecta must be significantly clumped, i.e.~the filling factor is $f \ll 1$.

\begin{figure}
\centering
\includegraphics[width=8.25cm]{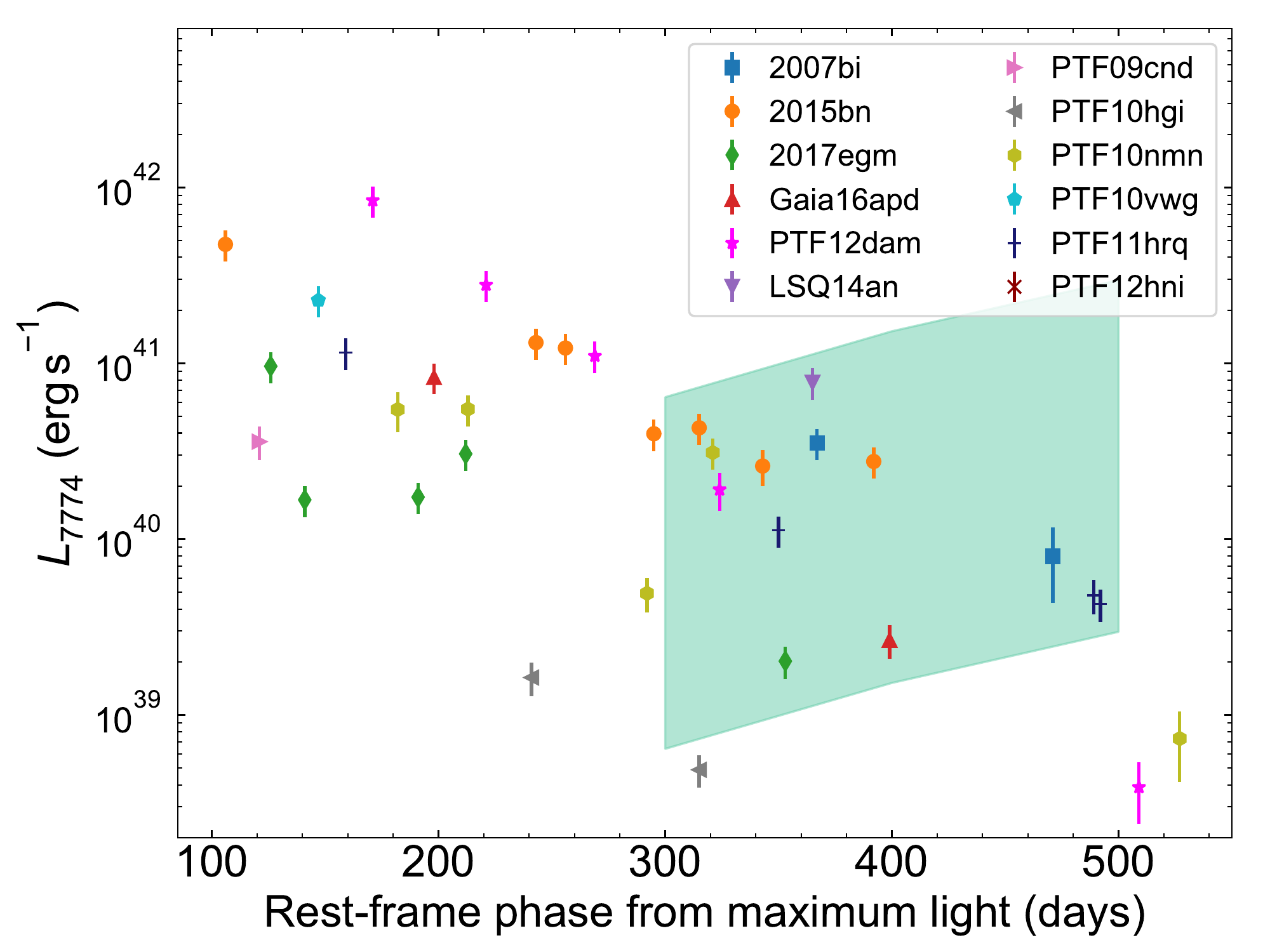}
\caption{Evolution of the \oi\ line. The shaded region indicates theoretical luminosities for electron densities and filling factors satisfying $10^{14} \lesssim n_{\rm e}^2 f \lesssim 10^{16}$\,cm$^{-6}$.}
\label{f:ne}
\end{figure}

\subsection{\oi\ and constraints on clumping}

Our velocity analysis suggested that \oi\ emission arises from deep in the ejecta, overlapping the calcium-rich zones. \oi\ is a recombination line and can therefore provide an independent check on the density and filling factor: $L_{\rm rec} \propto n_{\rm e}^2 f$. However, this line can remain optically thick for hundreds of days, and thus at earlier epochs there may a significant contribution from scattering \citep{jer2015}.

Using recombination coefficients from \citet{jer2015} and \citet{mau2010}, and assuming a velocity of 6000\,\kms, we show in Figure \ref{f:ne} that the \oi\ luminosity at 300--500 days is consistent with $10^{14} \lesssim n_{\rm e}^2 f \lesssim 10^{16}$\,cm$^{-6}$. For modest filling factors $f<0.1$, this translates to a density $n_{\rm e}>10^{7.5}$\,cm$^{-3}$. Or fixing $n_{ \rm e}>10^8$\,cm$^{-3}$ based on the calcium lines, we have $f<1$, i.e.~the observed lines cannot be reproduced without some degree of clumping. 

We can verify this with a simple calculation: for a fiducial 5\,\M\ expanding at 6000\,\kms\ for 400 days, the number density of ions is $\sim 10^7$\,cm$^{-3}$ (assuming an oxygen-dominated composition). This is the maximum possible electron density for un-clumped ejecta, if the ejecta are singly ionised on average. Thus to achieve $n_{ \rm e}>10^8$\,cm$^{-3}$ requires $f<0.1$ (i.e., the volume occupied by the ejecta is only 10\%\ of the available expansion volume). If the oxygen is instead mostly neutral, as may well be the case based on the dominant \o\ line, the situation requires more extreme clumping.

\mg\ is also a recombination line and may be a useful indicator of the electron density at higher velocity coordinate; however, this line also contributes significant cooling, and thus detailed spectral models are needed to disentangle these two components \citep{jer2017a}.

\section{Discussion}
\label{s:diss}

\begin{figure*}
\centering
\includegraphics[width=14cm]{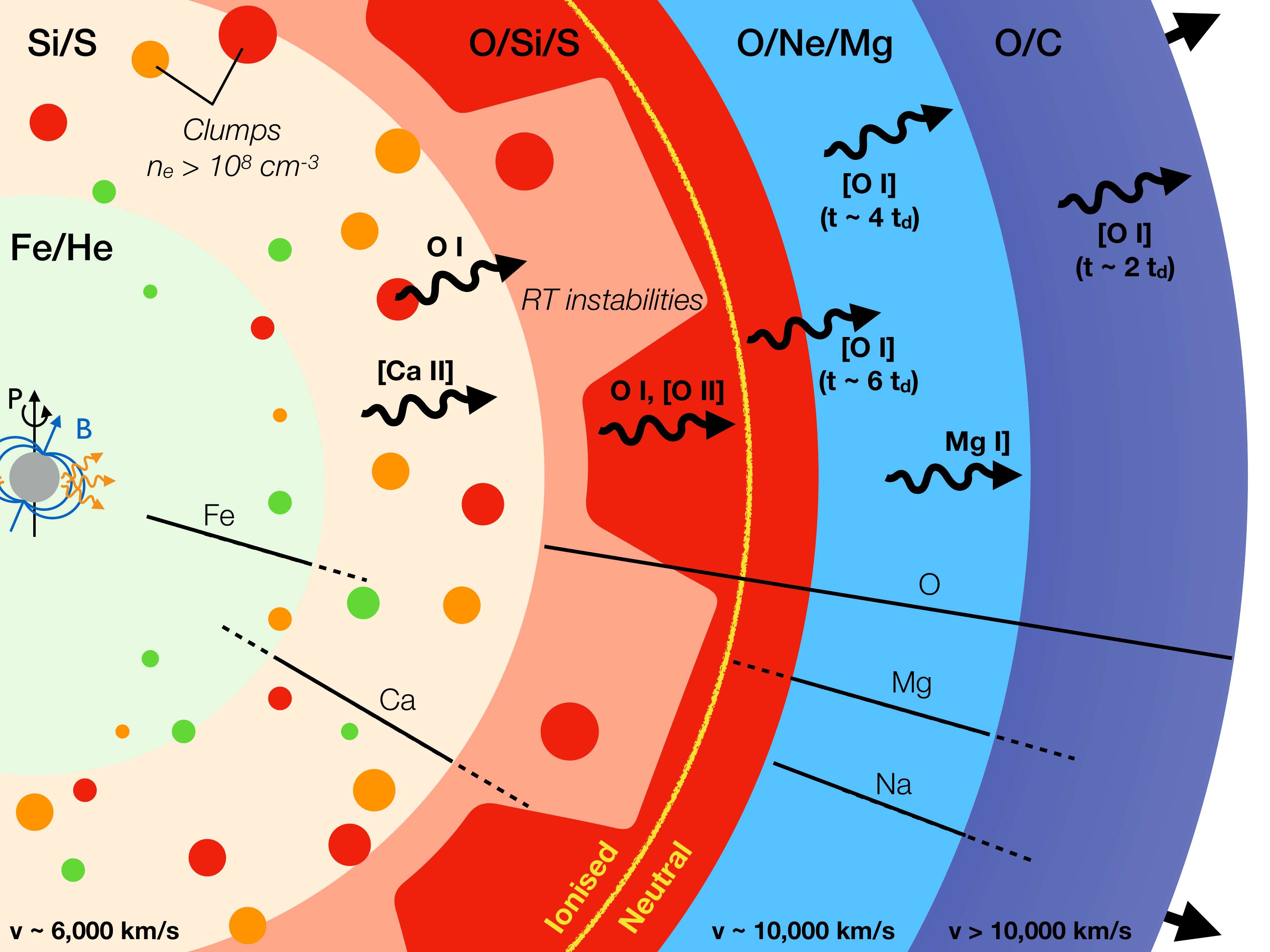}
\caption{Cartoon showing the structure of a SLSN in the nebular phase. Zone velocities and the regions responsible for key emission lines are labelled. \o\ emission comes from progressively deeper layers over time. The inner zones have a low filling factor due to an engine-blown bubble and resultant clumping by fluid instabilities -- this region likely produces the \ca\ and \oi\ lines that are prominent in SLSNe, while an engine-driven oxygen ionisation front, frozen in mass coordinate, allows production of [\ion{O}{2}] and [\ion{O}{3}]. If the ionised region extends very far out, the spectrum can be dominated by [\ion{O}{2}], as seems to be the case in SN\,2017egm, LSQ14an and PTF10hgi.}
\label{f:schem}
\end{figure*}

Here we combine the results from the previous sections to piece together a plausible physical picture of SLSNe evolving through the nebular phase. A schematic diagram is shown in Figure \ref{f:schem} and elaborated upon below.

\subsection{Density and ionisation structure as a clue to the power source}

The two strongest optical lines, \o\ and \ca, have similar critical densities, $n_e \sim 10^6$\,cm$^{-3}$ \citep[e.g.][]{li1993,fes1999}, so it is interesting that the \ca\ line often develops earlier. One way to suppress \o\ emission would be if oxygen was mainly concentrated in a dense region where collisional de-excitation is dominant, but we know this is not the case since we see \o\ at higher velocities (where relative density should be lowest) compared to other lines, including \ca\ (Figure \ref{f:vave}).

Another possibility is that most of the oxygen is initially ionised, and \o\ develops as the neutral fraction becomes substantial. This interpretation is also problematic, since the ionisation potential of \ion{O}{1} is greater than that of \ion{Ca}{2}, so in this scenario \ca\ might also be suppressed. Moreover, for the long-lived power input needed to sustain SLSN luminosities, \citet{margalit2018} recently found that the ionization structure remains frozen into the ejecta, such that a dramatic change in the neutral fraction of oxygen is unlikely on the timescales probed here.

The absolute flux in the \o\ line also argues against the above scenarios, since the line luminosity requires a large neutral oxygen mass, and decreases smoothly in time even as the line becomes strongly peaked. This suggests that \ca\ is enhanced, rather than \o\ suppressed. Indeed, it has long been known that some SLSNe exhibit strong \ca\ emission even before the onset of the nebular phase \citep{gal2009}.

In Figure \ref{f:ltd}, we showed that the evolution in line luminosities is strongly coupled to the decline timescale of the light curve, suggesting that the lines are reprocessing the input from a continuous and decaying power source. Calcium resides at relatively low velocity coordinate, likely in the Si/S and O/Si/S zones from comparison to explosion models, whereas oxygen is abundant in the O/Si/S zone as well as the outer O/Ne/Mg and O/C zones. Therefore the luminous \ca\ line appearing early in the evolution can be explained if a large amount of energy is being reprocessed by the inner regions of the ejecta. This is a natural consequence of a central engine power source, the most popular model for interpreting SLSN light curves.

Furthermore, two-dimensional simulations of SLSNe show that a central engine can hollow out the interior of the ejecta and create a complex density structure due to instabilities in the compressed fluid \citep{kchen2016}. This could potentially provide the low filling factor required by certain line ratios, and rarified inter-clump regions can more quickly reach electron densities below the critical density for \ca -- effectively, allowing parts of the innermost ejecta to become `nebular' before the outer layers.

A strong \ca\ line during the early or pseudo-nebular phase may therefore be a signature of a central engine. However, this does raise a further question of how emission from low velocity coordinate can be observed while the SN still has a photospheric component (i.e. continuum). \citet{jer2015} show wavelength-dependent optical depths and photon escape probabilities for one of their stripped SN models, covering all the ejecta zones of interest here. While not tuned specifically for SLSNe, their results demonstrate that for reasonable parameters, photons with $\lambda \gtrsim 6000$\,\AA\ can easily traverse the ejecta by 200 days, even while the optical depth remains high at bluer wavelengths. Future radiative transfer calculations should help to test this hypothesis.

In cases where \ca\ turns on even earlier, such as SN\,2007bi, SN\,2015bn and LSQ14an \citep{gal2009,nic2016b,ins2017}, additional effects may be required. For example, a non-spherical geometry could also serve to produce paths through the ejecta with a wide range of optical depth. The line profiles of LSQ14an support a significantly aspherical explosion, while (spectro-) polarimetry of SN\,2015bn has also revealed asymmetry \citep{ins2016,leloudas2017}. However, the line profiles for most SLSNe suggest asymmetry is probably modest.

\oi\ emission arises at similar velocity coordinate to \ca. This is also readily interpretable in the central engine scenario, as the engine can both compress and ionise the inner parts of the ejecta, creating high-density ionised regions that are favourable for oxygen recombination. \citet{margalit2018} showed that for an $L \propto t^{-2}$ magnetar engine, the ionisation front remains at roughly constant mass coordinate as the ejecta expand, consistent with our finding that the \oi\ recombination line comes predominantly from the same velocity zone at all times.

This is in contrast to the \o\ emission, which clearly comes from progressively deeper lying regions in the ejecta over time. We suggest that this emission initially comes from outer ejecta that is relatively undisturbed by the central engine overpressure, and that the gradual movement to lower velocity coordinate is simply a result of the density decreasing under free expansion. If the \ion{O}{1} ionisation front is located in the O/Si/S layer, \o\ eventually reaches a comparable velocity to \oi, as we observe in some of our latest spectra.

The presence of this ionisation front can have other effects on the spectra of SLSNe. In particular, the unusually strong [\ion{O}{3}] emission in SLSNe compared to SNe Ic can be explained by persistent oxygen ionisation by the central engine. We can also account for some of the diversity among SLSNe by appealing to the location of the ionisation fronts. If some engines succeed in ionising oxygen all the way through the ejecta, this would allow for suppression of \o\ throughout the nebular phase, and provide a source of [\ion{O}{2}] to enhance the 7300\,\AA\ line. Since \ion{Ca}{2} should also be significantly ionised in this scenario, the 7300\,\AA\ line in LSQ14an, SN\,2017egm and PTF10hgi is likely dominated by [\ion{O}{2}].

Why does this occur only in a few SLSNe? \citet{jer2017a} showed that in low-mass models of SLSNe a process of `runaway ionisation' can occur when a large amount of energy is deposited in a small ejecta mass. \citet{nic2017d} suggested that SN\,2017egm may be among the lowest-mass SLSNe, making it a candidate for this process. LSQ14an, on the other hand, is more likely an SLSN with a large ejecta mass -- in this case, the asymmetry suggested by the line profiles may result in lower-density regions that are easier to ionise. PTF10hgi is one of the least luminous SLSNe, so may be an example of a low-mass event, though \citet{qui2018} detected helium in the photospheric spectrum, so envelope stripping is not as extreme as SN\,2017egm. In both LSQ14an and PTF10hgi, strong [\ion{O}{3}] emission may be an indication of low density.

While this narrative seems consistent with the evolution in velocity and luminosity for the key lines, as well as the diversity in ionisation state between events, a central engine may have more difficulty explaining the shape of the emission line profiles. As discussed in section \ref{s:profiles}, a spatial distribution of ions with a hollow cavity is predicted analytically to emit flat-topped lines, whereas most of the lines we see are close to Gaussian in appearance. Velocity and density gradients, mixing, or ejecta asymmetries could help to mitigate this problem. However, solving it fully will require radiative transfer calculations for ejecta with a central cavity, to compare with the observations.

The simple schematic picture sketched out here should be used as a starting point for detailed modelling of SLSN nebular spectra. \citet{jer2017a} have produced model spectra for pure oxygen ejecta and carbon-burning compositions, and their results have informed much of the analysis here. To model all of the ejecta zones simultaneously is a complex computational feat, but we hope that our results can inform plausible choices for the velocities, densities, filling factors and energy deposition in future calculations of synthetic spectra.

We conclude this subsection by noting that SLSNe powered by circumstellar interaction could also produce dense regions (a cold dense shell and resultant instabilities) and ionising photons (from the shock fronts) to potentially reproduce the line ratios we see here. However, such a model may not so naturally account for the velocity structure we observe.

\subsection{Comments on progenitors}

Since nebular spectroscopy probes the innermost layers of the explosion, this technique provides the most direct fingerprints of the pre-supernova progenitor star. In section \ref{s:lums}, we showed that for many events the \ca/\o\ ratio was reasonably well matched by helium cores of $\sim 3.5 - 5.9$\,\M, though this ignored various complicating factors in deriving theoretical ratios from explosion models. On the observational side, we have found that [\ion{O}{2}] emission can have a large effect on the 7300\,\AA\ luminosity, such that the true core mass may often be $> 6$\,\M. 

The light curve models of SLSNe from \citet{nic2017c} suggested an average ejecta mass of 4.8\,\M\ -- if we assume that the helium core mass is equal to the ejecta mass plus a typical neutron star mass of 1.4\,\M, this suggests a helium core of $\sim 6.2$\,\M, in broad agreement with our findings here. However, we note that the distribution of ejecta masses inferred from light curve modelling is quite broad, including several events with masses of 2--3\,\M. These estimates are complicated by the use of an uncertain grey opacity \citep{arn1982}, which may lead to systematic offsets between masses inferred from light curves and spectroscopy.

There is also likely a selection effect favouring nebular spectroscopy of high-mass events: since slower-fading SLSNe tend to be more luminous at peak \citep{ins2014}, they are easier to observe at nebular times even when corrected for the longer $t_d$. Nebular follow-up of larger samples of SLSNe, spanning a range of decline rates, will be required to determine if the overall mass range is as broad as that implied by light curve models.

Many events have \o\ luminosities that exceed $10^{41}$\,\ergs\ for long periods of time. \citet{jer2017a} showed that in their models this required an oxygen mass of $\sim 10$\,\M, or even greater depending on the energy deposition.
In our measurements, SN\,2017egm, PTF10hgi and PTF12hni have significantly lower \o\ luminosities and therefore may be examples of SLSNe from lower mass progenitors, though ionisation of oxygen also likely plays a role in setting the line luminosity, as we discussed above.

Another important clue for diagnosing progenitor mass is the abundance of iron group elements. In section \ref{s:mean}, we found that SLSNe have an enhanced flux around $\sim 5000$\,\AA\ compared to typical SNe Ic, indicating a larger mass of iron-group elements. In fact, this region of the spectrum looks similar to broad-lined SNe Ic such as SN\,1998bw, as previously noted for specific SLSNe by \citet{nic2016c} and \citet{jer2017a}. 

A larger iron mass is generally linked to a more massive progenitor, as very massive stars are expected to synthesise more heavy elements in explosive burning \citep[e.g.][]{ume2008}. Thus SLSNe and broad-lined SNe Ic appear to come from more massive stripped stars than do normal SNe Ic. This is also supported by analytic light curve fits, which have found larger ejecta masses in broad-lined SNe Ic and SLSNe than in SNe Ic \citep[e.g.][]{dro2011,cano2013,taddia2015,nic2015b,nic2017c}. Differences in initial metallicity of the progenitor cannot account for the increased metal abundance in SLSNe and broad-lined SNe Ic, since these explosions prefer \emph{lower}-metallicity environments than do normal SNe Ic.

However, our principal component analysis showed that the iron plateau exhibits more variation between SLSNe than any other spectral feature, which may reflect a diversity in iron group abundance between SLSNe (though temporal effects are also responsible for much of the variation). Therefore even if SLSNe could result from stars that are \emph{on average} of a similar mass to the progenitor of SN\,1998bw, or at least produce a comparable iron mass, there is likely significant scatter.

Since SLSNe seem to come from relatively massive stripped-envelope stars, one important question is how they lose their hydrogen layers prior to explosion. Possibilities include line-driven winds, eruptive outbursts, interaction with a binary companion, or chemically homogeneous evolution. A useful observational handle is the distance to any hydrogen-rich material, presumably lost by the star prior to explosion.

\begin{figure}
\centering
\includegraphics[width=8.25cm]{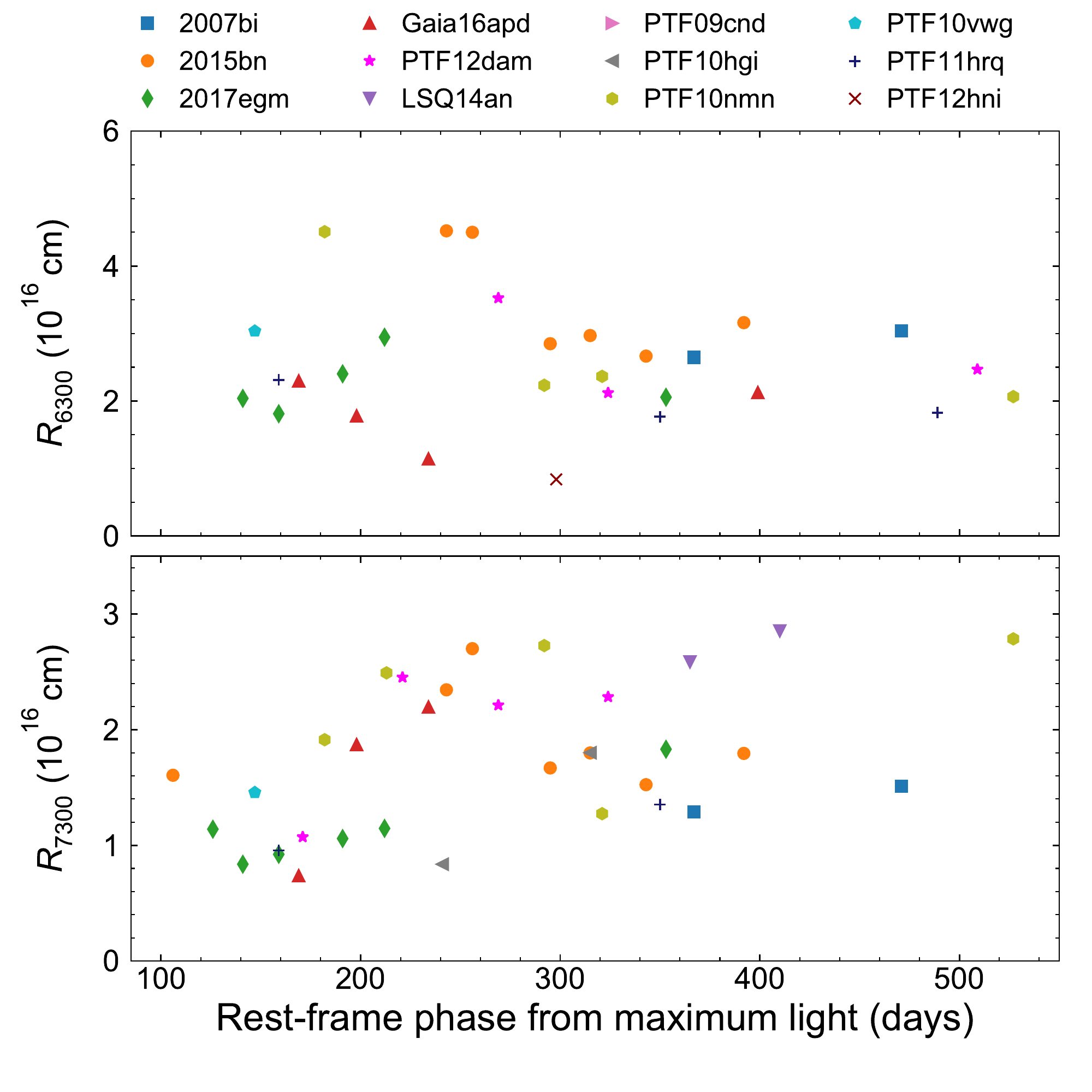}
\caption{Radii calculated from emission line velocities.}
\label{f:rad}
\end{figure}

\citet{yan2017} determined that the few SLSNe that do have late-time hydrogen signatures in their spectra typically reach this material at radii $\sim 10^{16}$\,cm. For our sample, we estimate the radii of emitting material for each SLSN at each epoch simply by multiplying observed line velocities by the phase from peak plus a rise time for each SLSN \citep{gal2009,nic2013,nic2016b,nic2017,ins2017,dec2018,bos2018}. We use velocities for both \o\ (the fastest material) and \ca\ (for a more conservative estimate). 

The results are shown in Figure \ref{f:rad}. Most of the SLSNe in our sample reach radii of at least a few $\times 10^{16}$\,cm over the timescales of the observations, comparable to the radii where hydrogen was detected by \citet{yan2017}. Therefore the fact that our spectra do not show hydrogen is not because the radii probed are smaller (they are not), but presumably because a fairly low fraction of SLSN progenitors, $\sim 3/15 = 20\%$, have significant hydrogen-rich material within $\sim 10^{16}$\,cm. If hydrogen is ejected explosively, e.g.~by pulsational pair-instability eruptions \citep{woo2007,woo2017}, a typical velocity is expected to be $\sim 1000$\,\kms. Thus to reach at least a few $\times 10^{16}$\,cm, the eruption must have occurred at least about a decade prior to explosion. In the case of a wind with velocity 10--100\,\kms, the hydrogen envelope must be completely lost by 100-1000 years before explosion (roughly the end of helium burning).

To summarise this section, SLSN progenitors are likely massive stars with helium cores $> 6$\,\M\ -- and often 10\,\M\ or more to produce enough oxygen for the observed \o\ luminosity \citep{jer2017a}. If the nebular-phase luminosity between $\sim 4000-5000$\,\AA\ is a good indication of the total iron-group mass, itself an indication of progenitor mass, SLSN progenitors appear to be on average similar to that of SN\,1998bw, though with considerable scatter. Envelope loss is complete at least a decade before explosion, and possibly much earlier.

\section{Conclusions}
\label{s:conc}

We have conducted a systematic study of the observed properties of SLSN spectra as they evolve through the nebular phase. Our sample comprised 41 spectra of 12 SLSNe, including both fast and slow evolving events.

After applying a consistent interpolation and smoothing procedure to all spectra, and normalising the observed phase by the different decay rates of the SLSN light curves, we found that all events could be reasonably well described in terms of a single spectral sequence, ordered by this normalised phase. Our GMOS spectra of Gaia16apd (399 days after peak) and SN\,2017egm (353 days after peak) are among the latest obtained for SLSNe, especially when considering the relatively fast light curve timescales of these events.

We analysed these spectra in terms of their statistical properties and the velocity and luminosity evolution of specific lines. For convenience, we summarise our main conclusions here:
\begin{itemize}
\item SLSN spectra become dominated by nebular features within 2--4 $e$-folding times after their light curve peaks. This provides a means early in the evolution of an individual event to plan the optimal time for nebular-phase follow-up.
\item The main emission lines are easily identified with well-known transitions of oxygen, calcium, magnesium, sodium and iron -- the same species typically seen in normal and broad-lined SNe Ic.
\item \o\ is initially weaker than \ca, and takes longer to develop a Gaussian line profile, but usually becomes the strongest optical line at later times. The ratio of the 6300\,\AA/6364\,\AA\ components indicates that this line is optically thin during most of the nebular phase.
\item Compared to SNe Ic, SLSNe are differentiated by a prominent \oi\ recombination line, often along with [\ion{O}{2}] and [\ion{O}{3}], indicating higher ionisation, and the presence of oxygen in regions with significant variation in electron density.
\item SLSNe also show elevated flux compared to normal SNe Ic, on average, over the iron-dominated part of the spectrum between 4000-5500\,\AA. However, the SLSN iron region is similar to some broad-lined SNe Ic such as SN\,1998bw.
\item Principal component analysis showed that $\gtrsim 70$\% of the variance in SLSN spectra could be attributed to 5 eigenspectra, corresponding roughly to emission from \ion{Fe}{2}, \ion{Ca}{2} NIR triplet, \o, \ca, and \oi, in order of decreasing variance.
\item We find no compelling evidence for clustering of SLSNe into sub-populations based on their nebular spectra. A K means clustering analysis with two assumed clusters separates the spectra as much by phase relative to explosion as by the actual SLSNe to which these spectra belong.
\item Most SLSNe show no strong asymmetry in their \o, \ca, or \mg\ line profiles; LSQ14an is a possible exception to this. However, at least half of the SLSNe in our sample exhibit an excess on the red side of \oi, likely attributable to \ion{Mg}{2}\,$\lambda$7877,7896.
\item The ejecta structure and composition inferred from the widths of the strongest lines is consistent with explosion models of massive stars, with calcium towards the centre and magnesium further out, and oxygen spanning a wide range of ejecta zones. \o\ arises from faster (further out) regions of the ejecta than \oi\ for several hundreds of days.
\item The luminosity evolution in all lines decreases with time. After normalising the spectral phase by the light curve decline timescale, all SLSNe show virtually the same decline rate in their line luminosities, indicating that the nebular lines are reprocessing the same power source responsible for the earlier luminosity evolution of SLSNe. The ratios between many lines look similar to SNe Ic, provided one takes into account the relatively slower light curve evolution of SLSNe.
\item The \ca/\o\ ratio matches some models with masses of $\sim 3.5-5.9$\,\M, but the measured ratio may be contaminated by [\ion{O}{2}]\,$\lambda$7320,7330 emission, such that the true core mass is likely even higher in some events. However, other events appear to be indicative of a population extending to lower masses, and selection effects may account for a large number of high-mass events in the nebular sample.
\item The \oi\ line and the ratio between calcium lines indicates a region of high electron density with low filling factor at low velocity coordinate. We suggest that this could be due to the hydrodynamic impact of a central engine.
\item The ionisation structure shows no evidence of changing over several hundred days. This is expected in the engine-powered models recently calculated by \citet{margalit2018}. In some SLSNe, oxygen may be ionised in a large fraction of the ejecta throughout the nebular phase -- these events are SN\,2017egm, LSQ14an and PTF10hgi.
\item The radii reached by the ejecta on the timescales of these data are comparable to the radii where some SLSNe encounter hydrogen-rich material \citep{yan2015,yan2017}. That our spectra show no sign of hydrogen indicates a diversity in when the hydrogen layer is lost by the progenitors.
\end{itemize}

There remain important outstanding questions, including an accurate calibration of the ejecta masses for better comparison with light curve models, and determining the processes in the pre-supernova stellar evolution that expel the stellar envelope, and likely allow the formation of a central engine.
It also remains to be demonstrated whether our suggested ejecta distribution is consistent with the observed smooth line profiles, and if \ca\ emission from deep-lying zones can escape the ejecta early in the transition to the nebular phase.
Thus, a full understanding of SLSN progenitors will require more detailed modelling of their late-time spectra. Our analysis here provides an observationally-motivated starting point for exploring the model parameter space.

\software{scikit-learn \citep{ped2011}, scipy \citep{scipy}, matplotlib \citep{matplotlib}, PyRAF, IRAF \citep{iraf}}

\acknowledgments

We thank an anonymous referee for many insightful comments that improved the paper.
M.N.~is supported by a Royal Astronomical Society Research Fellowship.
We thank Ashley Villar for comments on the manuscript, Daniel Fabricant and Igor Chilingarian for help with Binospec observations, and Ting-Wan Chen for providing a spectrum of PTF12dam.
The Berger Time-Domain Group at Harvard is supported
in part by the NSF under grant AST-1714498 and by NASA
under grant NNX15AE50G. This paper is based upon work
supported by the National Science Foundation Graduate Research
Fellowship Program under Grant No.~DGE1144152.
R.C.~acknowledges support from NASA Chandra Grant Award number GO7-18046B.
Based on observations (Proposal IDs GN-2017A-FT-15 and
GN-2018A-FT-109) obtained at the Gemini Observatory
acquired through the Gemini Observatory Archive and
processed using the Gemini IRAF package, which is operated
by the Association of Universities for Research in Astronomy,
Inc., under a cooperative agreement with the NSF on behalf
of the Gemini partnership: the National Science Foundation (United States), the National Research Council (Canada), CONICYT (Chile), Ministerio de Ciencia, Tecnolog\'{i}a e Innovaci\'{o}n Productiva (Argentina), and Minist\'{e}rio da Ci\^{e}ncia, Tecnologia e Inova\c{c}\~{a}o (Brazil).
This paper uses data products
produced by the OIR Telescope Data Center, supported by the
Smithsonian Astrophysical Observatory. Some observations
reported here were obtained at the MMT Observatory, a joint
facility of the Smithsonian Institution and the University of
Arizona. This paper includes data gathered with the 6.5 meter
Magellan Telescopes located at Las Campanas Observatory,
Chile.
This work is based in part on observations obtained at the MDM Observatory, operated by Dartmouth College, Columbia University, Ohio State University, Ohio University, and the University of Michigan. We thank J.~Rupert for obtaining the MDM data.
STSDAS and PyRAF are products of the Space Telescope Science Institute, which is operated by AURA for NASA.

\bibliography{mybib}


\appendix


\section{Evaluating $k$ in K-means clustering}

\renewcommand\thefigure{\thesection\arabic{figure}}
\setcounter{figure}{0}

\begin{figure*}[t!]
\centering
\includegraphics[width=7cm]{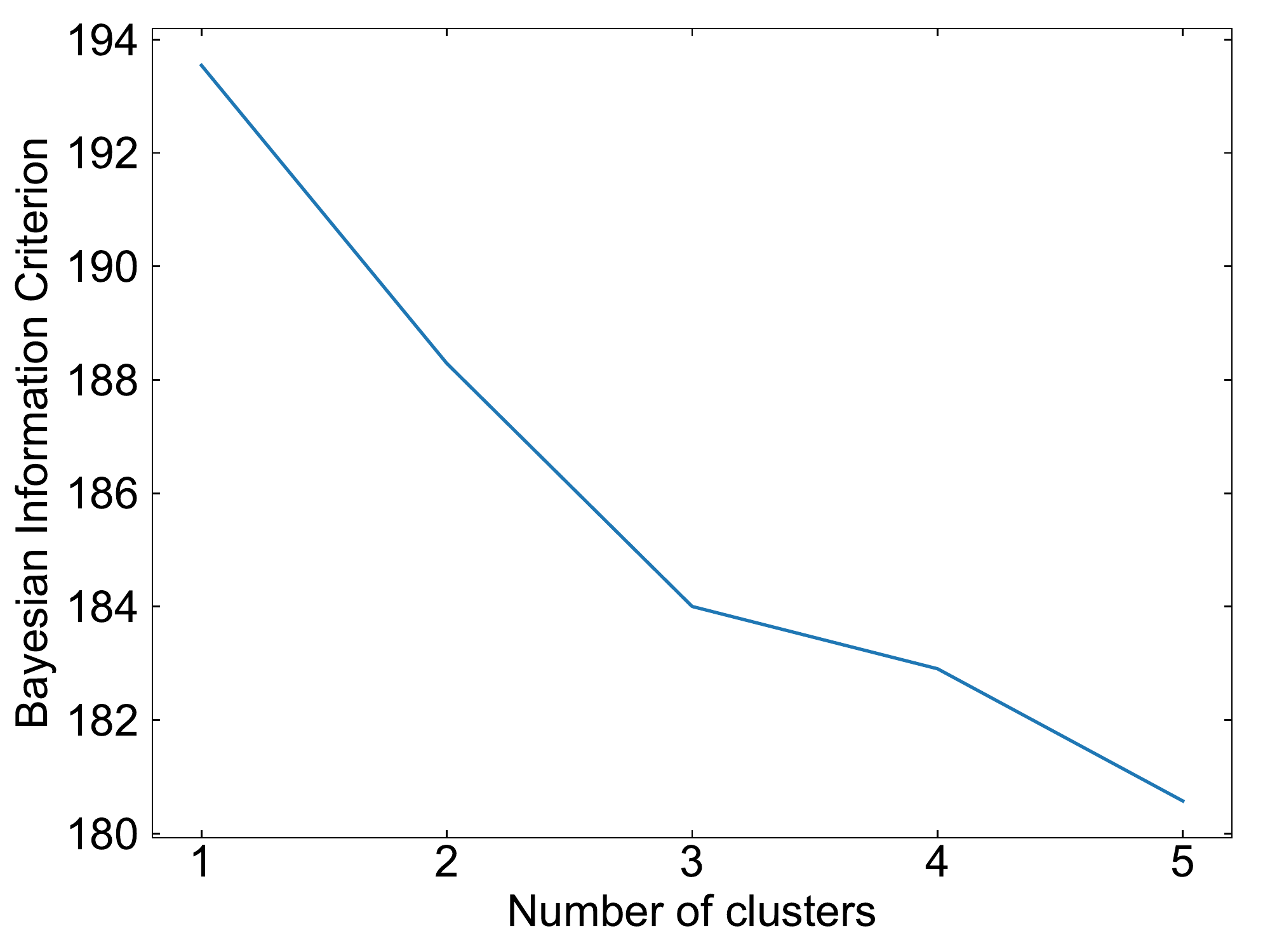}
\includegraphics[width=7cm]{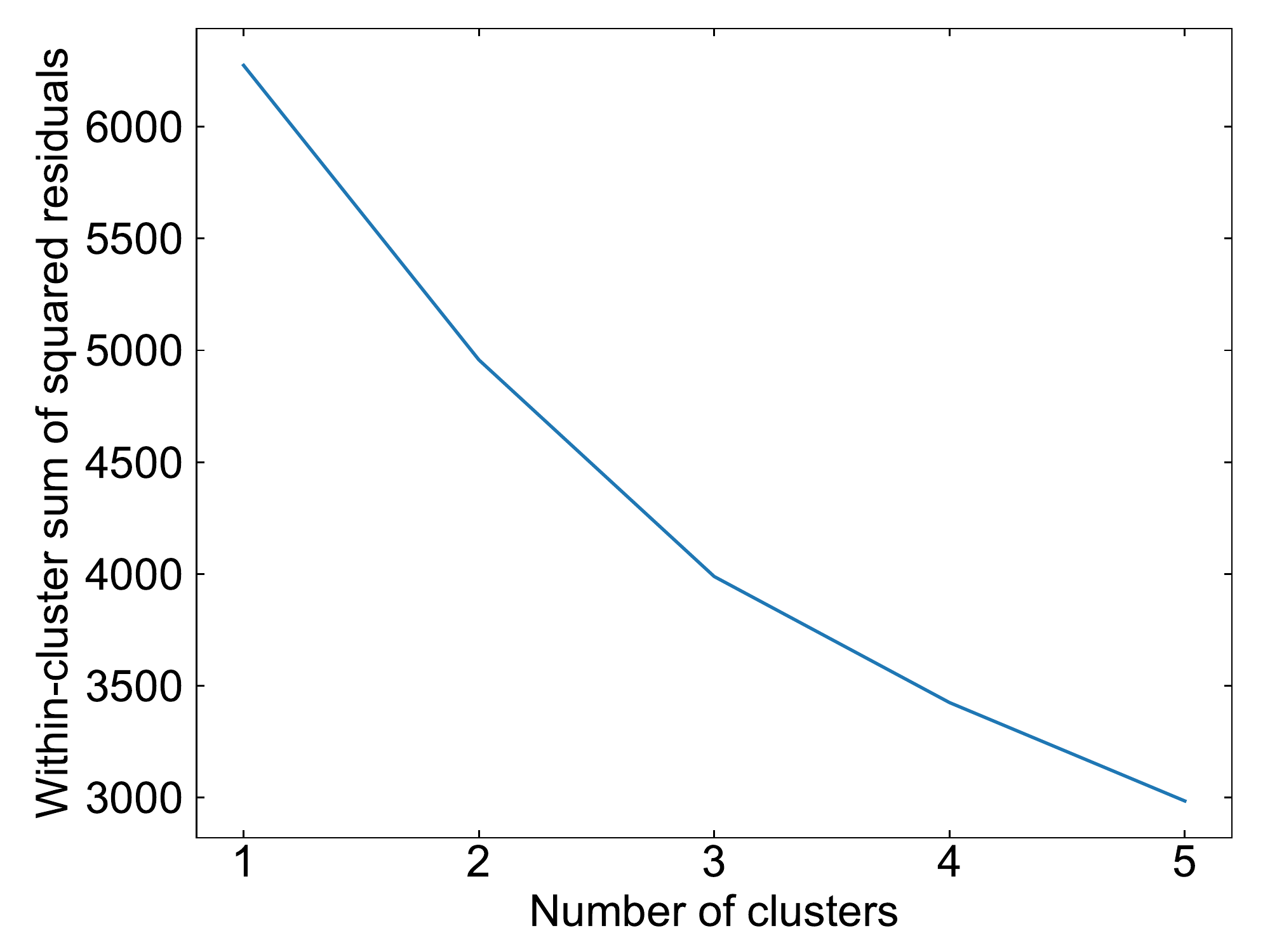}
\includegraphics[width=14.5cm]{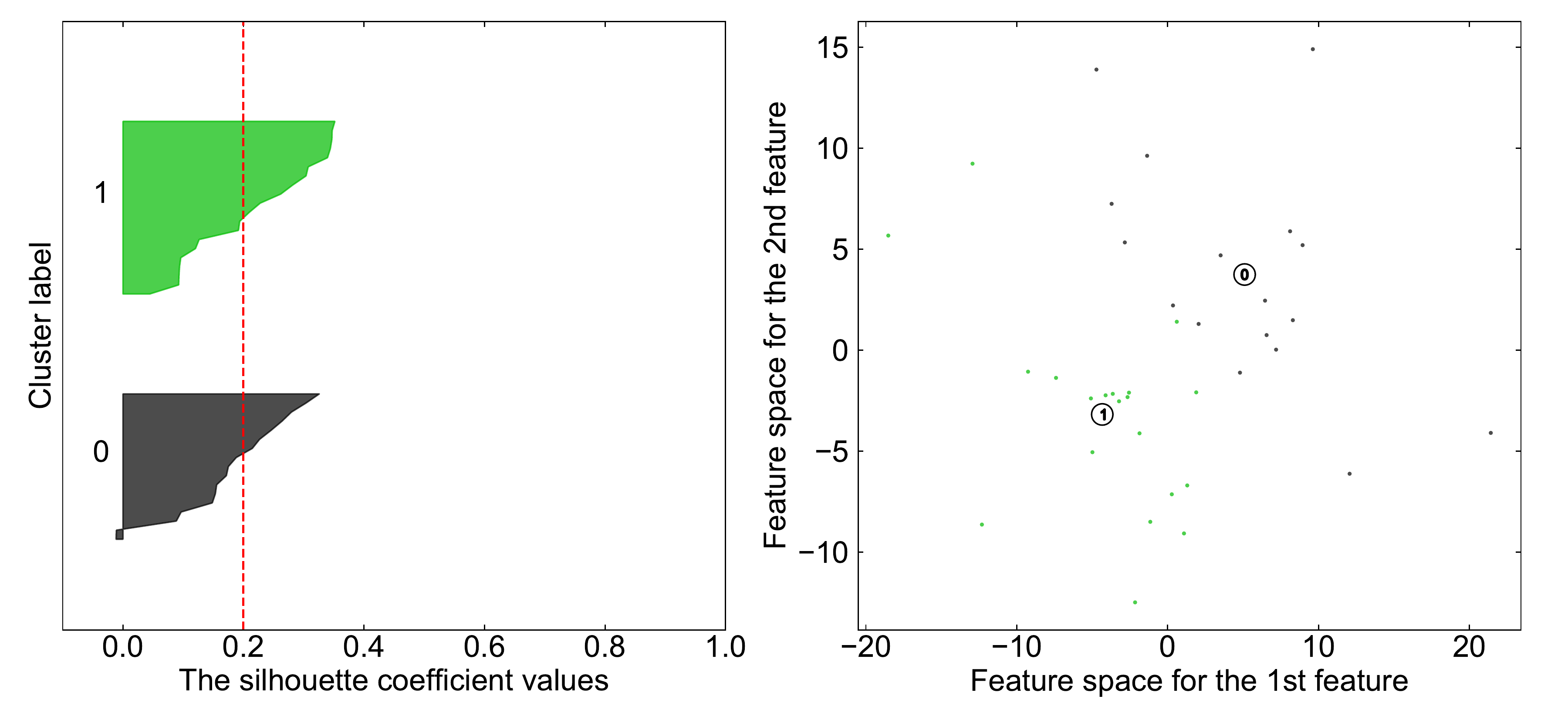}
\includegraphics[width=14.5cm]{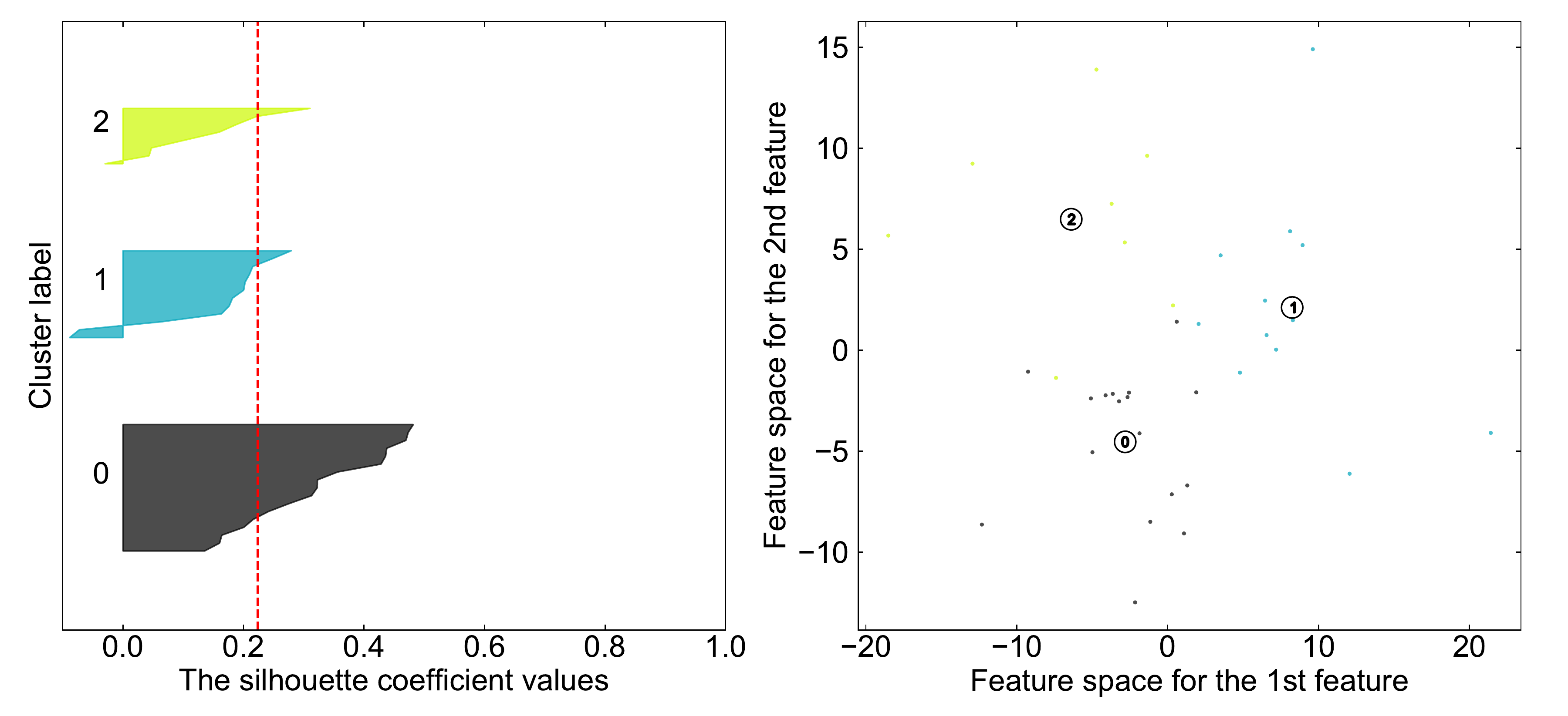}
\caption{Metrics for evaluating the number ($k$) of clusters in our K means analysis. Top left: Bayesian Information Criterion (BIC) as a function of $k$. No minima are present in the BIC as we increase the number of clusters. A slight inflection may indicate a mild preference for $k=3$. Top right: Within-cluster sum of squares (variance). Again, no flattening with $k$ is observed. Middle row: silhouette plots for $k=2$. The left shows the silhouette coefficient for each data point, parameterising the distance from each of the other clusters, while the right shows a representation of the data in the first two PCA components. The two clusters are approximately equal in size, but are poorly separated. Bottom row: silhouette plots for $k=3$.  One cluster is significantly larger than the other two, while the silhouette score remains low.}
\label{f:tests}
\end{figure*}

The K-means clustering algorithm requires that the number of clusters, $k$, in which the data should be partitioned must be specified in advance. Therefore it is essential to use some metric to evaluate whether these $k$ clusters are separated in a meaningful way, and ultimately to determine the optimal number of clusters. Three such methods are described here, and shown in Figure \ref{f:tests}. 

First we evaluate the Bayesian information criterion \citep[BIC;][]{sch1978} for different values of $k$. This behaves similar to the Bayes factor (the likelihood ratio between models, if neither model is favoured a priori), but the BIC is not normalised. It strongly penalises additional model complexity (i.e.~more clusters), such that a lower BIC score is better \citep{kas1995}. We find that the BIC decreases quite smoothly as $k$ increases, with no local minima indicating a strongly preferred number of clusters, though there is a slight inflection at $k=3$. We plot the BIC only up to $k = 5$, as a much larger number of clusters would not be meaningful in a dataset of this size.

We next employ the elbow criterion. This method looks at the variance in the data explained as a function of $k$, parameterised as the sum of squared residuals from the mean within each cluster. A flattening in this function indicates a preferred value of $k$. The function plotted in Figure \ref{f:tests} does not flatten up to $k = 5$, leaving no strong incentive for choosing any particular $k$.

Finally, we conduct a silhouette analysis using \textsc{sklearn.metrics}. This analysis calculates a score for each data point based on how close it lies to the mean of the \emph{other} clusters. The silhouette score ranges from -1 to 1, with 1 indicating the clusters are well separated. A score of 0 indicates that a point is on the boundary between clusters. For $k > 3$, we find that some clusters have low membership and score negatively, thus a larger number of clusters is ruled out. We show silhouette plots for $k = 2$ and $k = 3$ in Figure \ref{f:tests}. In both cases, the average scores are low, $\approx 0.2$. With $k = 3$, we find one large cluster and two rather smaller clusters, whereas for $k = 2$ the clusters are approximately equal in size.

Overall, there is no strong evidence for $k > 1$ clusters in the data.

\pagebreak

\section{Line measurements}

Velocities were measured by Gaussian fits. For some of spectra in the sample, particularly at earlier times, the \o\ line is not well represented by a Gaussian. These are not included in velocity plots for \o. Specifically we exclude: SN\,2015bn at 106 days, SN\,2017egm at 126 days, PTF12dam at 171-221 days, PTF09cnd at 121 days, and LSQ14an and PTF10hgi (at all phases). An example is shown in Figure \ref{f:gauss}.

In section \ref{s:cont} we described the process by which we approximated the continuum level for each spectral line before making measurements. Figure \ref{f:continuum} shows an example of this process. As the continuum placement can be uncertain for blended lines such as \mg\ and [\ion{O}{1}]\,$\lambda$5577, we include an additional 20\% systematic error in all line measurements to account for this.

\renewcommand\thefigure{\thesection\arabic{figure}}
\setcounter{figure}{0}

\begin{figure*}[h!]
\centering
\includegraphics[width=8.cm]{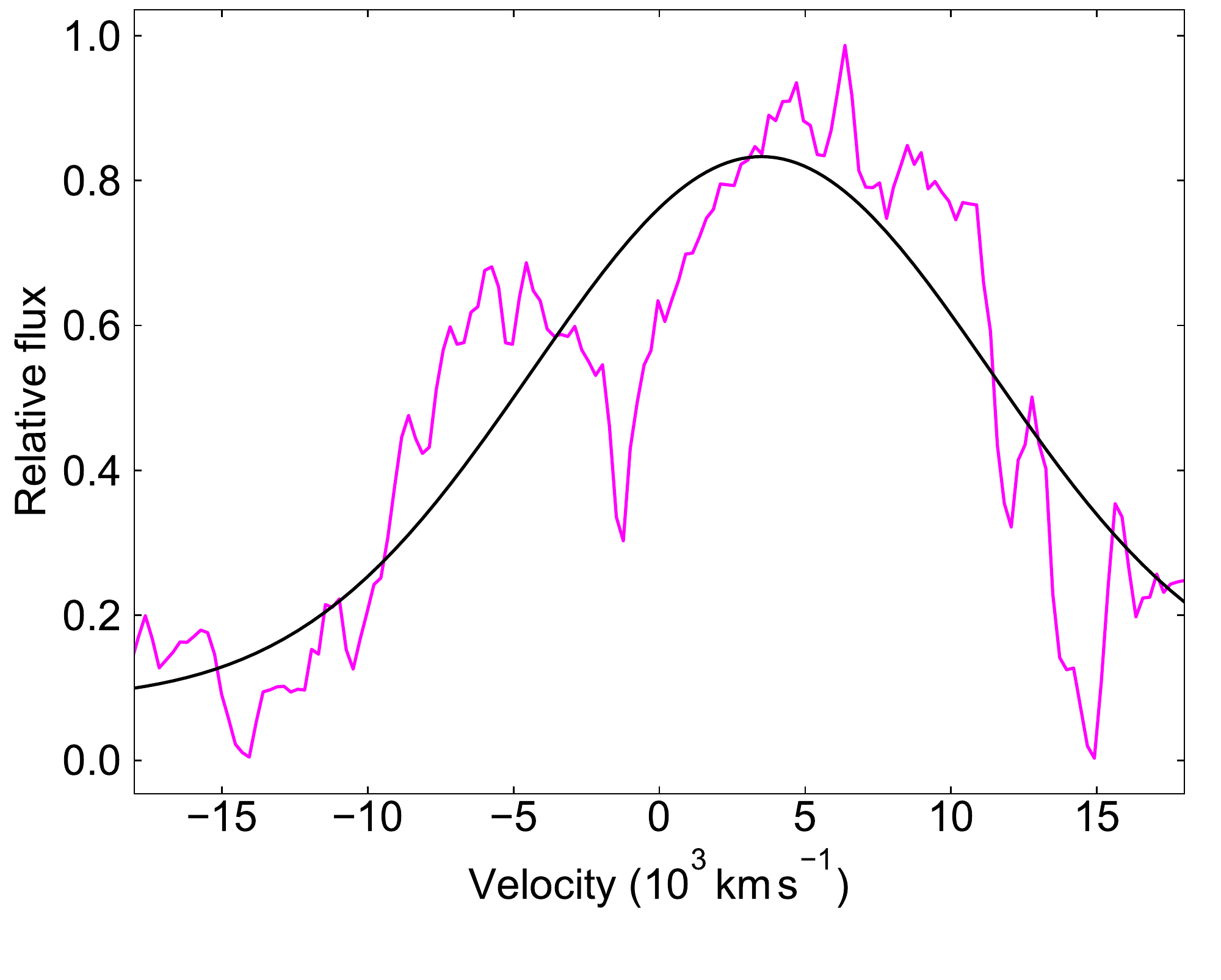}
\includegraphics[width=8.cm]{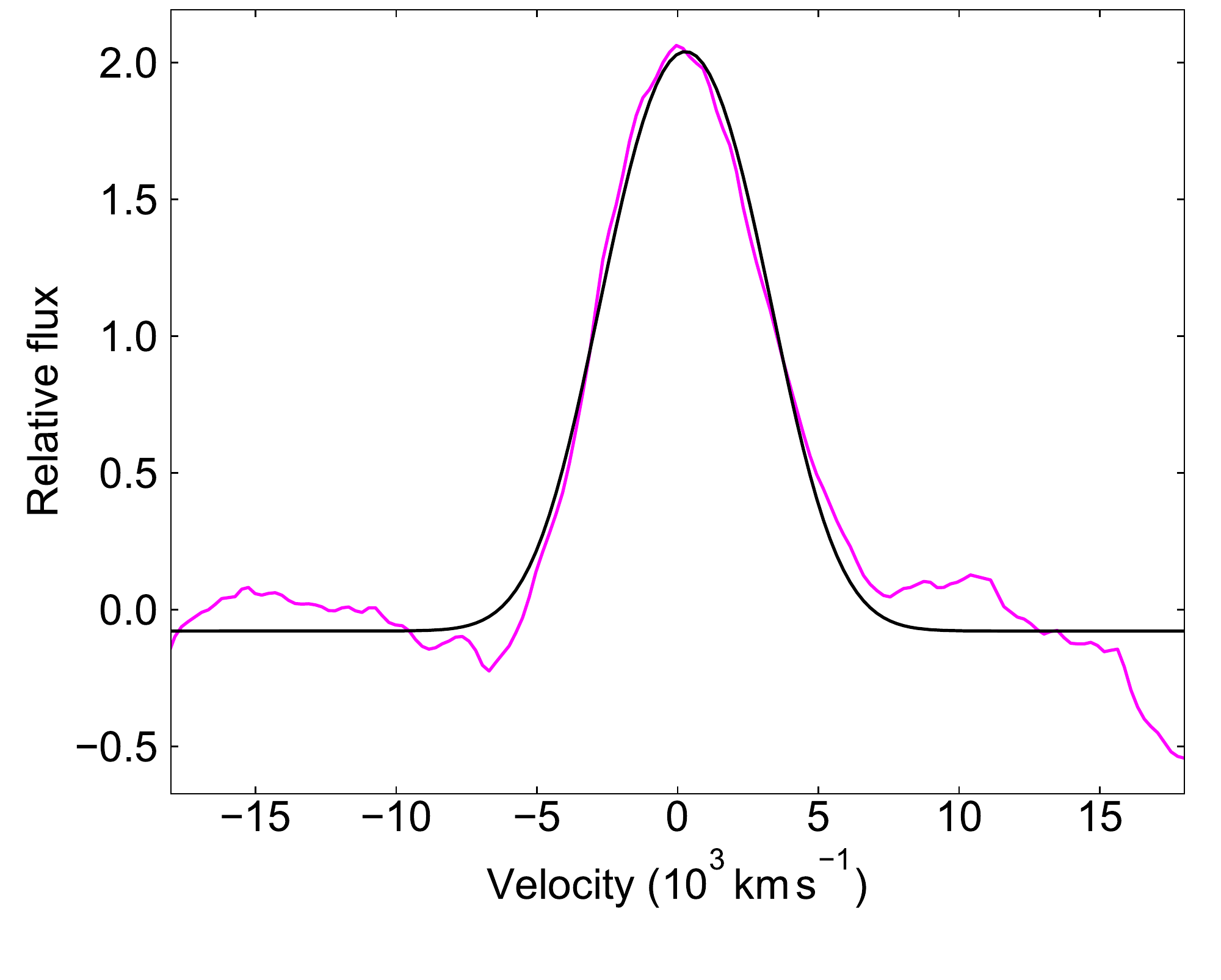}
\caption{Example of Gaussian fits to the \o\ line. The panel on the left shows PTF12dam at 221 days. At this phase the line has not yet developed a Gaussian profile, and we exclude it from the velocity analysis. The panel on the right shows the same SLSN at 324 days. At later phases, \o\ can be well fit with a Gaussian.
\bigskip}
\label{f:gauss}
\end{figure*}

\begin{figure*}[b!]
\centering
\includegraphics[width=12cm]{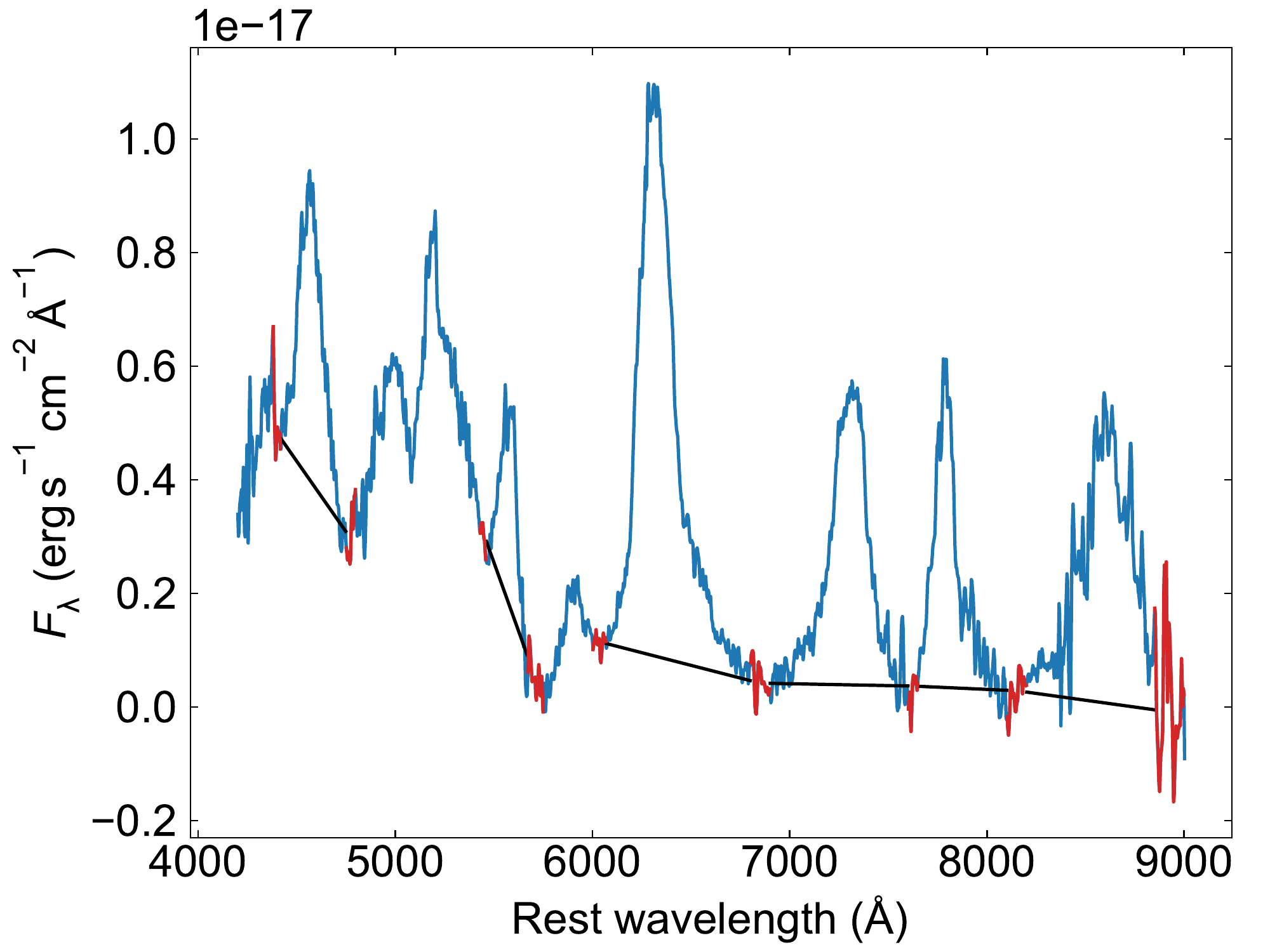}\\
\includegraphics[width=8cm]{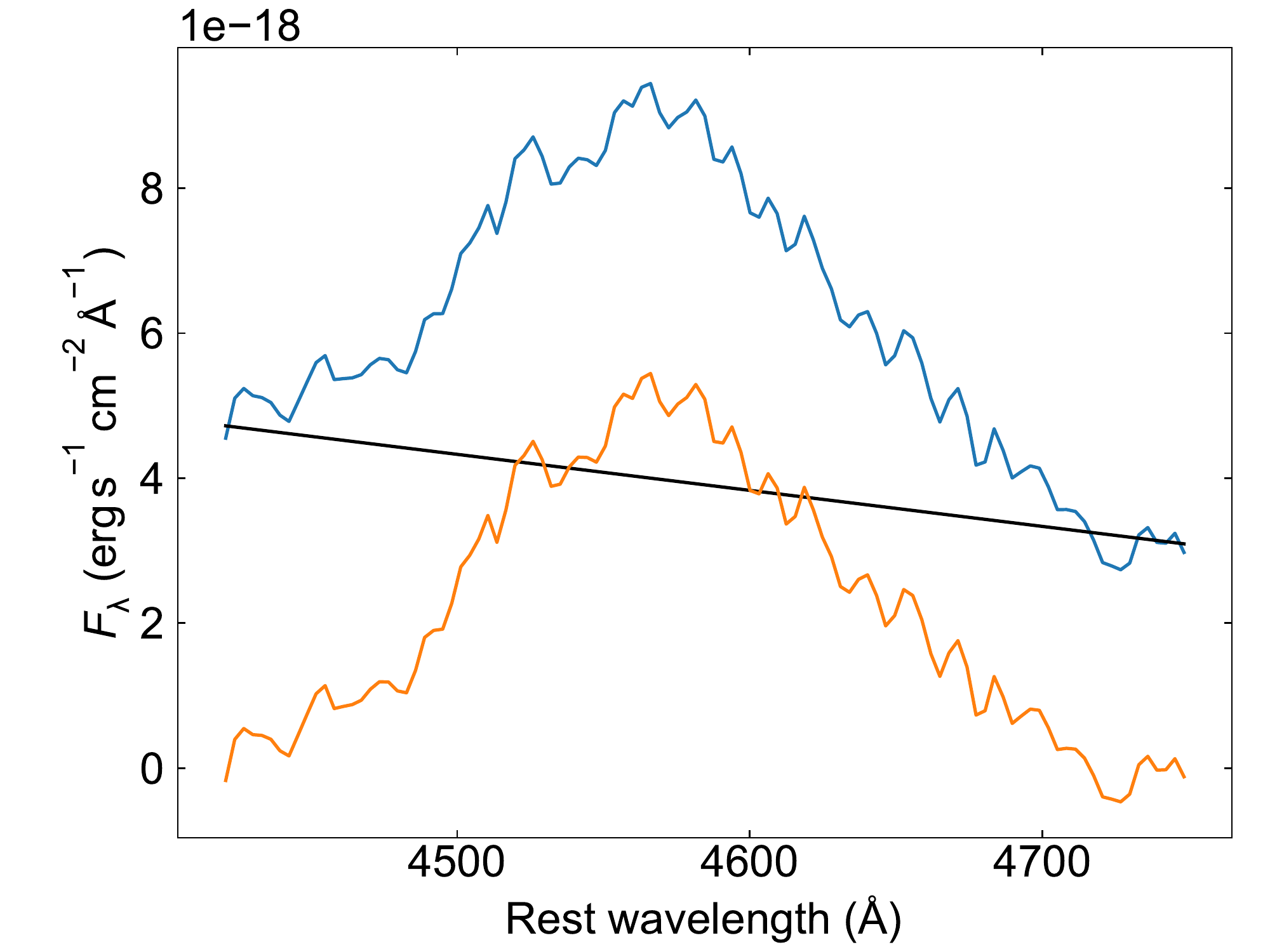}
\includegraphics[width=8cm]{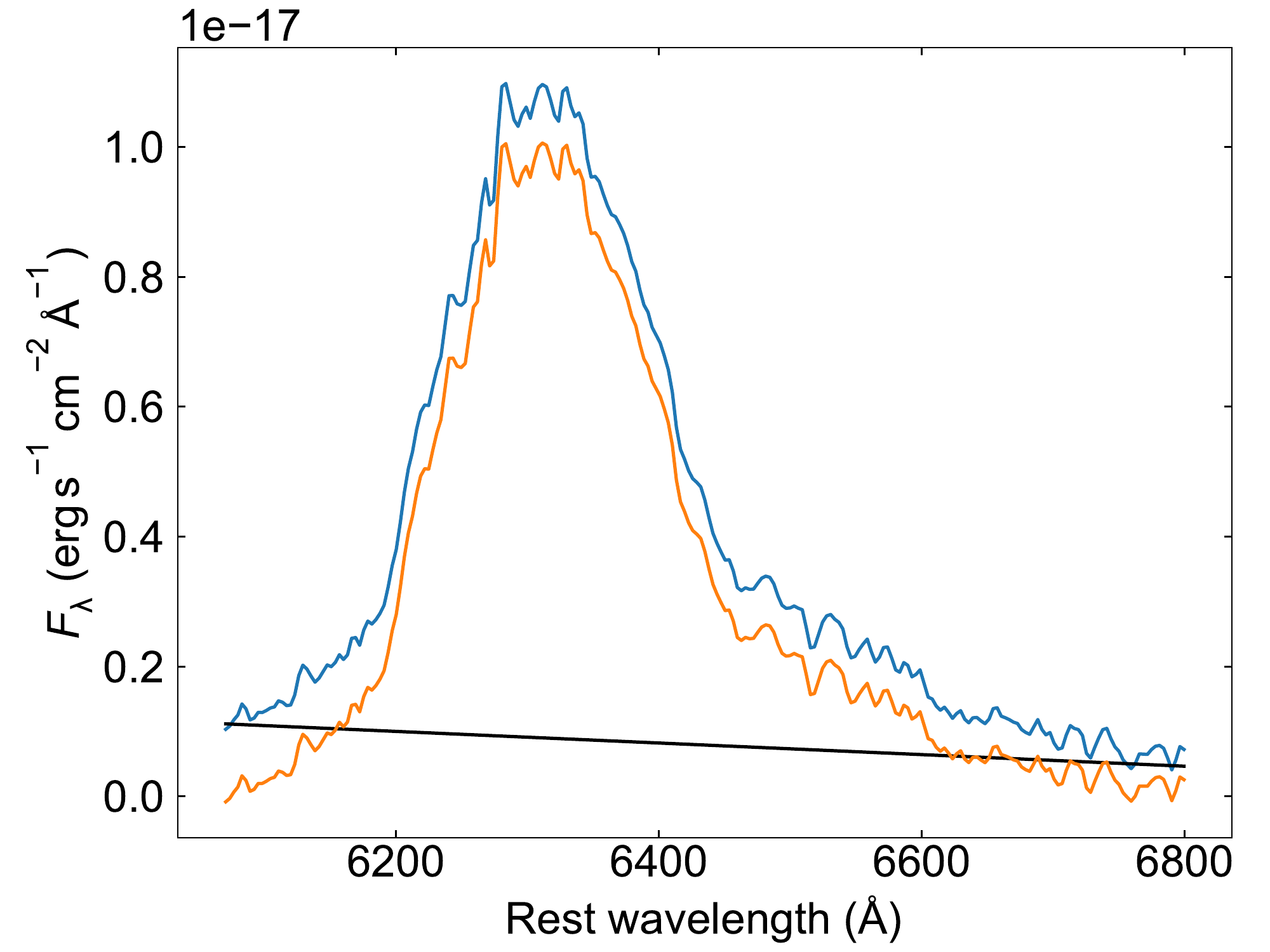}
\includegraphics[width=8cm]{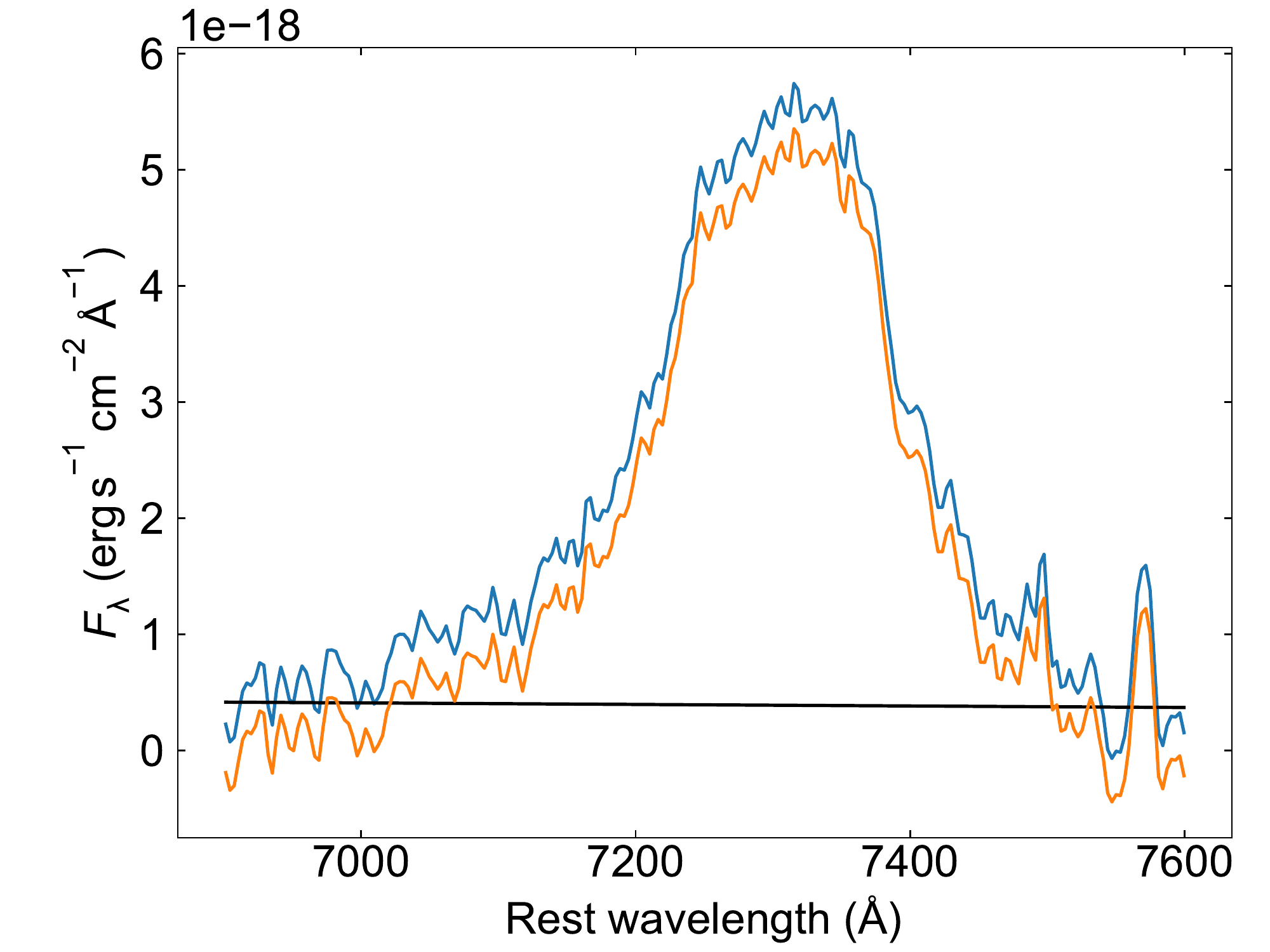}
\includegraphics[width=8cm]{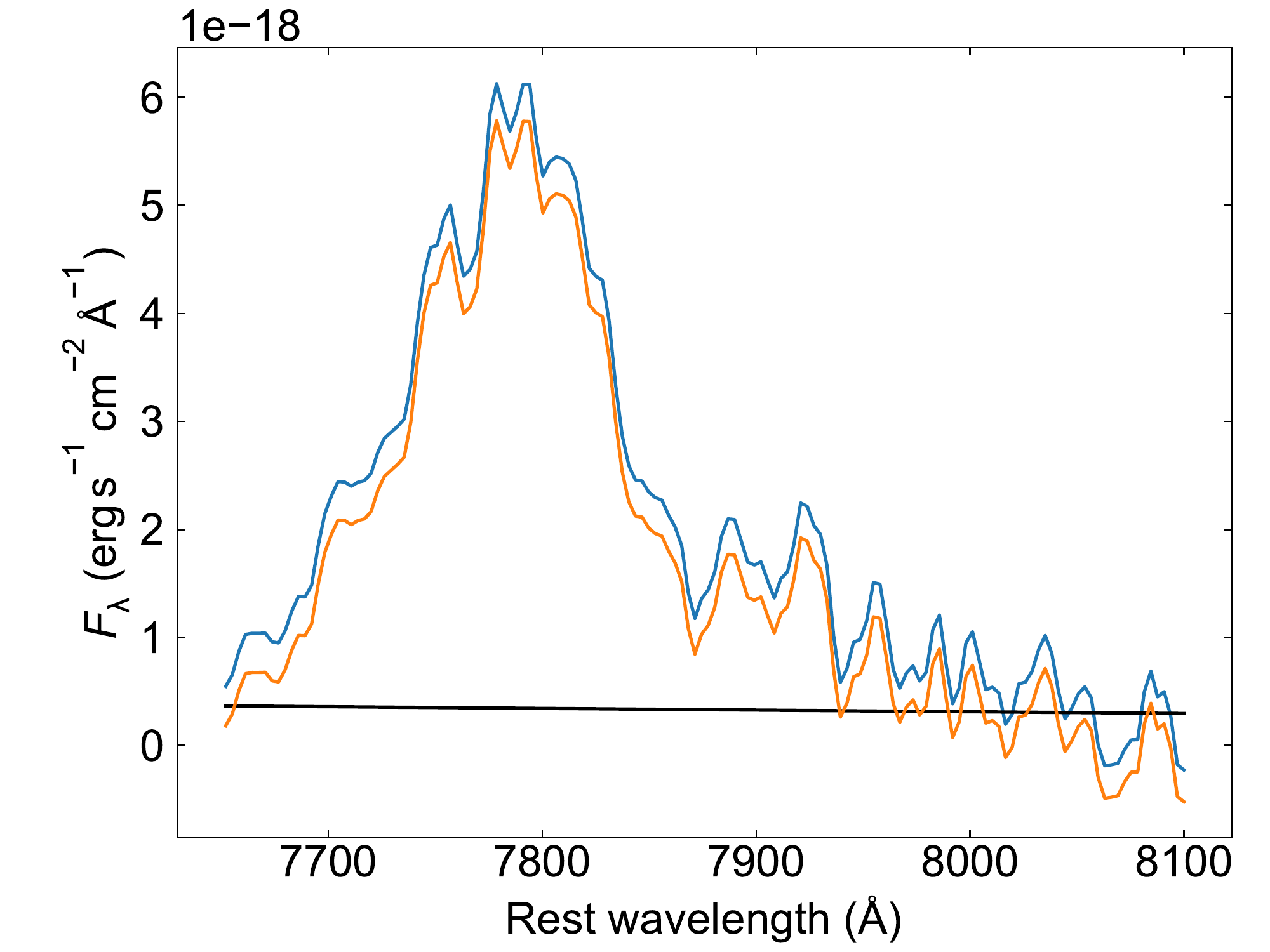}
\caption{Example of continuum subtraction, for SN\,2015bn at 392 days. Regions marked in red were used to fit a linear continuum to each line. Blue (orange) lines correspond to the spectrum before (after) continuum subtraction.}
\label{f:continuum}
\end{figure*}

\pagebreak 

\section{Additional plots}

Most of the spectra shown in this paper had been smoothed using the process described in section \ref{s:smooth}. In Figure \ref{f:raw}, we plot the original unsmoothed spectra for reference. The analysis also made use of a comparison sample of SNe Ic, obtained from the Open Supernova Catalog. These spectra are shown in Figure \ref{f:ic}. 

\renewcommand\thefigure{\thesection\arabic{figure}}
\setcounter{figure}{0}

\begin{figure*}
\centering
\includegraphics[width=16cm]{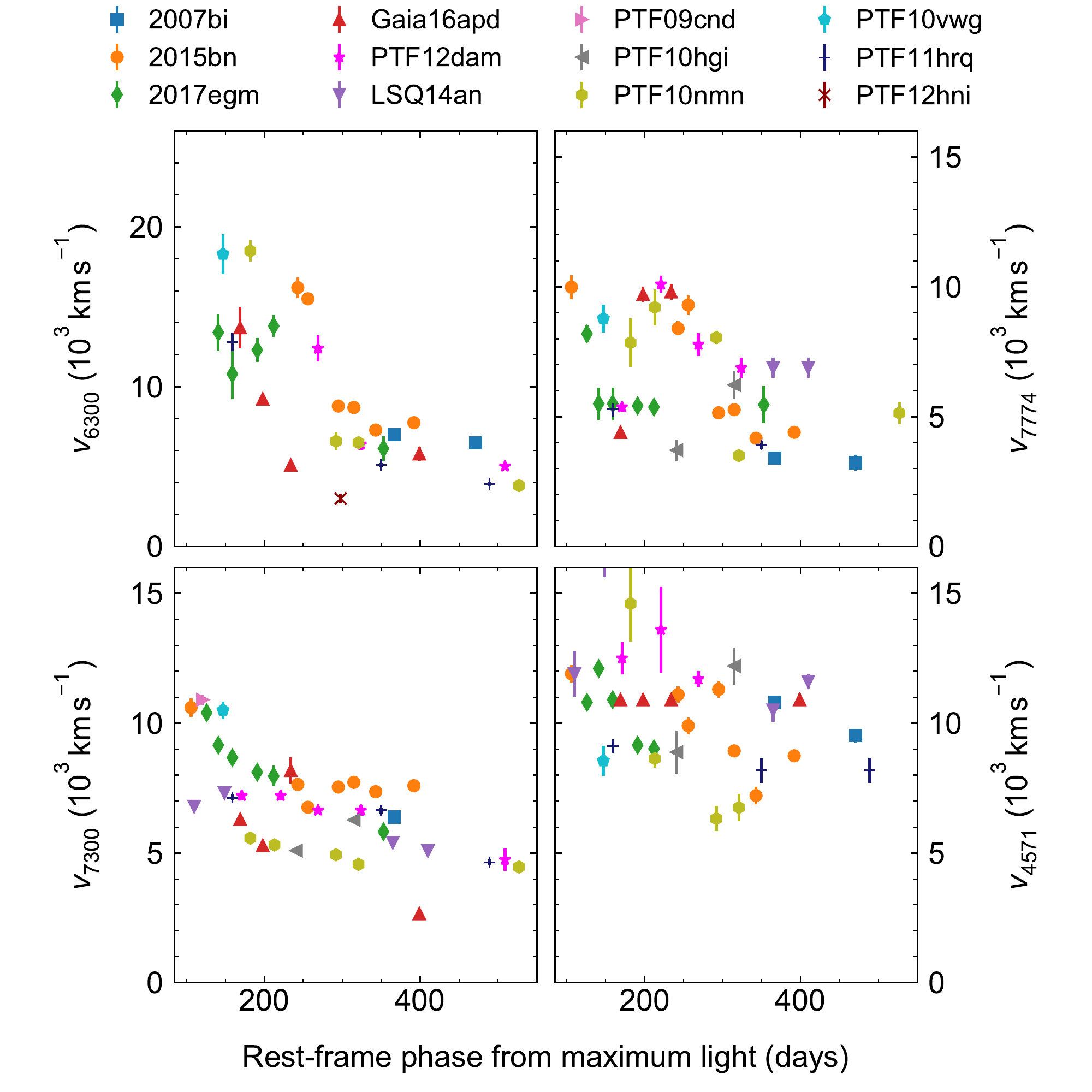}
\caption{All velocity measurements plotted against phase from maximum light.}
\label{f:v}
\end{figure*}

\begin{figure*}
\centering
\includegraphics[width=16cm]{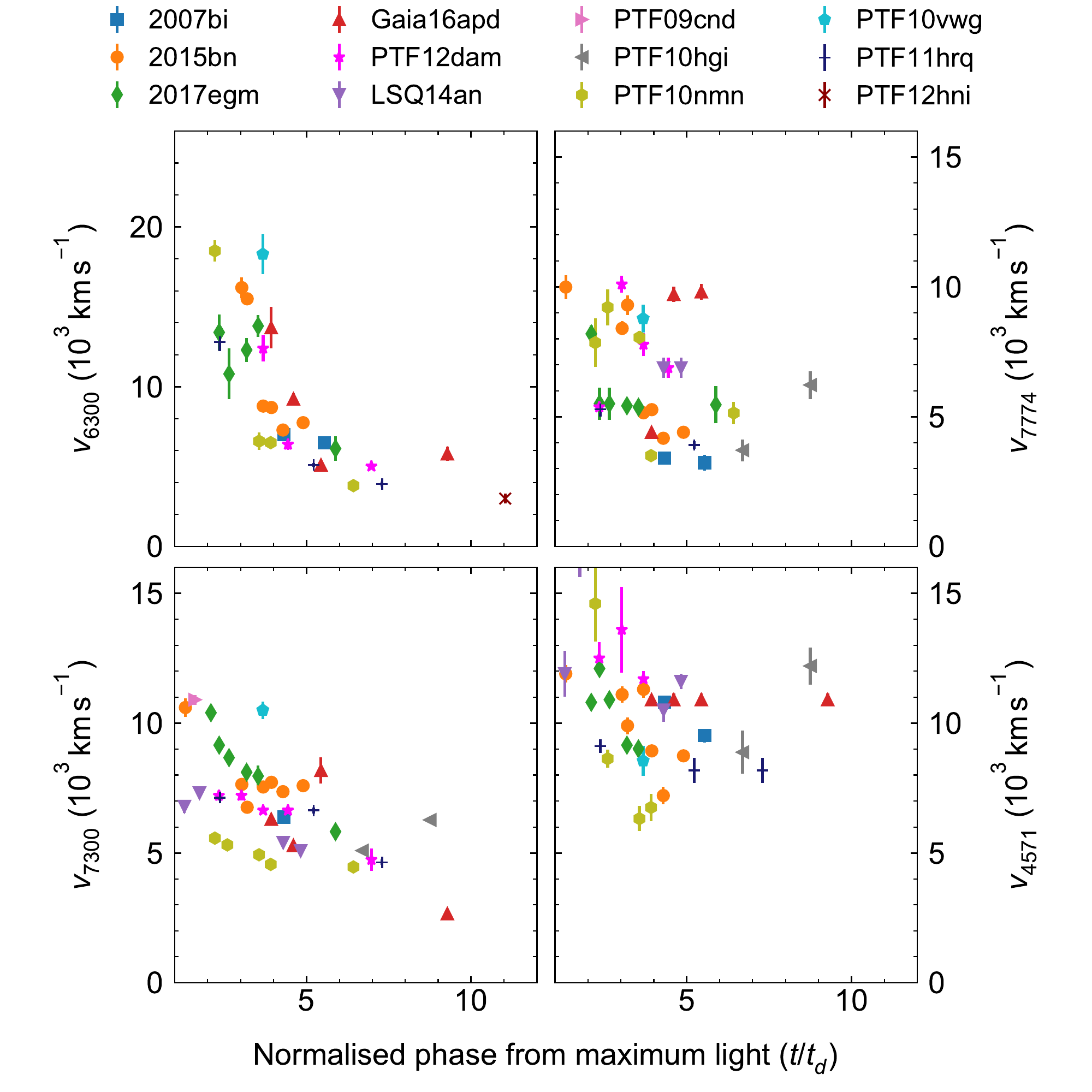}
\caption{Same as Figure \ref{f:v}, but plotted in terms of normalised phase, $t/t_d$.}
\label{f:vtd}
\end{figure*}

\begin{figure*}
\centering
\includegraphics[width=16cm]{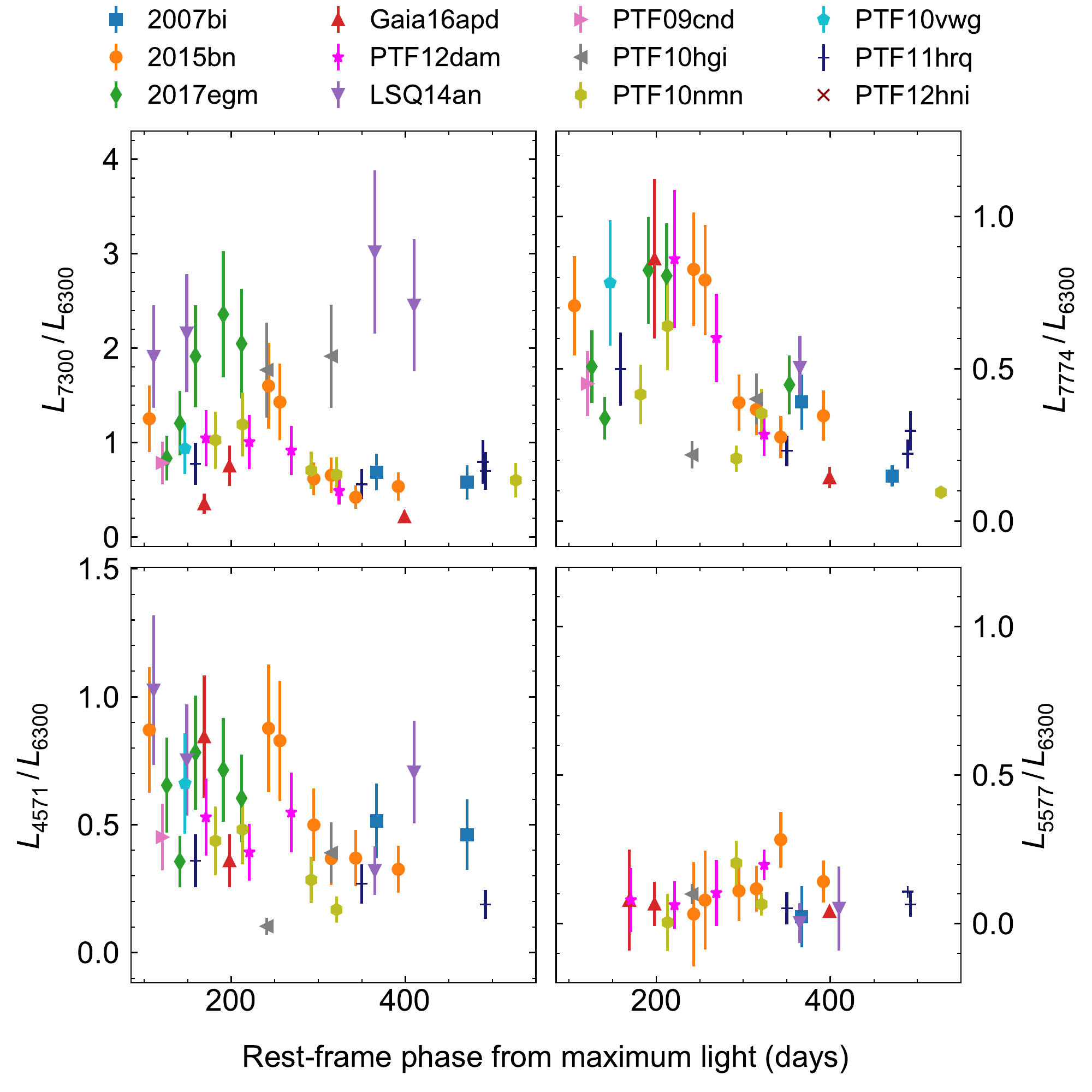}
\caption{Ratios of selected spectral lines to \o.}
\label{f:r-all}
\end{figure*}

\begin{figure*}
\centering
\includegraphics[width=16cm]{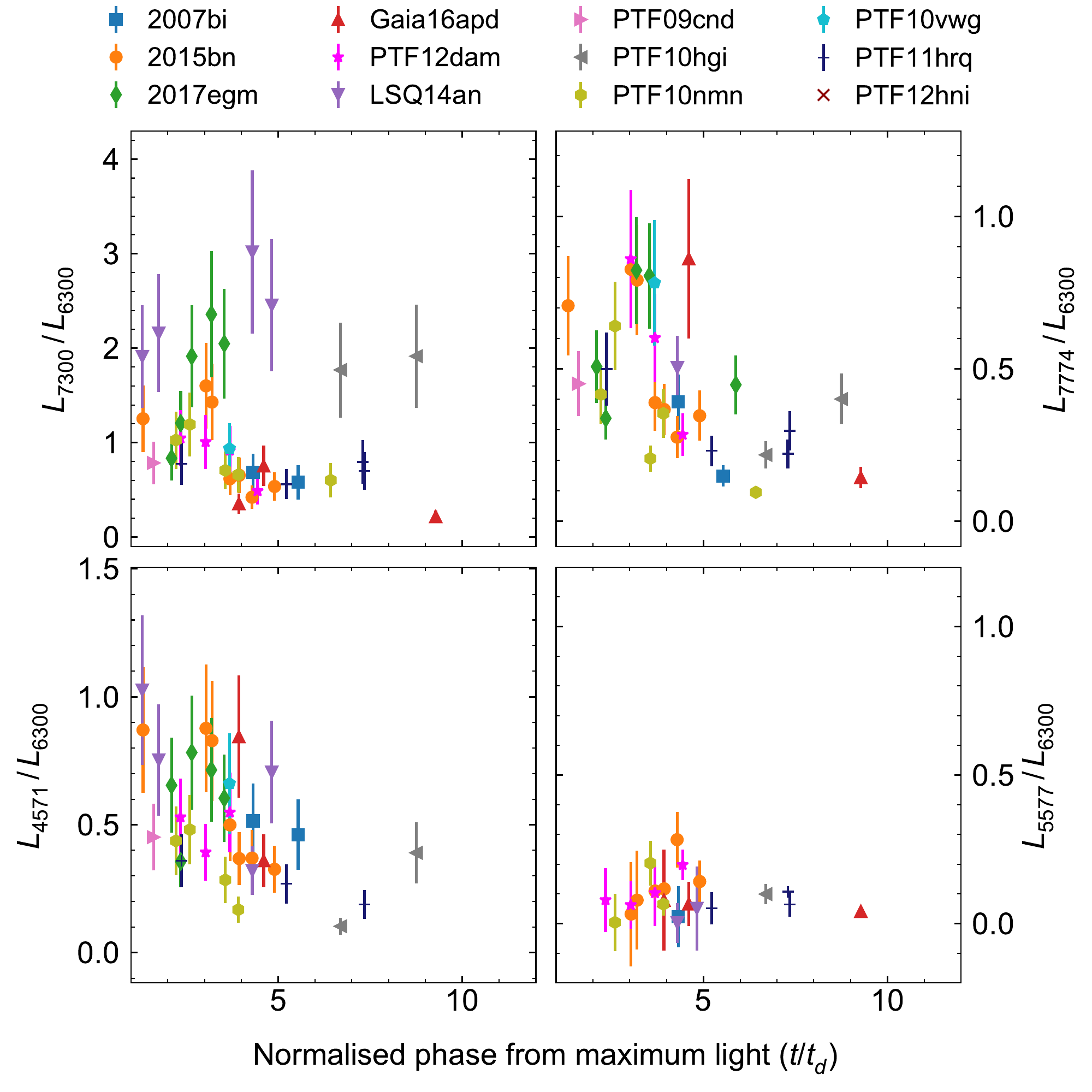}
\caption{Same as Figure \ref{f:r-all}, but plotted in terms of normalised phase, $t/t_d$.}
\label{f:r-td}
\end{figure*}

In sections \ref{s:profiles} and \ref{s:lums}, we plotted selected velocities and line ratios against the phase from maximum light, with and without phase normalisation. For completeness, we here show how all velocities (Figures \ref{f:v} and \ref{f:vtd}) and ratios (Figures \ref{f:r-all} and \ref{f:r-td}) in our sample evolve as a function of both rest-frame phase and the normalised phase $t/t_d$, where $t_d$ is the exponential decline timescale of the light curve.

\begin{figure*}[h!]
\centering
\vspace{5em}
\includegraphics[width=16cm]{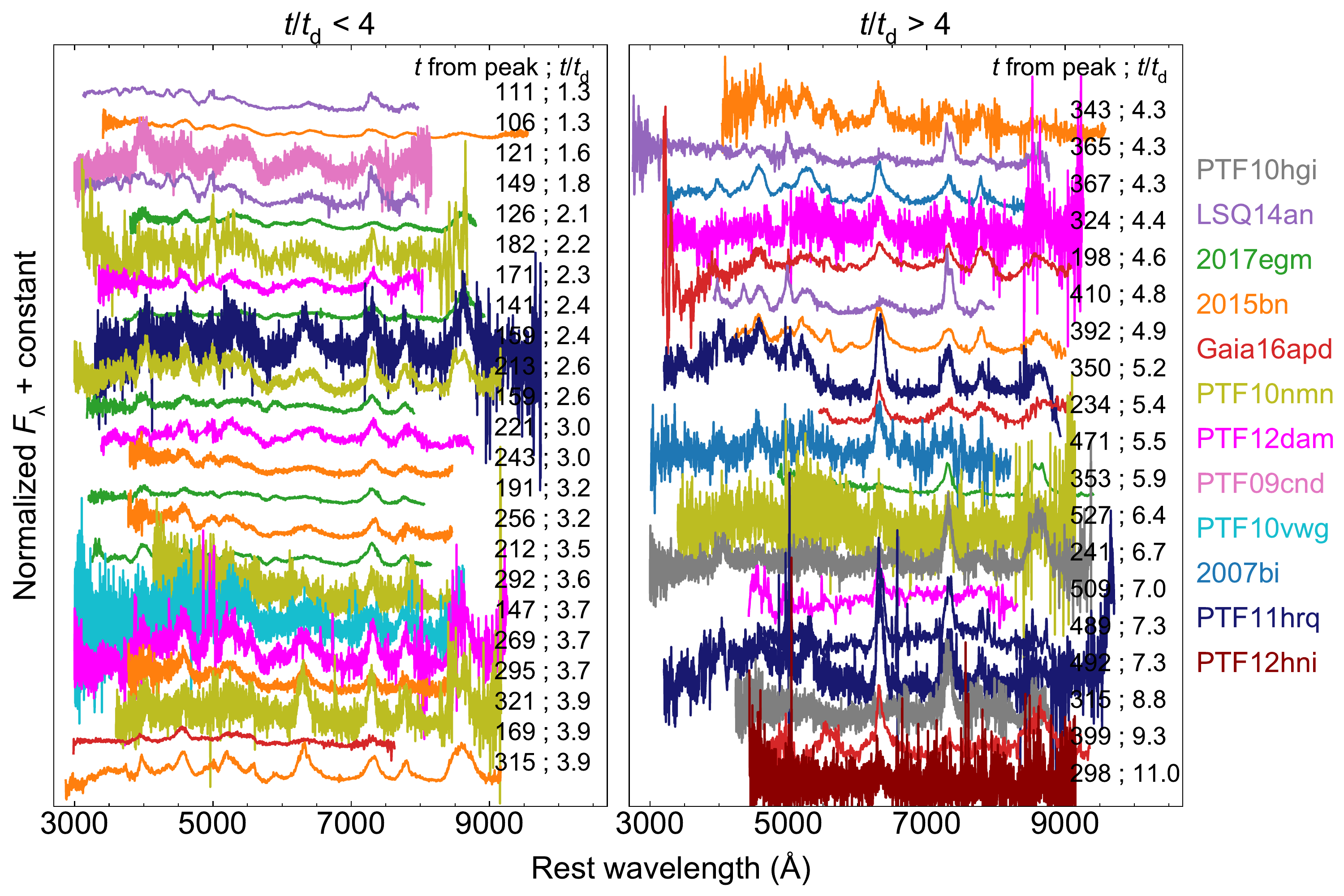}
\caption{Unsmoothed spectra of SLSNe. The only processing here is removal of host galaxy light and correction to rest-frame.}
\label{f:raw}
\end{figure*}

\begin{figure*}
\centering
\includegraphics[width=16cm]{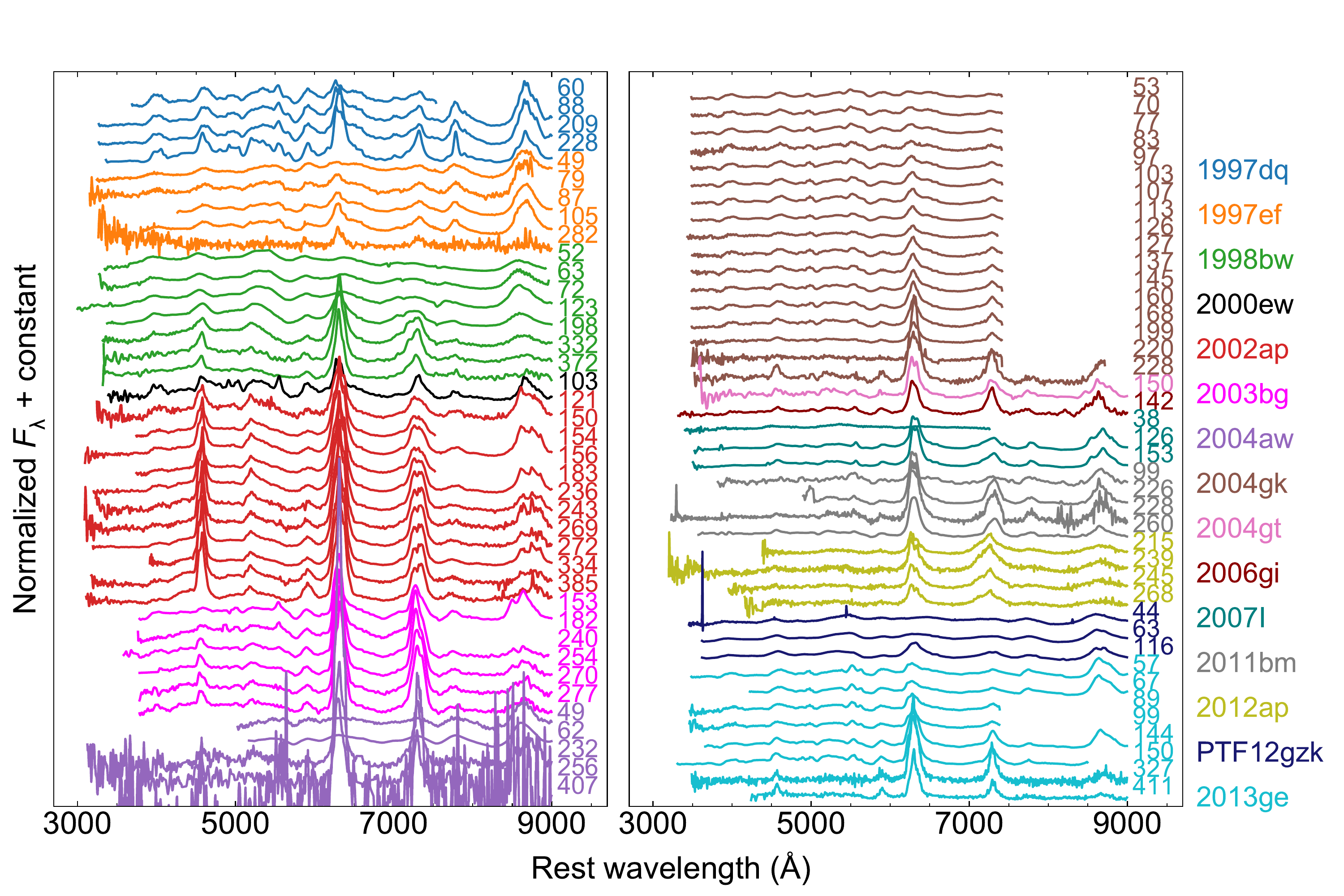}
\caption{Comparison sample of normal and broad-lined Type Ic SNe.}
\label{f:ic}
\end{figure*}

\end{document}